\documentclass[a4paper,11pt]{article}

\pdfoutput=1 % if your are submitting a pdflatex (i.e. if you have
\usepackage{jheppub} % for details on the use of the package, please
\usepackage[T1]{fontenc} % if needed

\allowdisplaybreaks

\title{\boldmath NNNLLA BFKL pomeron eigenvalue in the planar ${\mathcal N}=4$ SYM theory}

\author{V.N. Velizhanin}

\affiliation{\it Theoretical Physics Division\\
NRC ``Kurchatov Institute''\\
Petersburg Nuclear Physics Institute\\
Orlova Roscha, Gatchina\\
188300 St.~Petersburg, Russia}

\emailAdd{velizh@thd.pnpi.spb.ru}

\newcommand{\lambdai}{\lambda_i}
\newcommand{\Pa}{{\mathbf{P}}_a(u)}
\newcommand{\PA}{{\mathbf{P}}^a(u)}
\newcommand{\bP}{{\mathbf {P}}}
\newcommand{\Qi}{{\mathbf {Q}}_i(u)}

\newcommand{\bQ}{{\mathbf {Q}}}
\newcommand{\ca}{c_a}
\newcommand{\cN}{{\mathcal{N}}}
\newcommand{\N}{N}
\newcommand{\Cai}{C_{a|i}}

\def\-{\scalebox{0.6}[1.0]{\( - \)}}
\newcommand{\w}{\mathit{w}}
\newcommand{\nN}{\mathfrak{n}}

\def\b#1{{{\mathbf{#1}}}}
\def\t#1{{{\underline{#1}}}}
\def\z#1{{{\zeta_#1}}}
\def\zz#1#2{{{\zeta_{#1#2}}}}

\def\h#1{{{{\mathrm h}_#1}}}
\def\hh#1#2{{{{\mathrm h}_{#1#2}}}}
\def\hhh#1#2#3{{{{\mathrm h}_{#1#2#3}}}}
\def\hhhh#1#2#3#4{{{{\mathrm h}_{#1#2#3#4}}}}
\def\hhhhh#1#2#3#4#5{{{{\mathrm h}_{#1#2#3#4#5}}}}
\def\hB#1{{{{\mathrm {h}}_{{\mathrm{B}}#1}}}}

\newcommand{\sign}{\mathrm{sgn}}
\newcommand{\MZV}{\mathrm{MZV}}
\newcommand{\Ctl}{G}
\newcommand{\Li}{\mathrm{Li}}
\newcommand{\kk}{k}
\newcommand{\HSU}{\mathcal S}
\newcommand{\HSB}{\mathbf S}
\newcommand{\HSBI}{\mathfrak S}
\newcommand{\HSH}{S}

\newcommand{\Qai}{{\mathcal {Q}}_{a|i}(u)}

\newcommand{\QAj}{{\mathcal {Q}}^{a|j}(u)}

\newcommand{\bij}{b_i^j(u)}

\newcommand{\be}{\begin{equation}}
\newcommand{\ee}{\end{equation}}
\newcommand{\bea}{\begin{eqnarray}}
\newcommand{\eea}{\end{eqnarray}}

\abstract{We find the eigenvalue of the kernel of BFKL equation in the next-to-next-to-next-to-leading logarithm approximation (NNNLLA) in the planar $\cN=4$ SYM theory by using the quantum spectral curve to compute values at fixed spin, and reconstructing the general result using the LLL-algorithm, which also was used for the reconstruction of the general result for the intercept function for arbitrary conformal spin at fourth order, computed  earlier in Ref.\cite{Alfimov:2018cms} up to the rational part. We have found, that a new type of harmonic sums enter into results. Those sums have the argument multiplied by factor two with compare to usual harmonic sums and contain the last imagine index, what reflect their relations with the multiple polylogarithms, generated by ``fourth root of unity''. The final result contains Catalan constant, which did not appear early in the calculations for the anomalous dimension of twist-2 operators.}

\begin{document} 
\maketitle
\flushbottom

\section{Introduction}

The Balitsky-Fadin-Kuraev-Lipatov (BFKL) equation~\cite{Lipatov:1976zz,Kuraev:1977fs,Balitsky:1978ic,Fadin:1998py} was obtained during the study of the Regge processes at high energies $\sqrt{s}$ in the non-abelian gauge theories.
In this kinematics, when a transferring momentum is very small, the large logarithms appear and they should be summed in all orders of perturbative theory.
Thus, the BFKL equation in the leading-logarithm approximation sums all leading logarithmic terms $(\alpha_s\ln 1/x)^\ell$ in all order of the perturbative theory. In this leading approximation only ladder diagrams give the contribution, which can be evaluated with the Sudakov decomposition order by order and it is possible to write such corrections in a general form with the help of Bethe-Salpeter equation for the partial wave, which is known as BFKL equation.
As described, for example, in Ref.~\cite{Fadin:1998py}, the BFKL equation allows to estimate the total cross-section $\sigma (s)$ for the high energy scattering of colourless particles $A,B$ 
\begin{equation}
\sigma (s)=
\int \frac{d^2q_1}{2\pi q_1^2}
\int \frac{d^2q_2}{2\pi q_2^2}
\,\Phi _A(\vec{q}_1)
\,\,\Phi _B(\vec{q}_2)
\,\int_{a-i\infty }^{a+i\infty }\frac{d\omega }{2\pi i}
\,\left(
\frac{s}{q_1\,q_2}
\right) ^\omega \,G_\omega (\vec{q}_1,\vec{q}_2)\,,  \label{CrossS}
\end{equation}
where $G_\omega (\vec{q}_1,\vec{q}_2)$ is the $t$-channel partial wave for the reggeized gluon scattering at $t=0$ and $\vec{q}_1$ and $\vec{q}_2$ are transverse momenta of gluons with the virtualities $-\vec{q}_1^{\;2}\equiv -q_1^{2}$ and $-\vec{q}_2^{\;2}\equiv -q_2^2$ correspondingly, $s=2p_Ap_B$ is the squared invariant mass of the colliding particles with momenta $p_A$ and $p_B$. 
The generalized BFKL equation for $G_\omega (\vec{q}_1,\vec{q}_2)$ can be written in the following form
\begin{equation}
\omega \,G_\omega (\vec{q}_1,\vec{q}_2)=\delta
^{D-2}(\vec{q}_1-\vec{q}_2)+\int d^{D-2}
{q}\ K(\vec{q}_1,\vec{{q}})\,G_\omega (\vec{{q}},\vec{q}_2)\,,
\label{l2}
\end{equation}
where 
\begin{equation}
K(\vec{q}_1,\vec{{q}}_2)=2\,\omega (q_1)\,\delta^{(D-2)}(\vec{q}_1-\vec{q}_2)
+K_r(\vec{q}_1,\vec{q}_2)\,.  \label{l3}
\end{equation}
%and the space-time dimension is $D=4+2\varepsilon $ for $\epsilon \rightarrow 0$. 
The gluon Regge trajectory $\omega (q)$ and the integral kernel $K_r(\vec{q}_1,\vec{q}_2)$ are expanded in the series over the QCD coupling constant 
\begin{equation}
\omega (q)=\omega _B(q)+\omega ^{(2)}(q)+...\,,\qquad 
K_r(\vec{q}_1,\vec{q}_2)=K_r^B(\vec{q}_1,\vec{q}_2)
+K_r^{(1)}(\vec{q}_1,\vec{q}_2)+...\,\,,\,.
\label{l4}
\end{equation}
The gluon Regge trajectory $\omega(q)$ and the integral kernel $K_r^{(1)}(\vec{q}_1,\vec{q}_2)$ can be found up to the next-to-leading logarithm approximation in Ref.~\cite{Fadin:1998py}.

As it was shown in~\cite{Balitsky:1978ic}, a complete and orthogonal set of eigenfunctions of the homogeneous BFKL equation in the leading-logarithm approximation (LLA) is 
\begin{equation}
G_{n,\gamma }(\vec{q},\vec{q}_2)\ =\ 
\left( \frac{q^{2}}{q_2^{2}}\right) ^{\gamma -1}\,.
\end{equation}
The BFKL kernel in this representation is diagonalised up to the effects
related with the running coupling constant $\alpha _{s}(q^{2})$: 
\begin{equation}
\omega =\frac{\alpha _{s}(q^{2})N_{c}}{\pi }\biggl[ \chi (n,\gamma )+\delta
(n,\gamma )\frac{\alpha _{s}(q^{2})N
_{c}}{4\pi }\biggr] \,.
\end{equation}
In this paper we will mainly consider the case, when the conformal spin $n=0$, only in Section~\ref{Section:ConfSpin}, where we consider $\gamma=0$ and $n\neq 0$ case, using results from Ref.~\cite{Alfimov:2018cms}.

To find the eigenvalue of the kernel of BFKL equation one can used the eigenfunctions $q_1^{2(\gamma -1)}$ of the Born kernel:
\begin{equation}
\omega=\int d^{D-2}q\ K(\vec{q}_1,\vec{q})
\left(\frac{q^2}{q_1^2}\right) ^{\gamma-1}=\frac{\alpha _s(q_1^2)\,N_c\,}\pi
\left( \chi (\gamma )+\delta (\gamma )\frac{\alpha _s(q_1^2)N_c}{4\,\pi }
\right) \,,\, \label{l12}
\end{equation}
The quantity $\chi (\gamma ) $ is proportional to the eigenvalue of the Born kernel
\begin{equation}
 \chi (\gamma )=2\Psi (1)-\Psi (\gamma )-\Psi (1-\gamma )\,,\,\,\,\Psi
 (\gamma )=\Gamma ^{\prime }(\gamma )/\Gamma (\gamma )\,,  \label{BFKLLLA}
\end{equation}
 and  the correction $\delta (\gamma)$ is given by~\cite{Fadin:1998py}
\begin{eqnarray}
\delta (\gamma )&=&-\left[ \left( \frac{11}3-\frac{2n_f}{3N_c}\right) \frac
12\left( \chi ^2(\gamma )-\Psi ^{\prime }(\gamma )+\Psi ^{\prime }(1-\gamma
)\right) -\left( \frac{67}9-\frac{\pi ^2}3-\frac{10}9\frac{n_f}{N_c}\right)
\chi (\gamma )\right. 
\nonumber\\[2mm]&&
\left. -6\zeta (3)+\frac{\pi ^2\cos(\pi \gamma )}{\sin^2(\pi \gamma
)(1-2\gamma )}\left( 3+\left( 1+\frac{n_f}{N_c^3}\right) \frac{2+3\gamma
(1-\gamma )}{(3-2\gamma )(1+2\gamma )}\right) \right. 
\nonumber\\[2mm]&&
\left. -\Psi ^{\prime \prime }(\gamma )-\Psi ^{\prime \prime }(1-\gamma )
-
\frac{\pi ^3}{\sin (\pi \gamma )}+4\phi (\gamma )\right] \,.  \label{l12a}
\end{eqnarray}
The function $\phi(\gamma )$ is 
\begin{eqnarray}
\phi (\gamma )&=&-\int_0^1\frac{dx}{1+x}\left( x^{\gamma -1}+x^{-\gamma
}\right) \int_x^1\frac{dt}t\ln (1-t) \nonumber\\[2mm]
&=&\sum_{n=0}^\infty (-1)^n\left[ \frac{\Psi (n+1+\gamma )-\Psi (1)}{(n+\gamma
)^2}+\frac{\Psi (n+2-\gamma )-\Psi (1)}{(n+1-\gamma )^2}\right] \,.
\label{l14}
\end{eqnarray}

The BFKL equation in the leading logarithm approximation is the same in any gauge theory and has a lot of remarkable properties.
For example, the integrability in the quantum field theory was firstly discovered by L.N. Lipatov during the study of the BFKL equation~\cite{Lipatov:1993yb,Lipatov:1994xy}. 

The generalisation of the computations of the BFKL equation in the next-to-leading-logarithm approximation, performed by  V.S. Fadin and L.N. Lipatov in QCD~\cite{Fadin:1998py}, to the maximally extended $\cN=4$ supersymetric Yang-Mills (SYM) theory shows~\cite{Kotikov:2000pm}, that a lot of terms in the QCD result~(\ref{l12a}) are cancelled and the final result contains the functions, which have the same property called later as a transcedentality\footnote{Their large $\gamma$ limit coincides with the special transcendental numbers such as zeta-numbers $\zeta_i$}:
\begin{eqnarray}
\delta (\gamma )&=&
\Psi ^{\prime \prime }(\gamma )+\Psi ^{\prime \prime }(1-\gamma )
+6\zeta (3)
+
\frac{\pi ^3}{\sin (\pi \gamma )}
-4\phi (\gamma ) \,.  
\label{BFKLNLLA}
\end{eqnarray}
Using the same suggestion the results for the anomalous dimension of the twist-2 operators in $\cN=4$ SYM theory was obtained without any computation~\cite{Kotikov:2002ab}, but argued from the relation between BFKL and Dokshitzer-Gribov-Lipatov-Altarelli-Parizi (DGLAP)~\cite{Gribov:1972ri,Altarelli:1977zs,Dokshitzer:1977sg} equations. The maximal transcedentality principle was confirmed by the direct diagrammatic calculations at two loops~\cite{Kotikov:2003fb} and then successfully used for the finding the three-loop anomalous dimension~\cite{Kotikov:2004er} from the corresponding result, computed directly in QCD~\cite{Moch:2004pa}.
This result help to confirm a general form of the asymptotic Bethe-ansatz~\cite{Beisert:2004hm}, which can be used for the computations of the anomalous dimension of composite operators in the $\cN=4$ SYM theory. Then this maximal transcedentality principle was used for the computations of the general form of the anomalous dimension for twist-2 operators as with the help of integrability~\cite{Staudacher:2004tk,Kotikov:2007cy,Bajnok:2008qj,Lukowski:2009ce,Marboe:2014sya,Marboe:2016igj} as from the constraints coming from the generalised double-logarithmic equation~\cite{Velizhanin:2011pb,Velizhanin:2013vla}. It is no doubt, that the maximal transcedentality principle works for the BFKL equation at higher orders too, but the direct diagrammatic computations were very cumbersome. In principle, the information from the result for six-loop anomlous dimension of twist-2 operators in planar $\mathcal{N}=4$ SYM theory from Ref.~\cite{Marboe:2014sya} is enough to reconstruct the eigenvalue of the kernel of BFKL equation in the next-to-next-to-leading approximation (NNLLA) considering Fadin-Lipatov function (see Ref.~\cite{Velizhanin:2015xsa}). However, this result was obtained earlier in Ref.~\cite{Gromov:2015vua} with the novel approach to the integrability, called the Quantum Spectral Curve~\cite{Gromov:2013pga,Gromov:2014caa}. We used the QSC approach for the computations of the analytical expression for the six- and seven-loop anomalous dimensions of twist-2 operators in planar $\mathcal{N}=4$ SYM theory~\cite{Marboe:2014sya,Marboe:2016igj}, but it can be used for the computations of anomalous dimension for any complex Lorentz spin of operators, that is, for example, directly in the BFKL case. Later was done in Refs.~\cite{Alfimov:2014bwa,Gromov:2015wca,Gromov:2015vua,Alfimov:2018cms}.

In order to relate the BFKL eigenvalue in $\mathcal{N}=4$ SYM theory with QSC approach one can remember, that $\gamma$ in Eq.~(\ref{BFKLLLA}) is the anomalous dimension of twist-2 gluon operator, which is analytically continued into $J=1$~\cite{Kwiecinski:1985cq}.
In $\mathcal{N}=4$ SYM theory usualy the dimension $\Delta(S)$ of following twist-two operator ${\cal O}=\textrm{Tr} ZD_+^SZ$ is considered~\cite{Alfimov:2014bwa,Gromov:2015vua} in the similar case.
The inverse function $S(\Delta)$ is known to approach $-1$ perturbatively for $\Delta$ and the relation to the BFKL regime is given by $\Delta=i\nu$ and $j=2+S(\Delta)$~\cite{Brower:2006ea} and one need to compute $j(\Delta)$ as a series expansion in $g^2$.
Indeed, from the QSC formalism it was shown in \cite{Alfimov:2014bwa} that one reproduces correctly the LLA result from Eq.~(\ref{BFKLLLA}).

Then the expansion of $j(\Delta)$ can be written as
\begin{equation}
j(\Delta)=1+\sum\limits_{\ell=1}^\infty g^{2\ell}\left[F_\ell\left(\frac{\Delta-1}{2}\right)+F_\ell\left(\frac{-\Delta-1}{2}\right)\right]
 \label{BFKLj}
\end{equation}
with the three first known orders given by \cite{Costa:2012cb,Gromov:2015wca}
\begin{eqnarray}
F_1&=&-4 S_1\,,\label{LLA}\\
\frac{F_2}{4}&=&-\frac{3}{2}\zeta_3+\pi^2\ln\!2+\frac{\pi^2}{3}S_1+2S_3+\pi^2 S_{-1}-4 S_{-2,1}\,,\label{NLLA}\\
\frac{F_3}{16}&=&
16\, \hh31 \ln\!2
-16\, \hhh311
+8\, \z5
-2 \pi ^2 \z3\nonumber\\&&
+\left(\HSH_{-1}-\HSH_{1}\right) \left(
16\, \hh31
-28\, \z3\ln\!2 
%-\frac{17 \pi ^4}{45}
\right)
-\frac{17\pi ^4}{45}\left(S_{-1}+ \ln\!2 \right)\nonumber\\&&
+\HSH_1 \left(
4 \left(
2 \HSH_{-3,1}
+\HSH_{-2,2}
+\HSH_{2,-2}
-4 \HSH_{-2,1,1}
%+4 \hh31
\right)
-\frac{\pi ^2}{3} \left(4 \HSH_{-2}-\HSH_2\right) 
%+28 \ln 2 \z3
-\frac{\pi ^4}{6}\right)\nonumber\\&&
-\frac{\pi ^2}{3}  \Big(
7 \HSH_{-3}-6 \HSH_{-2,1}
%-32 \z3
\Big)
-\z3 \left(
-28 \HSH_{1,-1}
+14 \HSH_{-2}
+\HSH_2
+14 \ln\!2^2
\right)\nonumber\\&&
-\frac{2}{3} 
\bigg(
3 \HSH_5
+\HSH_{-4,1}-15 \HSH_{-3,2}
-12 \HSH_{-2,3}
+16 \HSH_{2,-3}
+36 \HSH_{-3,1,1}
+30 \HSH_{-2,1,2}
\nonumber\\&&
\qquad\
+11 \HSH_{-2,2,1}
-13 \HSH_{2,-2,1}
-19 \HSH_{2,1,-2}
-72 \HSH_{-2,1,1,1}
-72 \HSH_{-2,1,1,1}
\bigg)\qquad\label{NNLLA}
\end{eqnarray}
where the harmonic sums can be defined recursively by (see \cite{Vermaseren:1998uu})
\begin{eqnarray}
S_a (M)=\sum^{M}_{j=1} \frac{(\mbox{sign}(a))^{j}}{j^{\vert a\vert}}\, ,\,\,\,\,\,\,\,\,\,\,\,\,\,\,\,\,\,\,\,\,
S_{a_1,\ldots,a_n}(M)=\sum^{M}_{j=1} \frac{(\mbox{sign}(a_1))^{j}}{j^{\vert a_1\vert}}
\,S_{a_2,\ldots,a_n}(j)\, .\label{vhs}
\end{eqnarray}
To each sum $S_{a_1,\ldots,a_n}$ we assign a weight $\ell$ (or transcendentality), which is given by the sum of the absolute values of its indices
\begin{equation}
\ell=\vert a_1 \vert +\ldots \vert a_n \vert\,,
\end{equation}
and the weight of a product of harmonic sums equals the sum of the weights of its factors.
The maximal transcendentality principle~\cite{Kotikov:2002ab} states that, at a given order of perturbative theory, the anomalous dimension of twist-2 operators contains only harmonic sums with maximal transcendentality (with weight $2\ell\-1$ for the $\ell$th-order).

In this paper we compute, following Ref.~\cite{Gromov:2015vua}, the BFKL eigenvalue at forth order (NNNLLA or next-to-next-to-next-to-leading logarithm approximation) in the planar  \mbox{$\mathcal{N}=4$} SYM theory. We reproduced the results from Ref.~\cite{Gromov:2015vua} and performed computations for fixing values of $\Delta$ at order $g^8$. Using maximally transcedentality principle and results for the computed fixed values we found with the help of the number theory the general result for the arbitrary value of $\Delta$. In Section~\ref{Sec:QSC} we give the general description of QSC-approach for our purposes. Section~\ref{Sec:Comp} contains all detailed information about procedure of computations, with description all steps and representation of the corresponding results. In Section~\ref{Sec:Rec} we describe the reconstruction procedure. We start with the direct extension of the previous general results and introduce the harmonic sums with double arguments and harmonic sums with the last imagine index, which extend the basis from the usual harmonic sums. Only extended basis allowed to obtain the general result for arbitrary $\Delta$. Using the obtained earlierly result from Ref.~\cite{Alfimov:2018cms} for the fixing values in the case $n\neq0$ and $\gamma=0$, we find in Section \ref{Section:ConfSpin}  with the help of number theory the general expression for the rational part at fourth order, which was missed in Ref.~\cite{Alfimov:2018cms}. In Conclusion we provide the our main results of this paper and discus some aspects of the obtained results.

\section{Quantum Spectral Curve for twist-2 operators in BFKL limit}\label{Sec:QSC}

A very detailed description of the Quantum Spectral Curve is given in Ref., while some more specific aspect related with our problem can be found in Ref.~\cite{Gromov:2015wca}. Below we will write down only features of QSC, required (needed) for our computations.

The QSC can be considered as the set of functional equation for the function $\Qai$, $\Qi$ and $\Pa$ of the complex variable $u$ (spectral parameter), the solution of which along with analytical properties of the obtained solutions allow to find quantity, in which we are interesting (looking) for.   

All these functions have the power-like asymptotics at large $u$, which for the basic $\bP_a$, $\bP^a$, $\bQ_i$ and $\bQ^i$ can be taken from \cite{Gromov:2014caa}
\begin{equation}
\bP_a \simeq A_a u^{-\tilde{M}_a}\;, \quad \bP^a \simeq A^a u^{\tilde{M}_a-1}\,, \quad \bQ_i \simeq B_i u^{\hat{M}_i-1}\;, \quad \bQ^i \simeq B^i u^{-\hat{M}_i}\,,\label{PQas}
\end{equation}
where for twist-2 operators $\tilde{M}_a$, $a=1,\ldots,4$ and $\hat{M}_i$, $i=1,\ldots,4$ are given by
\begin{eqnarray}
\tilde{M}_a&=&\Big\{-2,-1,0,1 \Big\}, \nonumber \\
\hat{M}_i&=&\left\{\frac{\Delta-S_1-S_2+2}{2}, \frac{\Delta+S_1+S_2}{2}, \frac{-\Delta-S_1+S_2+2}{2}, \frac{-\Delta+S_1-S_2}{2} \right\}.
\label{charges}
\end{eqnarray}
For the twist-2 operators in BFKL regime $S_2=0$ and $S_1\equiv S\equiv-1+\w$. These operators belong to the so called left-right symmetric sector
for which \cite{Gromov:2014caa}
\begin{equation}
\bP^a=\chi^{ac} \bP_c,\qquad  \bQ^i=\chi^{ij} \bQ_j,\label{LRred}
\end{equation}
where \(\chi\) is the antisymmetric constant \(4\times 4\) matrix with the only nonzero entries \(\chi^{23}=\chi^{41}=-\chi^{14}=-\chi^{32}=1\).

There are the simple relations between $A_a$, $B_i$, $\tilde{M}_a$ and $\hat{M}_i$, which have the following form \cite{Gromov:2014caa,Gromov:2017blm}:
\begin{equation}
A_{a_0} A^{a_0}=i\frac{\prod\limits_{j=1}^4 (\tilde{M}_{a_0}-\hat{M}_j)}
{\prod\limits_{b=1 \atop b \neq a_0} (\tilde{M}_{a_0}-\tilde{M}_b)}\,, \quad 
B_{i_0} B^{i_0}=-i\frac{\prod\limits_{a=1}^4 (\hat{M}_{i_0}-\tilde{M}_a)}
{\prod\limits_{j=1 \atop j \neq i_0}^4 (\hat{M}_{i_0}-\hat{M}_j)}\,, 
\quad a_0,i_0=1,\ldots,4\,,\label{AA_BB}
\end{equation}
where there is no summation over the indices $a_0$ and $i_0$ implied.
The asymptotics (\ref{PQas}) can be written as
\begin{eqnarray}
\bP_a&\simeq&\left\{A_1 u^{-2},\ A_2 u^{-1},\ A_3,\ A_4u\right\}\label{Plarge}\\[2mm] 
\bQ_i&\simeq&\left\{B_1 u^{\frac{\Delta+1-\w}{2}},\ B_2 u^{\frac{\Delta-3+ \w}{2}},\ B_3u^{\frac{-\Delta+1-\w}{2}},\ B_4u^{\frac{-\Delta-3+\w}{2 }}\right\}
\label{Qlarge}
\end{eqnarray}
and if we choose the usual normalisation conditions $A_1=A_2=B_1=B_3=1$ in (\ref{Plarge}) and (\ref{Qlarge}) we obtain
\begin{eqnarray}
A_4&=&\frac{1}{96i}((5-\w)^2-\Delta^2)((1+\w)^2-\Delta^2), \\
A_3&=&\frac{1}{32i}((1-\w)^2-\Delta^2)((3-\w)^2-\Delta^2),\\
B_2&=&\frac{1}{16i}\frac{(4-(\Delta-1+\w)^2) (4-(\Delta-3+\w)^2)}{\Delta  (2-\w) (\Delta +2-\w)},\\
B_4&=&\frac{1}{16i}\frac{(4-(\Delta+1-\w)^2) (4-(\Delta+3-\w)^2)}{\Delta  (2-\w) (\Delta +2-\w)}.
\label{A4A3B2B4}
\end{eqnarray}

Moreover, it is suggested, that the following ansatze should work for $\Pa$
\begin{eqnarray}
&& \bP_1=\frac{1}{\Lambda \w x^2}+\sum_{k=1}^{+\infty}\frac{c_{1,k}}{x^{2k+2}}, 
\quad 
%\bP^1=A^1 \sqrt{\Lambda w}\left(x+\frac{1}{x}\right)+\sum_{k=1}^{+\infty}\frac{c^{1,k}}{x^{2k-1}}\;, \\
%&& 
\bP_2=\frac{1}{\sqrt{\Lambda \w}x}+\sum_{k=2}^{+\infty}\frac{c_{2,k}}{x^{2k+1}}\;, \nonumber\\ 
%\bP^2=A^2+\sum_{k=2}^{+\infty}\frac{c^{2,k}}{x^{2k}}\;, \notag \\
&& \bP_3=A_3+\sum_{k=2}^{+\infty}\frac{c_{3,k}}{x^{2k}}\;, 
\quad 
%\bP^3=-\frac{1}{\sqrt{\Lambda w} x}+\sum_{k=1}^{+\infty}\frac{c^{3,k}}{x^{2k+1}}\;, 
%\notag \\
%&& 
\bP_4=A_4 \sqrt{\Lambda \w}\left(x+\frac{1}{x}\right)+\sum_{k=1}^{+\infty}\frac{c_{4,k}}{x^{2k-1}}\;, 
%\quad 
%\bP^4=\frac{1}{\Lambda w x^2}+\sum_{k=1}^{+\infty}\frac{c^{4,k}}{x^{2k+2}}\;. \notag
\label{Pansatz}
\end{eqnarray}
where
\begin{equation}
x(u)=\frac{u+\sqrt{u-2g}\sqrt{u+2g}}{2g},\qquad\qquad \Lambda\w=g^2
\label{xu}
\end{equation}
%\Lambda=g^2/\w$
and some coefficients of $\Pa$ in the expansion over $\w$
\begin{equation}
\bP_a=\sum\limits_{k=0}^{+\infty}\bP^{(k)}_a \w^k,
%\quad \bP^a=\sum\limits_{k=0}^{+\infty}\bP^{(k)a}\w^k\;.
\label{Pscaling}
\end{equation}
can be found in \cite{Alfimov:2018cms}.
The scaling of $\bP$-functions (\ref{Pscaling}) suggests that the coefficients should have the expansion
\begin{equation}
c_{m,n}=\sqrt{\!\Lambda \w}^{\;2n-m-1}\sum_{k=0}^{+\infty}c_{n,m}^{(k)}\w^k\;, 
%\quad c^{m,n}=\sqrt{\Lambda w}^{2n+m-6}\sum_{k=0}^{+\infty}c^{n,m(k)}w^k\;.
\label{cmn_rescale}
\end{equation}
The relation between $\w$ and $g^2$ for given $\Delta=N+\delta$ can be written in the following form
\begin{equation}
\w_{\Delta=N+\delta}=\sum_\ell \frac{g^{2\ell}}{\displaystyle{\sum\limits_k\lambda^\nN_{\ell,k}\delta^k}}
%{\lambda_\ell^\Delta}
\end{equation}
while $g^2$ is expanded in $\w$ for given $\Delta=N+\delta$ as
\begin{equation}
g^2=\sum_\ell  \w_{N}^\ell \sum_k r^\nN_{\ell,k}\delta^k
%{\lambda_\ell^\Delta}
\label{g2overw}
\end{equation}
with the obvious relations between $\lambda_\ell$ and $r_\ell$. In our computations we used expansion over $\w$ as more natural and simple. The expansion over $g^2$ is contained in the expressions for $\Pa$ in Eq.~(\ref{Pansatz}) and in $x(u)$
in Eq.~(\ref{xu}), moreover, we replace all combinations $\Lambda\w$ by $g^2$ and after expansion $\Pa$ in $g^2$ we apply rescaling for $c_{m,n}$ from Eq.~(\ref{cmn_rescale}). After this we perform the expansion of $g^2$ over $\w$ with Eq.~(\ref{g2overw}) up to the necessary order in $\w$ and $\delta$.

In the lowest order $\Qi$ can be found from $4$th-order finite-difference equation~\cite{Alfimov:2014bwa} 
\begin{equation}
0=\bQ^{[+4]}D_0-\bQ^{[+2]}
\left[D_1-\bP_a^{[+2]}\bP^{a[+4]}D_0\right]+
\frac{\bQ}{2}
\left[D_2-\bP_a\bP^{a[+2]}D_1+\bP_a\bP^{a[+4]}D_0 \right]+\text{c.c.}
\label{QPeq}
\end{equation}
with
\begin{eqnarray}{}
& D_0=\det\left(\begin{array}{llll}
\bP^{1[+2]} & \bP^{2[+2]} & \bP^{3[+2]} & \bP^{4[+2]} \\
\bP^{1} & \bP^{2} & \bP^{3} & \bP^{4} \\
\bP^{1[-2]} & \bP^{2[-2]} & \bP^{3[-2]} & \bP^{4[-2]} \\
\bP^{1[-4]} & \bP^{2[-4]} & \bP^{3[-4]} & \bP^{4[-4]}
\end{array}\right),\quad
D_1=\det\left(\begin{array}{llll}
\bP^{1[+4]} & \bP^{2[+4]} & \bP^{3[+4]} & \bP^{4[+4]} \\
\bP^{1} & \bP^{2} & \bP^{3} & \bP^{4} \\
\bP^{1[-2]} & \bP^{2[-2]} & \bP^{3[-2]} & \bP^{4[-2]} \\
\bP^{1[-4]} & \bP^{2[-4]} & \bP^{3[-4]} & \bP^{4[-4]}
\end{array}\right),\nonumber \\
& D_2=\det\left(\begin{array}{llll}
\bP^{1[+4]} & \bP^{2[+4]} & \bP^{3[+4]} & \bP^{4[+4]} \\
\bP^{1[+2]} & \bP^{2[+2]} & \bP^{3[+2]} & \bP^{4[+2]} \\
\bP^{1[-2]} & \bP^{2[-2]} & \bP^{3[-2]} & \bP^{4[-2]} \\
\bP^{1[-4]} & \bP^{2[-4]} & \bP^{3[-4]} & \bP^{4[-4]}
\end{array}\right),
\end{eqnarray}
where we use the usual shorthand notation for the shift in the variable $u$: $f(u+ik/2)=f^{[k]}(u)$.
Then, using
\begin{equation}
\Qai^+-\Qai^-=\bP_a \bQ_i %\;, \quad \textrm{Im}\;u>0
\label{Qaieq}
\end{equation}
we can find $\Qai$ in the lowest order.

To find the perturbative expansion of $\Qai$, $\Qi$ and $\Pa$ in $\w$ and $\delta$ we following the procedure, described in Ref.~\cite{Gromov:2015vua}. 
Let $dS$ be the mismatch in the equation
\begin{equation}
{\cal Q}_{a|i}^{(0)}(u+\tfrac{i}{2}) -
{\cal Q}_{a|i}^{(0)}(u-\tfrac{i}{2})
+\bP_a\bP^b {\cal Q}_{b|i}^{(0)}(u+\tfrac{i}{2})=dS_{a|i},
\label{eqdS}
\end{equation}
$dS_{a|i}$ is small in $\epsilon$ ($\epsilon$ is some small expansion parameter, i.e. $\w$ and $\delta$ in our case). We can always represent the exact solution in the form
\begin{equation}
{\cal Q}_{a|i}(u)={\cal Q}_{a|i}^{(0)}(u)+{b_i^{\;j}}(u+\tfrac{i}{2})\;{\cal Q}_{a|j}^{(0)}(u)
\label{Qc}
\end{equation}
where the unknown functions $b_i^{\;j}$ are also small.
After plugging this ansatz into the equation~(\ref{eqdS}) we get
\begin{equation}
\left(b_i^{j}(u)-b_i^{j}(u+i)\right){\cal Q}_{a|j}^{+(0)}=dS_{a|i}+dS_{a|j}b_i^{j}.
\end{equation}
Since $b_i^{\;j}$ is small it can be neglected in the r.h.s. where it multiplies another small quantity.
Finally multiplying the equation by ${\cal Q}^{(0)a|k}$ and using normalisation condition for $\Qai$ 
\begin{equation}
\Qai \QAj=-\delta^j_i
\end{equation}
we arrive at
\begin{equation}
{b_i^k}(u+i)-b_i^k(u)=-dS_{a|i}(u){{\cal Q}^{(0){a|k}}}\left(u+\tfrac{i}{2}\right)+{\cal O}(\epsilon^{2n})\,.\label{bbdSQai}
\end{equation}
We see that the r.h.s. contains only the known functions $dS_{a|i}$and ${\cal Q}^{(0)}_{a|k}$ and does not contain
$b_i^{\;j}$ and can be easily solved.
As was show in Refs.~ \cite{Leurent:2013mr,Marboe:2014gma} the QSC-approach involves into solutions the $\eta$-functions
\begin{equation}
\eta_{a_1,a_2,\dots,a_k}(u)=\sum_{0\le n_1<\ldots< n_k<\infty}\frac{1}{(u+i n_1)^{a_1}\cdots (u+i n_k)^{a_k}},\label{Eta_function}
\end{equation}
which are related in a simple way to the harmonic sums~(\ref{vhs}) when $u=i$. 
As it was explained in~\cite{Leurent:2013mr,Marboe:2014gma}  the product of any two $\eta$-functions can be written as a sum of $\eta$-functions,
and most importantly one can easily solve equations of the type, which appear, for example, in~Eq.~(\ref{Qaieq}) or Eq.~(\ref{bbdSQai})
\begin{equation}
f(u)-f(u+i)=u^n \eta_{a_1,\dots,a_k}(u)
\end{equation}
for any integer $n$ again in terms of a sum of powers of $u$ multiplying $\eta$-functions.
The last equation have the same form, as Eq.~(\ref{bbdSQai}) and actually used for its solution.
After ${\mathcal Q}_{a|i}$ is found one can use 
\begin{equation}
\Qi=-\Qai^+\PA
\end{equation}
to find $\bQ_i$.

In the next step we demand, that the large $u$ asymptotic of $\Qi$ should satisfy Eq.~(\ref{A4A3B2B4}). This allow to find the relations between free parameters and to fix some of them.

The last step is so-called gluing conditions, which is a relation for small $u$ expansions of $\Qi$, which have found in the following form~\cite{Gromov:2015vua}:
\begin{equation}
\bQ_1(u)+\alpha \bQ_3(-u)={\rm reg}\,,\qquad\qquad
\frac{\bQ_1(u)-\alpha \bQ_3(-u)}{\sqrt{u^2-4g^2}}={\rm reg}\,.
\end{equation}
Requiring the absence of the negative powers will fix $\alpha$, the coefficients $c_{a,n}$ and the function $\Delta(S)$

\section{Computations}\label{Sec:Comp}

All calculations subdivided into several parts according to steps, described above. Namely, we compute the general expression for $\Qai$ in the given order in $g$ and $\delta$ and construct $\Qi$ from it, find the large $u$ asymptotic for these $\Qi$ to fix some coefficients in the obtained general expressions for $\Qai$ and $\Qi$ and fix the rest coefficients using analyticity of the certain combinations of the components $\Qi$ near $u=0$. In each step there is one operation, which can be precomputed in the form of database, to speed up all computations considerable. We give below more details. 

The computation of the $\Qai$ is the most time-consuming part. According to the Eq.~(\ref{eqdS}) we compute first of all $dS_{a|i}(u)$, up to necessary order in $w$ and in $\delta$ using \texttt{MATHEMATICA} function \texttt{CoefficientList} with the third argument, which allow to control the number of necessary coefficients in the list. 
To find $\bij$ we used so called $\Psi$-operation. The code, which realised this $\Psi$-operation, is available in the ancillary files of Ref.~\cite{Marboe:2014gma}. However, $\Psi$-operation is applied only to single $\eta$-function, so, we need to linearised the expression over $\eta$-functions, that is rewrite the products of $\eta$-functions in the terms of the combination of the single $\eta$-functions. For this purpose we produce database for all necessary such expansions with the help of \texttt{MATHEMATICA}-package \texttt{HarmonicSums}~\cite{Ablinger:2014rba} using a clear relation between $\eta$-functions and Euler-Zager sums. As $\Qai$ in the right-hand side of Eq.~(\ref{bbdSQai}) is in low order in $w$ and $\delta$ such database restricted only with product for the given $\eta\vec{a}$ by one $\eta_{i}$ with one or two indices and we produced database, which contains the rules for the linearisation of all such products. After linearisation is performed we apply $\Psi$-operation using code, extracted from the supplemented math-file of Ref.~\cite{Marboe:2014gma}. We check, if our database does not include any substitutions we generate them and recompute $\Psi$-operation. The very important thing in this place, that the $\Psi$-operation is performed up to $\mathcal P$-periodic function, which in this case is just a constant, so, we add constant matrix to $\Qai$ with given order of $w$ and $\delta$.
Then we use observation, that the expansion of $\bij$ over $u$ can be restricted only with some several powers, namely, up to $u^2$ for $\Delta=1+\delta$ and up to $u^{(N-1)/2+2}$ for $\Delta=N+\delta$. Finally, we multiply to $\Qai$ again only in the low orders, perform expansion up to $u^{(N-1)/2+2}$ and perform the linearisation for the products of (two) $\eta$'s. In the end of this step we have $\Qai$ at given order in $w$ and $\delta$. From the obtained expression for $\Qai$ we produce $\Qi$ with
\begin{equation}
\Qi=-\Qai^+\PA
\end{equation}
using the results for already computed $\Qi$ in low orders. $\Qai$ and $\Qi$, which we found in this way, contain the arbitrary constants $C_{a|i}$ from $\Psi$-operation, $c_a$ from $\Pa$ and $\lambda_{\ell,i}$ from the expansion of $x$. 

In the next step we perform the asymptotic expansion of $\Qi$ to find relations between $\Cai$ and $\ca$ or even fix some of them. The asymptotic expansion of $\eta$'s is a rather simple recurrent procedure, which start from the most simple $\eta_1$\footnote{We ignore logarithmic terms in the real expansion.}
\be
\eta_1^{{}^{\mathrm{{asym}}}}\hspace*{-4.5mm}{\scriptstyle{(}}u{\scriptstyle{)}}=(-i)\sum_{k=1}\frac{(-1)^k\,(k-1)!\,B_k}{k!\,(-i\, u)^{k}}+\frac{\pi}{2}-i\ln\!\left(\frac{1}{u}\right).\label{S1asym}
\ee
The code for the asymptotic expansion can be extracted from the auxilliry files in \texttt{arXiv} version of Ref.~\cite{Gromov:2015wca}. However, we have found the more fast expansion, which extend the result for $\eta_1$  with the following formula\footnote{I thanks Ivan Surnin, who find this method}
\bea
\eta_a^{{}^{\mathrm{{asym}}}}\hspace*{-4.5mm}{\scriptstyle{(}}u{\scriptstyle{)}}&=&\frac{(-i)^a}{(a-1)!}\sum_{k=0}\frac{(-1)^k\,(a+k-2)!\,B_k}{k!\,(-i\, u)^{k+a-1}}\ ,\label{Sasymsingle}\\
\eta_{k,\vec{a}}^{{}^{{}{\mathrm{asym}}}}\hspace*{-1mm}{\scriptstyle{(}}u{\scriptstyle{)}}&=&
\mathbb{M}\bigg[ 
\frac{\eta_{\vec{a}}^{{}^{{}{\mathrm{asym}}}}\hspace*{-4.5mm}{\scriptstyle{(}}z+i{\scriptstyle{)}}}{z^k}\bigg |_{z=u}
\ ;\ 
\eta_k^{{}^{{}{\mathrm{asym}}}}\hspace*{-4.5mm}{\scriptstyle{(}}u{\scriptstyle{)}}
\bigg],
\eea
where function $\mathbb{M}\,[u^{-k}\,;\,\eta_k^{{}^{{\mathrm{\,asym\,}}}}\hspace*{-5.5mm}{{{\scriptstyle{(}}u{\scriptstyle{)}}}}]$ maps the negative power of $u$ into the asymptotic expansion of the $\eta$-function with the single index equal to the power of $u$
\be
\mathbb{M}\bigg[ 
\frac{1}{u^k}
\ ;\ 
\eta_k^{{}^{{}{\mathrm{asym}}}}\hspace*{-4.5mm}{\scriptstyle{(}}u{\scriptstyle{)}}
\bigg]\quad\;\Rightarrow\quad\frac{1}{u^k}\ \mapsto\ \eta_k^{{}^{{}{\mathrm{asym}}}}\hspace*{-4.5mm}{\scriptstyle{(}}u{\scriptstyle{)}}
\ee
given by (\ref{Sasymsingle}).
We generate the database for the $\eta$-functions with weights $\ell=1,\, \ldots,\, 15$ up to $u^{-100}$, which is enough for the computation up to $\Delta=97+\delta$. 

The asymptotic expansion will provide some conditions if we suggest, that it should have the following leading coefficients
\be
%\bP_a&\simeq&(A_1 u^{-2},A_2 u^{-1},A_3,A_4u)_a\label{Plarge}\\ 
\Qi\simeq(B_1 u^{\frac{\Delta+1}{2}},B_2 u^{\frac{\Delta-3}{2}}, B_3u^{\frac{-\Delta+1}{2}}, B_4u^{\frac{-\Delta-3}{2 }})
\label{Qlarge2}
\ee
with
\bea
B_1&=&1,\qquad B_2\ =\ \frac{1}{16i}\frac{(4-(\Delta-1+\w)^2) (4-(\Delta-3+\w)^2)}{\Delta  (2-\w) (\Delta +2-\w)},\\
B_3&=&1,\qquad B_4\ =\ \frac{1}{16i}\frac{(4-(\Delta+1-\w)^2) (4-(\Delta+3-\w)^2)}{\Delta  (2-\w) (\Delta +2-\w)}
\label{B1B2B3B4}
\eea
and contain only even powers in the expansion over $u$ due to party symmetry~\cite{Gromov:2015wca}

The last step is so called gluing conditions, which are related to the requirement of regularity at the origin
for the following combinations~\cite{Gromov:2015wca}
\be
{\mathbf Q}_1(u)+\alpha {\mathbf Q}_3(-u)={\rm reg}\,,\quad
\frac{{\mathbf Q}_1(u)-\alpha {\mathbf Q}_3(-u)}{\sqrt{u^2-4g^2}}={\rm reg}\,,\quad
\alpha=\sum_{n}\sum_{m}\alpha_{n,m}\,g^{2n}\delta^m\,.
\label{Gluing}
\ee
In these relations one first expands in $g$ (or in $\w$) and $\delta$ the l.h.s. and then in $u$ around the origin and
${\mathbf Q}_3(-u)$ can be obtained from the small-$u$ expansion of ${\mathbf Q}_3(u)$ and the replacement \mbox{$u\to(-u)$}. The requirement for the cancellation of the negative powers will fix $\alpha_{n,m}$, all the coefficients $\Cai$, $\ca$ and $\lambdai$, except $c_{1,1}$, which we fixed from the additional constraints.

The gluing conditions demand the expansions of $\eta$'s near $u=0$, which can be performed with precomputed database. The code for the computations of this database can be taken from the auxiliary file in \texttt{arXiv} version of Ref.~\cite{Marboe:2014gma}, but for out calculations we have found the following general expression for the small-$u$ expansion for $\eta_{\vec{a}}(u)$: 

\bea
&& \eta_{i_1,i_2,i_3,i_4,\ldots,i_\ell}
=(-1)^{\pmb{i}}
\sum_{\kappa=0}^{\begin{array}{cc}
\scriptstyle\mathfrak{n}+i_1 & \scriptstyle \mathrm{for\ s.t.}\\[-2mm]
\scriptstyle\mathfrak{n} &\scriptstyle \mathrm{for\ f.t.}\scriptstyle %\mathfrak{u} &\scriptstyle \mathrm{for\ f.t.} 
\end{array}}
i^{\left(\kappa+{\pmb{i}}\right)}
u^\kappa
\sum_{k_1=0}^{\kappa}
\sum_{k_2=0}^{\hat{k}_1}
\sum_{k_3=0}^{\hat{k}_2}
\cdots
\sum_{k_{\ell-2}=0}^{\hat{k}_{\ell-3}}\times
\nonumber\\
&&
%\cdots
\times
\quad\Bigg(\qquad\qquad\quad\
\sum_{k_{\ell-1}=0}^{\hat{k}_{\ell-2}}
\Bigg[\prod_{l=1}^{\ell}
\binom{\hat{\imath}k_{l,{l-1}}}{\hat{\imath}_l}
\Bigg]
\zeta(ik_{1,0},ik_{2,1},ik_{3,2},\ldots,ik_{\ell,\ell-1})\nonumber%\zeta(ik_{1,0},ik_{2,1},ik_{3,2},\ldots,ik_{\ell,\ell-1})\nonumber\\
\\
&&\qquad+
(-1)^{\pmb{i}}
\big(i u\big)^{-i_1}
\!\!\!\!\sum_{k_{\ell-1}=\hat{k}_{\ell-2}}^{\hat{k}_{\ell-2}}
\!\Bigg[\prod_{l=1}^{\ell}
\binom{\hat{\imath}k_{l,{l-1}}}{\hat{\imath}_l}
\Bigg]
\zeta(ik_{2,1},ik_{3,2},\ldots,ik_{\ell,\ell-1})\quad\Bigg),\nonumber\\
& & \hat{\imath}_n=i_n-1,\quad {\pmb{i}}=\sum_{n=0}^{\ell}i_{n},\quad \hat{k}_i=\sum_{j=0}^{i}k_{j},\quad \hat{k}_0=\kappa-\hat{k}_\ell,\quad ik_{j_1,j_2}=i_{j_1}k_{j_2}.\quad\label{smallu}
\eea
where first and second terms are differ by the common factor, the lower limit for the last summation and absence of the first argument in the multiple zeta-values, moreover, the first summation (over $\kappa$) for the second term (s.t.) should be taken up to $({\mathfrak{n}}+i_1)$ to obtain expansion up to $u^{{\mathfrak{n}}}$.

We store all obtained expressions for $\Qai$ and $\Qi$ along with all values for coefficients $\Cai$, $\ca$, $\alpha_{n,m}$ and $\lambdai$ obtained after each step in expansion over $\w$ and $\delta$ to use them in the next steps. The computations for $\Delta=1+\delta$ and for $\Delta=N +\delta$ with $N=7, 9, 11,...$ are slightly different as in the first case the expressions for $\Qai$, $\Qi$ and $\Pa$ contain the poles in $\delta$. We give all details for these cases below.

\subsection{$\Delta=1+\delta$ case}

As we mention above in this case $\Qai$, $\Qi$ and $\Pa$ can contain the poles in $\delta$-expansion. 
Take the lower orders $\Pa$ ($c_{4,1,1,0}=c_{4,1,1,1}=c_{4,1,1,2}=0$)~\cite{Alfimov:2014bwa}
\be
{\mathbf P}_1=\frac{1}{u^2},\qquad {\mathbf P}_2=\frac{1}{u},\qquad {\mathbf P}_3=\frac{i \delta}{2},\qquad {\mathbf P}_4=\frac{i \delta}{2}u
\ee
we received the following solution of the fourth order equation~(\ref{QPeq})
\bea
{\mathbf Q}_1(u)&=&u+\frac{i \delta }{4}+ u\,\delta  \left(\frac{i \pi }{4}-\frac{i \eta _1}{2}\right),\quad
{\mathbf Q}_2(u)\ =\ \frac{i \delta }{2 u}\, ,\nonumber\\
{\mathbf Q}_3(u)&=&1+\delta  \left(\frac{i \eta_1}{2}-\frac{i \pi }{4}+\frac{1}{2}-\frac{i}{2} u\eta _2\right),\quad
{\mathbf Q}_4(u)\ =\ -\frac{i \delta }{2 u^2}\,.\label{QiN1LO}
\eea
To satisfy the normalization condition
\be
\Qai \QAj=-\delta^j_i\label{NormConQai}
\ee
we should add to the expressions, obtained from the above $\Qi$ and $\Pa$ and Eq.~(\ref{Qaieq}), the arbitrary constants, divided by $\delta$ along with the ordinary one, which gives
\bea
&&{\mathcal Q}_{1|1}= 
 -\frac{2 i}{\delta } 
 -\eta_1
 +\frac{\pi }{2}
-\frac{i}{4}\delta\left(
 \pi  \eta_1
- \eta_2
-2 \eta_{1,1}
- \frac{\pi ^2}{4} 
\right),
\quad{\mathcal Q}_{1|3}= \frac{1}{4} u\left(u-i\right) \delta, 
\nonumber\\  
% +\frac{1}{16} i \pi ^2 \delta 
&&{\mathcal Q}_{1|2}=-\frac12(1+2ui)-\frac{i}{8}\delta(1+2ui)(2 i+\pi -2 \eta_1),\quad
{\mathcal Q}_{1|4}= -\frac{i}{12} (u-i)(1+2ui) \delta,  \nonumber\\&&
{\mathcal Q}_{2|1}= -\frac{i}{2} \delta  \eta_3 ,\qquad
{\mathcal Q}_{2|2}= -\frac{i}{2} \delta  \eta_2 ,\qquad
{\mathcal Q}_{2|3}= \frac{i }{2}  \delta ,\qquad
{\mathcal Q}_{2|4}= 0,\nonumber\\&& 
{\mathcal Q}_{3|1}= -\eta_2 +\frac{i}{4}\delta  (\pi  \eta_2+2 i \eta_2+2 \eta_{1,2}-2 \eta_{2,1}),\quad
{\mathcal Q}_{3|3}= -\frac{i}{4}\delta(1+2ui) ,
\nonumber\\&& 
{\mathcal Q}_{3|2}= \frac{2 i}{\delta }-\eta_1+ \frac{\pi }{2}
-\frac{i}{4} \delta \left(2-\pi  \eta_1-2 i u \eta_2+2\eta_{1,1}+\frac{\pi^2}{4}\right) ,\quad
{\mathcal Q}_{3|4}= \frac{1}{4} u(u-i) \delta ,
\nonumber\\&& 
{\mathcal Q}_{4|1}= \frac{i}{2} \delta  \eta_4,\qquad 
{\mathcal Q}_{4|2}= \frac{i}{2} \delta  \eta_3 ,\qquad
{\mathcal Q}_{4|3}= 0 ,\qquad
{\mathcal Q}_{4|4}=\frac{i \delta }{2}.\label{QaiN1LO}
\eea
The presence of the poles in ${\mathcal Q}_{1|1}$ and ${\mathcal Q}_{3|2}$  will demand to know the expression for $\Qai$ in the first order in $\delta$ to fulfill normalisation condition~(\ref{NormConQai}) up to constant term in the $\delta$-expansion. This means, that when we compute $\Qai$ in order $\delta^{n+1}$ we will realy fix them in one order less. Going in the higher orders in $\w$ (or $g^2$) we will produce the poles of higher orders in $\delta$, which will demand the knowledge of $\Qai$ in the higher orders in $\delta$ for low orders in $\w$. Solve QSC-system step by step for $\Qi$ and $\Qai$ we obtained the results up to $r_{4,10}$, which contains multiple zeta values $\zeta_{\vec{a}}$ up to weight $12$ ($\zeta_{12}$ and similar), restricted by the available database for the relations between MZV~\cite{Blumlein:2009cf}. 
%To fix $c_{1,1}$
The final result is the following:
\begin{align}
\w_{1+\delta}&=
\frac{20480}{\delta ^7}
-\frac{8192 \pi ^2}{3 \delta ^5}
+\frac{16896 \z{3}}{\delta ^4}
-\frac{1568 \pi ^4}{9 \delta ^3}
+\frac{64}{\delta ^2}\left(357 \z{5}
-\frac{26 \pi ^2 \z{3}}{3}\right)\nonumber\\&
+\frac{640}{\delta }\left( \z{3}^2
-\frac{44672 \pi ^6}{2835}\right)
-\frac{212 \pi ^4 \z{3}}{9}
-560 \pi ^2 \z{5}
+30198 \z{7}\nonumber\\&
+\delta  \bigg(
-\frac{2048 \hh53}{19}
+\frac{30720 \hh71}{19}
+\frac{416 \pi ^2 \z{3}^2}{3}
-\frac{26824 \z{3} \z{5}}{19}
-\frac{9794059 \pi ^8}{6463800}\bigg)\nonumber\\&
+\delta ^2 \bigg(
-648 \z{3}^3
-\frac{5713 \pi ^6 \z{3}}{5670}
-\frac{4919 \pi ^4 \z{5}}{90}
-\frac{1117 \pi ^2 \z{7}}{2}
+\frac{163495 \z{9}}{6}\bigg)\nonumber\\&
+\delta ^3 \bigg(\frac{28288 \pi ^2 \hh53}{1539}
-\frac{141440 \pi ^2 \hh71}{513}
-\frac{37248 \hh73}{1583}
+\frac{1042944 \hh91}{1583}
+\frac{178 \pi ^4 \z{3}^2}{45}\nonumber\\&
+\frac{70876}{171} \pi ^2 \z{3} \z{5}
-\frac{26726079 \z{3} \z{7}}{6332}
-\frac{2039938 \z{5}^2}{1583}
-\frac{445486835443 \pi ^{10}}{4051949378400}\bigg)\nonumber\\&
+\delta ^4 \bigg(
\frac{188580330112 \hh53\z{3}}{1567962333}
+\frac{2541228510971 \pi ^8 \z{3}}{13547194557120}
-\frac{390134568811 \pi ^2 \z{9}}{895978476}\nonumber\\&
-\frac{792266806357 \pi ^6 \z{5}}{2634176719440}
+\frac{336109772735957 \zz11}{22299908736}
-\frac{107690308055 \z{3}^2 \z{5}}{174218037}\nonumber\\&
+\frac{11575203840 \hhh911}{8296097}
-\frac{227457106121 \pi ^4 \z{7}}{4778551872}
-\frac{962833600640 \hh71 \z{3}}{522654111}
+\frac{26 \pi ^2 \z{3}^3}{3}\nonumber\\&
-\frac{11419980800 \hhh731}{58072679}
-\frac{8467776000 \hhh713}{58072679}
+\frac{6852394496 \hhh533}{522654111}
+\frac{3410471936 \hhh551}{58072679}
\bigg)\nonumber\\&
+\delta ^5 \bigg(\frac{6952 \pi ^4 \hh53}{7695}
-\frac{6952 \pi ^4 \hh71}{513}
+\frac{112400 \pi ^2 \hh73}{14247}
-\frac{3147200 \pi ^2 \hh91}{14247}\nonumber\\&
-\frac{1657984 \hh93}{9495}
+\frac{10710528 \hB{1}}{5275}
+\frac{133 \z{3}^4}{2}
+\frac{3329 \pi ^6 \z{3}^2}{7560}
+\frac{170291 \pi ^4 \z{3} \z{5}}{6840}\nonumber\\&
+\frac{3035729 \pi ^2 \z{3} \z{7}}{9498}
-\frac{348296669 \z{3} \z{9}}{50640}
+\frac{1474101 \pi ^2 \z{5}^2}{12664}
-\frac{308761781 \z{5} \z{7}}{84400}\nonumber\\&
+\frac{2195648 \hh75}{79125}
-\frac{1762420922909626739 \pi ^{12}}{424371709170864000000}\bigg).\label{Delta1}
\end{align}

%\subsection{Leading order solution for $\Delta=N+\delta$}
\subsection{$\Delta=N+\delta$ case}
\label{sec:LONSol}

For $N=7,9,11,\ldots$ the solutions of the QSC-system for $\Qi$, $\Pa$ and $\Qai$ do not contain the poles over $\delta$, which considerable simplify the general procedure.

The solution in the leading order for $\Qi$ can be found again with the help of the fourth order difference equation~(\ref{QPeq}).
The ansatz, which can be used to solve this equation, has the following general form
\bea
{\mathbf Q}_i^{\Delta=N+\delta}(u)&=& \Bigg\{
\sum_{k=\kappa}^{\mathcal{K}}a^1_{0,2k}\,u^{2k}\ ,\quad
\sum_{k=\kappa}^{\mathcal{K}}a^2_{0,2k}\,u^{2k}\ ,\nonumber\\&&
\sum_{k=0}^{2\mathcal{K}-1}a^3_{0,k}\,u^k
+\eta_2(u)\sum_{k=\kappa}^{\mathcal{K}}a^3_{2,2k}\,u^{2k}\ ,\nonumber\\&&
\sum_{k=-2}^{2\mathcal{K}-1}a^4_{0,k}\,u^k
+\eta_2(u)\sum_{k=\kappa}^{\mathcal{K}}a^4_{2,2k}\,u^{2k}
+\eta_4(u)\sum_{k=\kappa}^{\mathcal{K}}a^4_{4,2k}\,u^{2k}
\Bigg\},
\label{AnsatzLO}
\eea
where $\mathcal{K}=(n+1)/4$ and summation start from $\kappa=0$ for $n=4\,N-1$ and  $\kappa=1$ for $n=4\,N+1$ for integers $N$.

Firstly, we use  the known asymptotic for $\Qi$ to fix some  coefficients $a^i_j$ demanding for ${\mathbf Q}_3(u)$ and ${\mathbf Q}_4(u)$ the absence of the lowest poles in $u$ and the positive powers of $u$. Then, plugging the ansatz into eq.~(\ref{QPeq}) and expand $\eta$'s up to some first terms, depending from the argument of the $\eta$-function as
\be
\eta_a(u+i\,k)=\sum_{m=0}^{8-k}\frac{1}{\left(u+i\,m+ i\,k\right)^a}
\ee
we obtain expressions, which contain the terms $(u\pm i m)$ in negative powers and polynomial in $u$ and all such terms should be equal to zero differently. This give a lot of equations, which fix a lot of coefficients. To find $\Qai$ in the leading order we substituting the obtained expressions for $\Qi$ and the expressions for $\Pa$ with $\lambda_{1,1}=-1/8$ and $c_{4,1,1,0}=0$ into equation
\be
\Qai^+-\Qai^-=\bP_a \bQ_i. %\;, \quad \textrm{Im}\;u>0
\label{Qaieq2}
\ee
The normalisation condition for $\Qai$
\be
\Qai \QAj=-\delta^j_i\label{NormConQaiN}
\ee
gives equations, which allow to fix almost all coefficients. The three unfixed coefficients can be found from the following equations for the asymptotic of $\Qi$
\bea
{\mathrm {Coef}}\left[ {\mathbf Q}_3^{\mathrm{asy}}(u),\frac{1}{u^{(N-1)/2}}\right]\equiv 1,\\
{\mathrm {Coef}}\left[ {\mathbf Q}_4^{\mathrm{asy}}(u),\frac{1}{u^{(N-1)/2}}\right]\equiv 0,\\
{\mathrm {Coef}}\left[ {\mathbf Q}_1^{\mathrm{asy}}(u),u^{(N+1)/2}\right]\equiv 1.
\eea

For the solution of the QSC-system in higher orders we demand, first of all, the absence of the poles in the expansions of $\Pa$ and then we apply the procedure, described above. In this way we obtained the results up to fourth order in $\w$ (or $g^8$) for $\Delta=7+\delta$ up to $\delta^1$, for $\Delta=9+\delta$ and $\Delta=13+\delta$ up to $\delta^2$ (the rational part up to $\delta^4$ and $\delta^3$ correspondingly), which are given in Appendix~\ref{Sec:CompVal} and  for $\Delta=N+\delta$ for $N=17,\ldots,29$ up to $\delta^{1}$ and only rational part for $\Delta=N+\delta$ for $N=33,\ldots,97$ up to $\delta^{0}$, which we used for the reconstruction of the general expression from these results for the fixed values with the help of the number theory, described in the next Section.

\section{Reconstruction}\label{Sec:Rec}

It turned out that the most problematic part of the full computations is the reconstruction of the general form of the BFKL from the calculated results for the fixed values. In principle this can be done with the help of the number theory, but if we know the basis functions, which will enter into the final answer. Unfortunately, in our case the simple generalisation from the low orders, which can be done with the maximal-transcedentality principle~\cite{Kotikov:2002ab} does not work. This was discovered after several unsuccessful attempts to reconstruct the general form from the known results for the fixed values. The maximal-transcedentality principle gives the basis with 2615 harmonic sums and we have about 500 constraints, what usually is enough for such procedure, while we obtain answer with the undesirable numbers. To study this problem we subdivided the reconstruction into several part according their transcendental factor, that is in the general expression
\begin{eqnarray}
F^{(4)}=
\sum_{\vec{a}}C_{\mathrm{Rat}}^{\vec{a}}S_{\vec{a}}^{\mathrm{weight}=7}
+\z2\sum_{\vec{a}}C_{\mathrm{\z2}}^{\vec{a}}S_{\vec{a}}^{\mathrm{weight}=5}
+\z3\sum_{\vec{a}}C_{\mathrm{\z3}}^{\vec{a}}S_{\vec{a}}^{\mathrm{weight}=4}
+\cdots
\end{eqnarray} 
we reconstruct separately the rational and transcendental ($\z2$, $\z3$, ...) parts.

\subsection{Usual harmonic sums}

We started the reconstruction procedure using the usual harmonic sums~(\ref{vhs}), assuming, that the basis for the result at fourth order will the same as in low orders~(\ref{LLA})-(\ref{NNLLA}).

\subsubsection{Rational part}

For the reconstruction of the rational part we compute the results for the general values of $\N$ up to the fourth order in $g^2$ and up to the third order in $\delta$-expansion which give the result for the poles up to $\w^{-3}$. 
Suggesting, that the general results will contain the harmonic sums we have found, that it is necessary to consider the harmonic sums not only with the usual argument (some quantity divided by two as in Eq.~(\ref{BFKLj})), but with twice  of its. In principle there are the relations between harmonic sums with the integer argument and the harmonic sums with this double argument, but only for the harmonic sums with all positive indices. In our case we have found the following general expression for the part, which is proportional to  $\w^{-4}$ pole in the fourth order:
\begin{eqnarray}
\big[F_4\big]_{\w^{-4},\mathrm{Rat}}(\kk)=&&8192(-1)^{\kk} \left(\HSH_{-2,1}(2\kk)+\HSH_{2,1}(2\kk)\right)
\nonumber\\&&\hspace*{-27mm}
+\ 1024 \left(
\HSH_3(\kk)
+(-1)^{\kk} \left(
16 \HSH_3(\kk)
-6 \HSH_{1,-2}(\kk)
-6 \HSH_{1,2}(\kk)
-12 \HSH_{2,1}(\kk)
-9 \HSH_{-3}(\kk)
\right)
\right)\!.\qquad\ \label{Ratw4}
\end{eqnarray}
The first terms can not be reduced to the sums with half argument. As such sums with double arguments will appear in the next results we introduce special notations for them as $\HSU_{\vec{a}}(k)=\HSH_{\vec{a}}(2k)$ and we will omit the argument of the harmonic sums.
For the lower pole $\w^{-3}$ the situation even more complicated, because we should include the harmonic sums with index $(-1)$, that is for example $\HSU_{-1,2,1}$ and so on and the basis growth rather fast, while the computations of the highest fixed values are became time-consuming. We suggest, that $\HSU_{-1}$ harmonic sum is factorised in the combination with $\HSU_{1}$ as $(\HSU_{1}+\HSU_{-1})$. This reduce the basis and we have found the following general expression for $\w^{-3}$ pole
\begin{eqnarray}
\big[F_4\big]_{\w^{-3},\mathrm{Rat}}=&&
%-16384  (-1)^\kk \HSU_{-1} 
%\left(\HSU_{-2,1}+\HSU_{2,1}\right)
%\nonumber\\&&\hspace*{-20mm}+8192  (-1)^\kk \Big(
%4 \HSU_{-3,1}+3 \HSU_{-2,2}+3 \HSU_{2,2}+4 \HSU_{3,1}-4 \HSU_{-2,1,1}-2 \HSU_{1,-2,1}-2 \HSU_{1,2,1}-4 \HSU_{2,1,1}\Big)
-8192  (-1)^\kk \Big(2\big(\HSU_{1}+\HSU_{-1}\big) \big(\HSU_{-2,1}+\HSU_{2,1}\big)
-2 \HSU_{-3,1}
-2 \HSU_{3,1}
-\HSU_{-2,2}
-\HSU_{2,2}
\Big)
\nonumber\\&&\hspace*{-20mm}
+512  \Big(4 \HSH_{-2,-2}-2 \HSH_{-2,2}-2 \HSH_{2,-2}+2 \HSH_{-4}-5 \HSH_4\Big)
-1024 (-1)^\kk \Big(
5 \HSH_{-4}+3 \HSH_{2,-2}+21 \HSH_{2,2}
\nonumber\\&&\hspace*{-20mm}
+10 \HSH_{1,-3}+10 \HSH_{1,3}
+4 \HSH_{-3,1}
+28 \HSH_{3,1}-8 \HSH_{1,-2,1}-12 \HSH_{1,2,1}-24 \HSH_{2,1,1}
-25 \HSH_4\Big)\ \label{Ratw3}
\end{eqnarray}
Suggesting similar factorisation properties for other poles we have found the following general expression for $\w^{-2}$ pole
\begin{eqnarray}
\big[F_4\big]_{\w^{-2},\mathrm{Rat}}=&&
8192 (-1)^{k} 
%\Big(\HSU_{-1}^2  (\HSU_{-2,1}+\HSU_{2,1})
%-6 \HSU_{-1}  (\HSU_{-3,1}+\HSU_{3,1})
%+4 \HSU_{-1} (\HSU_{-2,-2}+\HSU_{2,-2})
%\nonumber\\&&\hspace*{-15mm}
%+2 \HSU_{-1}  (\HSU_{1,-2,1}+\HSU_{1,2,1})
%+4 \HSU_{-1} (\HSU_{-2,1,1}+\HSU_{2,1,1})
%+4 (\HSU_{-2,3}+\HSU_{2,3})
%+10 (\HSU_{-3,2}+\HSU_{3,2})
%\nonumber\\&&\hspace*{-15mm}
%+12 (\HSU_{-4,1}+\HSU_{4,1})
%+4 (\HSU_{1,-2,-2}+\HSU_{1,2,-2})
%-6 (\HSU_{1,-3,1}+\HSU_{1,3,1})
%-12 (\HSU_{-3,1,1}+\HSU_{3,1,1})
%\nonumber\\&&\hspace*{-15mm}
%+2 (\HSU_{-2,1,-2}+\HSU_{2,1,-2})
%-3 (\HSU_{-2,1,2}+\HSU_{2,1,2})
%-5 (\HSU_{-2,2,1}+\HSU_{2,2,1})
%-(\HSU_{2,-2,1}+\HSU_{2,2,1})
%\nonumber\\&&\hspace*{-15mm}
%+2 (\HSU_{1,1,-2,1}+\HSU_{1,1,2,1})
%+4 (\HSU_{1,-2,1,1}+\HSU_{1,2,1,1})
%+6 (\HSU_{-2,1,1,1}+\HSU_{2,1,1,1})
%\Big)
\Big(
(\HSU_{ 1} + \HSU_{ -1})^2 (\HSU_{ -2, 1} + \HSU_{ 2, 1}) 
+ 2 (\HSU_{ 1} + \HSU_{ -1}) (\HSU_{ -2, 2} + \HSU_{ 2, 2})\nonumber\\&&\hspace*{-20mm}
- 4 (\HSU_{ 1} + \HSU_{ -1}) (\HSU_{ -3, 1} + \HSU_{ 3, 1}) 
+ 4 (\HSU_{ 1} + \HSU_{ -1}) (\HSU_{ -2, -2} + \HSU_{ 2, -2}) \nonumber\\&&\hspace*{-20mm}
- 2 (\HSU_{ 2} + \HSU_{ -2}) (\HSU_{ -2, 1} + \HSU_{ 2, 1}) 
+ 3 \HSU_{ -4, 1} + 
 4 \HSU_{ -3, -2} + 6 \HSU_{ -3, 2} + 
 2 \HSU_{ -2, -3} + 3 \HSU_{ -2, 3} + \nonumber\\&&\hspace*{-20mm}
 2 \HSU_{ 2, -3} + 3 \HSU_{ 2, 3} + 
 4 \HSU_{ 3, -2} + 6 \HSU_{ 3, 2} + 3 \HSU_{ 4, 1}
 \Big)
\nonumber\\&&\hspace*{-20mm}
-512 \Big(
-2 \HSH_{-4,1}
+5 \HSH_{-3,-2}
-3 \HSH_{-3,2}
+5 \HSH_{-2,-3}
-4 \HSH_{-2,3}
-3 \HSH_{2,-3}
-2 \HSH_{3,-2}
\nonumber\\&&\hspace*{-15mm}
-2 \HSH_{-2,1,-2}
+2 \HSH_{-2,1,2}
+2 \HSH_{-2,2,1}
+2 \HSH_{2,-2,1}
+5 \HSH_{-5}
-8 \HSH_5\Big)
\nonumber\\&&\hspace*{-20mm}
+512 (-1)^{k} \Big(
71 \HSH_5
-15 \HSH_{-4,1}
-4 \HSH_{-3,2}
-25 \HSH_{1,-4}
-35 \HSH_{1,4}
-6 \HSH_{2,-3}
-45 \HSH_{2,3}
\nonumber\\&&\hspace*{-15mm}
-6 \HSH_{3,-2}
-76 \HSH_{3,2}
-92 \HSH_{4,1}
+12 \HSH_{1,-3,1}
+8 \HSH_{1,-2,2}
-4 \HSH_{1,1,-3}
+8 \HSH_{1,1,3}
\nonumber\\&&\hspace*{-15mm}
+20\HSH_{1,2,2}
+56 \HSH_{1,3,1}
+4 \HSH_{2,-2,1}
+44 \HSH_{2,1,2}
+68 \HSH_{2,2,1}
+104 \HSH_{3,1,1}
\nonumber\\&&\hspace*{-15mm}
+8 \HSH_{1,1,-2,1}
-24 \HSH_{1,1,2,1}
-48 \HSH_{1,2,1,1}
-72 \HSH_{2,1,1,1}
\Big).\label{Ratw2}
\end{eqnarray}
For the $\w^{-1}$ pole we have found the general expression for the rational part only for the even values of $k=(N-1)/2$, which has the following form
\begin{align}
\big[F_4\big]_{\w^{-1},\mathrm{Rat}}=&
4096\Big(- 2 (\HSU_{-1} + \HSU_{1}) (\HSU_{-2, 1, -2} + \HSU_{-2, 1, 2} + \HSU_{2, 1, -2} + \HSU_{2, 1, 2}) \nonumber\\&
- 4 (\HSU_{-1} + \HSU_{1}) (\HSU_{-2, -2, 1} + \HSU_{-2, 2, 1} + \HSU_{2, -2, 1} + \HSU_{2, 2, 1}) \nonumber\\&
- (\HSU_{-1} + \HSU_{1})^2 (\HSU_{-2, 2} +  \HSU_{2, 2})
+ 2 (\HSU_{-1} + \HSU_{1}) (\HSU_{-2, 3} + \HSU_{2, 3}) \nonumber\\&
- 2 (\HSU_{-1} + \HSU_{1})^2 (\HSU_{-3, 1} + \HSU_{3, 1}) 
+ 4 (\HSU_{-1} + \HSU_{1}) (\HSU_{-3, 2} + \HSU_{3, 2}) \nonumber\\&
+ 10 (\HSU_{-1} + \HSU_{1}) (\HSU_{-4, 1} + \HSU_{4, 1}) 
+ 2 (\HSU_{-3} + \HSU_{3}) (\HSU_{-2, 1} + \HSU_{2, 1}) \nonumber\\&
+ 2 (\HSU_{-2} + \HSU_{2}) (\HSU_{-2, 2} + \HSU_{2, 2}) 
+ 4 (\HSU_{-2} + \HSU_{2}) (\HSU_{-3, 1} + \HSU_{3, 1}) \nonumber\\&
-4 \HSU_{-5, 1} 
- 3 \HSU_{-4, 2} 
+ 4 \HSU_{-3, -3} 
+ 2 \HSU_{-3, 3} 
+ 4 \HSU_{-2, -4} 
+ 3 \HSU_{-2, 4} \nonumber\\&
+ 4 \HSU_{2, -4} 
+ 3 \HSU_{2, 4} 
+ 4 \HSU_{3, -3} 
+ 2 \HSU_{3, 3} 
- 3 \HSU_{4, 2} 
- 4 \HSU_{5, 1}\Big)\nonumber\\&
+256 \Big(37 \HSH_{-6} 
+ 151 \HSH_{6} 
- 64 \HSH_{-5, 1} 
+ 11 \HSH_{-4, -2} 
- 29 \HSH_{-4, 2} 
+ 16 \HSH_{-3, -3} \nonumber\\&
- 18 \HSH_{-3, 3} 
+ 15 \HSH_{-2, -4} 
- 11 \HSH_{-2, 4} 
- 68 \HSH_{1, -5} 
- 92 \HSH_{1, 5} 
- 13 \HSH_{2, -4} 
- 104 \HSH_{2, 4} \nonumber\\&
- 14 \HSH_{3, -3} 
- 138 \HSH_{3, 3} 
- 21 \HSH_{4, -2} 
- 182 \HSH_{4, 2} 
- 220 \HSH_{5, 1} 
+ 12 \HSH_{-4, 1, 1} 
- 4 \HSH_{-3, -2, 1} \nonumber\\&
- 8 \HSH_{-3, 1, -2} 
+ 4 \HSH_{-3, 1, 2} 
+ 4 \HSH_{-3, 2, 1} 
- 4 \HSH_{-2, -3, 1} 
+ 4 \HSH_{-2, -2, -2} 
- 12 \HSH_{-2, 1, -3} \nonumber\\&
+ 4 \HSH_{-2, 1, 3} 
- 6 \HSH_{-2, 2, -2} 
+ 4 \HSH_{-2, 2, 2} 
+ 4 \HSH_{-2, 3, 1} 
+ 28 \HSH_{1, -4, 1} 
+ 4 \HSH_{1, -3, 2} \nonumber\\&
+ 16 \HSH_{1, -2, 3} 
- 8 \HSH_{1, 1, -4} 
+ 24 \HSH_{1, 1, 4} 
- 12 \HSH_{1, 2, -3} 
+ 52 \HSH_{1, 2, 3} 
+ 4 \HSH_{1, 3, -2} \nonumber\\&
+ 88 \HSH_{1, 3, 2} 
+ 132 \HSH_{1, 4, 1} 
- 8 \HSH_{2, -3, 1} 
- 2 \HSH_{2, -2, 2} 
- 12 \HSH_{2, 1, -3} 
+ 88 \HSH_{2, 1, 3} \nonumber\\&
+ 120 \HSH_{2, 2, 2} 
+ 160 \HSH_{2, 3, 1} 
+ 16 \HSH_{3, -2, 1} 
+ 4 \HSH_{3, 1, -2} 
+ 160 \HSH_{3, 1, 2} 
+ 196 \HSH_{3, 2, 1} \nonumber\\&
+ 240 \HSH_{4, 1, 1} 
+ 8 \HSH_{-2, 1, -2, 1} 
+ 8 \HSH_{-2, 1, 1, -2} 
+ 16 \HSH_{1, -3, 1, 1} 
+ 32 \HSH_{1, 1, -3, 1} \nonumber\\&
+ 24 \HSH_{1, 1, -2, 2} 
- 24 \HSH_{1, 1, 2, 2} 
- 48 \HSH_{1, 1, 3, 1} 
+ 8 \HSH_{1, 2, -2, 1} 
- 48 \HSH_{1, 2, 1, 2} 
- 72 \HSH_{1, 2, 2, 1} \nonumber\\&
- 96 \HSH_{1, 3, 1, 1} 
+ 16 \HSH_{2, -2, 1, 1} 
+ 8 \HSH_{2, 1, -2, 1} 
- 72 \HSH_{2, 1, 1, 2} 
- 96 \HSH_{2, 1, 2, 1} 
- 120 \HSH_{2, 2, 1, 1} \nonumber\\&
- 144 \HSH_{3, 1, 1, 1} 
- 32 \HSH_{1, 1, -2, 1, 1}
\Big).\label{Ratw1}
%
%- 7 \HSU_{4, 2} 
%- 16 \HSU_{5, 1} 
%- 2 \HSU_{3, 3} 
%- \HSU_{2, 4} 
%-16 \HSU_{-5, 1} 
%- 7 \HSU_{-4, 2} 
%- 2 \HSU_{-3, 3} 
%- \HSU_{-2, 4} 
%+ 2 \HSU_{2, 1, -3} 
%+ 2 \HSU_{2, 1, 3} 
%+ 2 \HSU_{2, 2, -2} 
%+ 6 \HSU_{2, 2, 2} 
%+ 6 \HSU_{2, 3, 1} 
%+ 6 \HSU_{3, -2, 1} 
%+ 4 \HSU_{3, 1, -2} 
%+ 4 \HSU_{3, 1, 2} 
%+ 6 \HSU_{3, 2, 1}
%+ 6 \HSU_{-3, -2, 1} 
%+ 4 \HSU_{-3, 1, -2} + 4 \HSU_{-3, 1, 2} + 6 \HSU_{-3, 2, 1} 
%+ 6 \HSU_{-2, -3, 1} + 4 \HSU_{-2, -2, 2} + 2 \HSU_{-2, 1, -3} 
%+ 2 \HSU_{-2, 1, 3} + 2 \HSU_{-2, 2, -2} + 6 \HSU_{-2, 2, 2} 
%+ 6 \HSU_{-2, 3, 1} + 6 \HSU_{2, -3, 1} + 4 \HSU_{2, -2, 2} 
\end{align}

Analise the structure of the expression involving the harmonic sums with double argument $\HSU_{\vec{a}}$ we suggested, that such harmonic sums should combine into the specific combination, in which all entered harmonic sums should have positive and negative indices except one, that is, for example
\begin{eqnarray}
{\mathbf{S}_{\b2,\t1,\b2}}=\HSU_{-2,1,-2}+\HSU_{2,1,-2}+\HSU_{-2,1,2}+\HSU_{2,1,2},
\end{eqnarray}
where the bold numbers in the indices in the left hand side should be positive or negative. Such combinations already appeared in our previous  computations of the anomalous dimension for twist-3 operators in $\mathcal {N}=4$ SYM theory~\cite{Velizhanin:2010cm}.
Then, the basis for the harmonic sums with double argument will consist of all harmonic sums with all positive indices, where all indices except one are bold. For example for weight~$3$ we have 4 harmonic sums with all positive indices
\begin{eqnarray}
{\mathrm{Weight}}_3=\Big\{\HSU_3,\HSU_{2,1},\HSU_{1,2},\HSU_{1,1,1}\Big\},
\end{eqnarray}
which produce the following basis according to our suggestion
\begin{eqnarray}
{\mathrm{Basis}}{\;\!}_{3}=\Big\{\HSB_{\b2,\t1},\HSB_{\t2,\b1},\HSB_{\b1,\t2},\HSB_{\t1,\b2},\HSB_{\t1,\b1,\b1},\HSB_{\b1,\t1,\b1},\HSB_{\b1,\b1,\t1}\Big\}.
\end{eqnarray}
The ansatz, which we used for the reconstruction of the rational part of $F^4$, is included 
255 combinations of the harmonic sums with double arguments $\HSB_{\vec{a}}$ and 1458 usual harmonic sums $\HSH_{\vec{a}}$. This ansatz, being analytically continued, should give the results~Eqs.~(\ref{Ratw4}), (\ref{Ratw3}), (\ref{Ratw2}) and (\ref{Ratw1}). Note one important feature of the analytic continuation of the harmonic sums with the double arguments. According to Eq.~(\ref{BFKLj}) $F^{4}$ should be analytically continued from the positive even values to both even and odd negative and positive values. Multiplying argument by two we obtain, that for the harmonic sums with double argument the analytical continuation should be performed from the positive even values to only \textit{even} negative and positive values. We use our database and code for the analytical continuation of the harmonic sums near integer positive and negative values from Ref.{Velizhanin:2020avm}. Using Eqs.~(\ref{Ratw4}), (\ref{Ratw3}), (\ref{Ratw2}) and (\ref{Ratw1}) for the general expression of the pole parts of the analytical continuation and the results for the fixed values up to $\Delta=97+\delta$ for the lowest regular part (proportional to $\w^0$) we have found the following general result for the rational part of the full expression for $F^{(4)}$
\begin{align}
F^{(4)}_{\mathrm{Rat}}=&
-4096 \Big(
 \HSB_{\b2, \t1, \b4} 
+ 4 \HSB_{\b3, \t1, \b3} 
+ 4 \HSB_{\b4, \t1, \b2} 
- 2 \HSB_{\b1, \b2, \t1, \b3} 
- 4 \HSB_{\b1, \b3, \t1, \b2} 
- 2 \HSB_{\b2, \t1, \b1, \b3} 
%- 2 \HSB_{\b2, \t1, \b2, \b2}  
+ 2 \HSB_{\b2, \t1, \b1, \b1, \b2}   
\nonumber\\&
- 2 \HSB_{\b2, \b1, \t1, \b3} 
- 4 \HSB_{\b3, \t1, \b1, \b2} 
- 4 \HSB_{\b3, \b1, \t1, \b2} 
+ 2 \HSB_{\b1, \b1, \b2, \t1, \b2} 
+ 2 \HSB_{\b1, \b2, \t1, \b1, \b2} 
+ 2 \HSB_{\b1, \b2, \b1, \t1, \b2} 
%+ 2 \HSB_{\b2, \t1, \b1, \b1, \b2}   
- 2 \HSB_{\b2, \t1, \b2, \b2}  
\nonumber\\&
+ 2 \HSB_{\b2, \b1, \t1, \b1, \b2} 
+ 2 \HSB_{\b2, \b1, \b1, \t1, \b2} 
+ 18 \HSB_{\b6, \t1} 
- 12 \HSB_{\b1, \b5, \t1} 
- 13 \HSB_{\b2, \b4, \t1} 
- 13 \HSB_{\b3, \b3, \t1} 
%- 12 \HSB_{\b4, \b2, \t1}   
- 2 \HSB_{\b1, \b2, \b1, \b2, \t1} 
\nonumber\\&
%- 16 \HSB_{\b5, \b1, \t1} 
- 2 \HSB_{\b1, \b3, \b1, \b1, \t1} 
+ 4 \HSB_{\b1, \b1, \b4, \t1} 
+ 6 \HSB_{\b1, \b2, \b3, \t1} 
+ 6 \HSB_{\b1, \b3, \b2, \t1} 
+ 8 \HSB_{\b1, \b4, \b1, \t1} 
+ 6 \HSB_{\b2, \b1, \b3, \t1} 
%- 4 \HSB_{\b2, \b2, \t1, \b2}   
- 4 \HSB_{\b1, \b2, \b2, \b1, \t1} 
\nonumber\\&
+ 8 \HSB_{\b2, \b2, \b2, \t1} 
+ 10 \HSB_{\b2, \b3, \b1, \t1} 
+ 6 \HSB_{\b3, \b1, \b2, \t1} 
+ 10 \HSB_{\b3, \b2, \b1, \t1} 
+ 8 \HSB_{\b4, \b1, \b1, \t1} 
- 2 \HSB_{\b1, \b1, \b2, \b2, \t1} 
- 2 \HSB_{\b1, \b1, \b3, \b1, \t1}   \nonumber\\&
- 12 \HSB_{\b4, \b2, \t1} 
%- 2 \HSB_{\b1, \b2, \b1, \b2, \t1} 
- 4 \HSB_{\b2, \b2, \t1, \b2}   
%- 4 \HSB_{\b1, \b2, \b2, \b1, \t1} 
- 16 \HSB_{\b5, \b1, \t1} 
%- 2 \HSB_{\b1, \b3, \b1, \b1, \t1} 
- 2 \HSB_{\b2, \b1, \b1, \b2, \t1} 
- 4 \HSB_{\b2, \b1, \b2, \b1, \t1} 
- 4 \HSB_{\b2, \b2, \b1, \b1, \t1} 
- 2 \HSB_{\b3, \b1, \b1, \b1, \t1}
\Big)  \nonumber\\&
+32 \Big(
5 \HSH_{7} 
+ 8 \HSH_{-6, 1} 
+ 4 \HSH_{-5,-2} 
+ 20 \HSH_{-5, 2} 
+ 4 \HSH_{-4,-3} 
- 2 \HSH_{-4, 3} 
+ 6 \HSH_{-3, 4} 
- 12 \HSH_{-2, 5}   \nonumber\\&
+ 8 \HSH_{2,-5} 
+ 16 \HSH_{3,-4} 
- 2 \HSH_{3, 4} 
+ 28 \HSH_{4,-3} 
- 4 \HSH_{4, 3} 
+ 44 \HSH_{5,-2} 
- 4 \HSH_{5, 2} 
- 36 \HSH_{6,-1} 
- 16 \HSH_{6, 1}   \nonumber\\&
- 80 \HSH_{-5, 1, 1} 
+ 8 \HSH_{-4,-2, 1} 
- 8 \HSH_{-4, 1,-2} 
- 12 \HSH_{-4, 1, 2} 
- 24 \HSH_{-4, 2, 1} 
+ 16 \HSH_{-3,-3, 1} 
+ 4 \HSH_{-3,-2, 2}   \nonumber\\&
- 16 \HSH_{-3, 1, 3} 
- 4 \HSH_{-3, 2,-2} 
- 16 \HSH_{-3, 3, 1} 
+ 24 \HSH_{-2,-4, 1} 
+ 4 \HSH_{-2,-3, 2} 
+ 8 \HSH_{-2,-2, 3} 
- 4 \HSH_{-2, 2,-3}   \nonumber\\&
- 8 \HSH_{-2, 3,-2} 
- 8 \HSH_{-2, 4, 1} 
- 64 \HSH_{1,-5, 1} 
- 16 \HSH_{1,-4, 2} 
- 16 \HSH_{1,-3, 3} 
- 12 \HSH_{1,-2, 4} 
- 12 \HSH_{1, 2,-4}   \nonumber\\&
- 16 \HSH_{1, 3,-3} 
- 56 \HSH_{1, 4,-2} 
+ 48 \HSH_{1, 5,-1} 
+ 24 \HSH_{2,-4, 1} 
- 4 \HSH_{2,-3,-2} 
+ 16 \HSH_{2,-3, 2} 
- 4 \HSH_{2,-2,-3}   \nonumber\\&
- 24 \HSH_{2, 1,-4} 
- 32 \HSH_{2, 2,-3} 
- 52 \HSH_{2, 3,-2} 
+ 52 \HSH_{2, 4,-1} 
- 12 \HSH_{3,-2, 2} 
- 64 \HSH_{3, 1,-3} 
- 84 \HSH_{3, 2,-2}   \nonumber\\&
+ 52 \HSH_{3, 3,-1} 
- 24 \HSH_{4,-2, 1} 
- 104 \HSH_{4, 1,-2} 
+ 48 \HSH_{4, 2,-1} 
+ 64 \HSH_{5, 1,-1} 
+ 16 \HSH_{5, 1, 1} 
+ 48 \HSH_{-4, 1, 1, 1}   \nonumber\\&
- 16 \HSH_{-3,-2, 1, 1} 
- 16 \HSH_{-3, 1,-2, 1} 
- 16 \HSH_{-2,-3, 1, 1} 
+ 16 \HSH_{-2,-2,-2, 1} 
- 32 \HSH_{-2, 1,-3, 1}   \nonumber\\&
- 8 \HSH_{-2, 1,-2, 2} 
+ 8 \HSH_{-2, 1, 2,-2} 
- 16 \HSH_{-2, 2,-2, 1} 
+ 8 \HSH_{-2, 2, 1,-2} 
+ 64 \HSH_{1,-4, 1, 1} 
- 16 \HSH_{1,-3, 1, 2}   \nonumber\\&
- 32 \HSH_{1,-3, 2, 1} 
+ 32 \HSH_{1,-2, 1, 3} 
+ 32 \HSH_{1,-2, 3, 1} 
+ 16 \HSH_{1, 1,-4, 1} 
- 16 \HSH_{1, 1,-3, 2} 
+ 32 \HSH_{1, 1, 3,-2}   \nonumber\\&
- 32 \HSH_{1, 1, 4,-1} 
- 32 \HSH_{1, 2,-3, 1} 
+ 48 \HSH_{1, 2, 1,-3} 
+ 64 \HSH_{1, 2, 2,-2} 
- 48 \HSH_{1, 2, 3,-1} 
+ 16 \HSH_{1, 3,-2, 1}   \nonumber\\&
+ 128 \HSH_{1, 3, 1,-2}  
- 48 \HSH_{1, 3, 2,-1} 
- 64 \HSH_{1, 4, 1,-1} 
- 64 \HSH_{2,-3, 1, 1} 
+ 8 \HSH_{2,-2, 1,-2} 
- 16 \HSH_{2,-2, 1, 2}   \nonumber\\&
- 32 \HSH_{2,-2, 2, 1} 
- 32 \HSH_{2, 1,-3, 1} 
+ 96 \HSH_{2, 1, 1,-3} 
+ 112 \HSH_{2, 1, 2,-2} 
- 48 \HSH_{2, 1, 3,-1} 
+ 160 \HSH_{2, 2, 1,-2}   \nonumber\\&
- 64 \HSH_{2, 2, 2,-1} 
- 80 \HSH_{2, 3, 1,-1} 
+ 48 \HSH_{3,-2, 1, 1} 
+ 16 \HSH_{3, 1,-2, 1} 
+ 224 \HSH_{3, 1, 1,-2} 
- 48 \HSH_{3, 1, 2,-1}   \nonumber\\&
- 80 \HSH_{3, 2, 1,-1} 
- 64 \HSH_{4, 1, 1,-1} 
+ 32 \HSH_{-2, 1,-2, 1, 1} 
+ 32 \HSH_{-2, 1, 1,-2, 1} 
+ 64 \HSH_{1,-3, 1, 1, 1}   \nonumber\\&
+ 64 \HSH_{1, 1,-3, 1, 1} 
+ 32 \HSH_{1, 1,-2, 1, 2} 
+ 64 \HSH_{1, 1,-2, 2, 1} 
- 96 \HSH_{1, 1, 2, 1,-2} 
+ 32 \HSH_{1, 1, 2, 2,-1}   \nonumber\\&
+ 32 \HSH_{1, 1, 3, 1,-1} 
- 192 \HSH_{1, 2, 1, 1,-2} 
+ 32 \HSH_{1, 2, 1, 2,-1} 
+ 64 \HSH_{1, 2, 2, 1,-1} 
+ 32 \HSH_{1, 3, 1, 1,-1}   \nonumber\\&
+ 64 \HSH_{2,-2, 1, 1, 1} 
- 288 \HSH_{2, 1, 1, 1,-2} 
+ 32 \HSH_{2, 1, 1, 2,-1} 
+ 64 \HSH_{2, 1, 2, 1,-1} 
+ 64 \HSH_{2, 2, 1, 1,-1}   \nonumber\\&
+ 32 \HSH_{3, 1, 1, 1,-1} 
- 128 \HSH_{1, 1,-2, 1, 1, 1}
\Big),\label{F4Rat}
\end{align}
but only if we suggest, that the analytical continuation to negative values and positive values for the harmonic sums with double arguments $\HSB_{\vec{a}}$ should enter with the different sign, on the contrary to the usual harmonic sums $\HSH_{\vec{a}}$. This suggestion leads to the necessity to multiply the definition of $\HSB_{\vec{a}}$ by $i^k$ as, for example
\begin{equation}
{\mathbf{S}_{\mathbf{2},\t1,\mathbf{2}}}(k)=i^k\big(\HSU_{-2,1,-2}+\HSU_{2,1,-2}+\HSU_{-2,1,2}+\HSU_{2,1,2}\big).\label{S_with_factor_I}
\end{equation}
Because $k$ in this case is always even, while analytical continuation to negative and positive values differs by $2$ we get the desired minus sign between two parts after analytical continuation of Eq.~(\ref{F4Rat}). 

\subsubsection{$\z2$ part}

To find $\z2$ part of $F^{(4)}$ we obtain the general expressions for all poles and for the regular part up to $\w^1$. In this case ansatz will consist of 48 harmonic sums with double arguments $\HSB_{a}$ and 162 usual harmonic sums $\HSU_{a}$. However this the most general ansatz did not give result. The solution was found by adding to the ansatz the new type of harmonic sums with double argument, which have plus/minus signs for all indices, i.e., for example
\begin{eqnarray}
{\HSB_{\b2,\b1,\b2}}(k)&=&i^k\big(\HSU_{-2,1,-2}+\HSU_{2,1,-2}+\HSU_{-2,1,2}+\HSU_{2,1,2}\nonumber\\&&\quad\ 
+\HSU_{-2,-1,-2}+\HSU_{2,-1,-2}+\HSU_{-2,-1,2}+\HSU_{2,-1,2}\big).\label{HSB0}
\end{eqnarray}
The combination of the harmonic sums in the brackets in the right-hand side can be reduced to the harmonic sum with all positive indices, but with half argument (up to common factor). However, as this combination in Eq.~(\ref{HSB0}) is multiplied by $i^k$, this is not true anymore, and such sums can not be expressed through other, that is should be considered separately. What is interesting, that there are combinations of $\HSB_{\vec{a}}$ sums in such extended basis, which for weight~$7$ did not give contribution to the rational part, but only to transcendental parts. This combination has the following form, for example
\begin{equation}
2\HSB_{\b2,\b1,\b2,\b1,\t2}-\HSB_{\b2,\b1,\b2,\b1,\b2}\overset{\mathrm{A.C.}}{\Rightarrow}
0\times\sum_{\vec{a}} C_{\mathrm{Rat}}^{\vec{a}}\HSH_{\vec{a}}^{\mathrm{weight}=7}
+\z2\sum_{\vec{a}} C_{\z2}^{\vec{a}}\HSH_{\vec{a}}^{\mathrm{weight}=5}
%+\z3\sum_{\vec{a}} C_{\z3}^{\vec{a}}\HSH_{\vec{a}}^{\mathrm{weight}=4}
+\cdots.
\end{equation} 
Adding all such combinations into the ansatz for $\z2$ part we have found the following general expression for the $\z2$-part of $F^{(4)}$
\begin{align}
 F^{(4)}_{\z2}=&
3072 \Big[
36 \HSB_{\b6, \t1} 
- 18 \HSB_{\b6, \b1} 
- 24 \HSB_{\b1, \b5, \t1} 
+ 12 \HSB_{\b1, \b5, \b1} 
- 26 \HSB_{\b2, \b4, \t1} 
+ 13 \HSB_{\b2, \b4, \b1} \nonumber\\&
- 26 \HSB_{\b3, \b3, \t1} 
+ 13 \HSB_{\b3, \b3, \b1} 
- 24 \HSB_{\b4, \b2, \t1} 
+ 12 \HSB_{\b4, \b2, \b1} 
- 32 \HSB_{\b5, \b1, \t1} 
+ 16 \HSB_{\b5, \b1, \b1} \nonumber\\&
+ 8 \HSB_{\b1, \b1, \b4, \t1} 
- 4 \HSB_{\b1, \b1, \b4, \b1} 
+ 12 \HSB_{\b1, \b2, \b3, \t1} 
- 6 \HSB_{\b1, \b2, \b3, \b1} 
+ 12 \HSB_{\b1, \b3, \b2, \t1} 
- 6 \HSB_{\b1, \b3, \b2, \b1} \nonumber\\&
+ 16 \HSB_{\b1, \b4, \b1, \t1} 
- 8 \HSB_{\b1, \b4, \b1, \b1} 
+ 12 \HSB_{\b2, \b1, \b3, \t1} 
- 6 \HSB_{\b2, \b1, \b3, \b1} 
+ 16 \HSB_{\b2, \b2, \b2, \t1} 
- 8 \HSB_{\b2, \b2, \b2, \b1} \nonumber\\&
+ 20 \HSB_{\b2, \b3, \b1, \t1} 
- 10 \HSB_{\b2, \b3, \b1, \b1} 
+ 12 \HSB_{\b3, \b1, \b2, \t1} 
- 6 \HSB_{\b3, \b1, \b2, \b1} 
+ 20 \HSB_{\b3, \b2, \b1, \t1} 
- 10 \HSB_{\b3, \b2, \b1, \b1} \nonumber\\&
+ 16 \HSB_{\b4, \b1, \b1, \t1} 
- 8 \HSB_{\b4, \b1, \b1, \b1} 
- 4 \HSB_{\b1, \b1, \b2, \b2, \t1} 
+ 2 \HSB_{\b1, \b1, \b2, \b2, \b1} 
- 4 \HSB_{\b1, \b1, \b3, \b1, \t1} 
+ 2 \HSB_{\b1, \b1, \b3, \b1, \b1} \nonumber\\&
- 4 \HSB_{\b1, \b2, \b1, \b2, \t1} 
+ 2 \HSB_{\b1, \b2, \b1, \b2, \b1} 
- 8 \HSB_{\b1, \b2, \b2, \b1, \t1} 
+ 4 \HSB_{\b1, \b2, \b2, \b1, \b1} 
- 4 \HSB_{\b1, \b3, \b1, \b1, \t1} 
+ 2 \HSB_{\b1, \b3, \b1, \b1, \b1} \nonumber\\&
- 4 \HSB_{\b2, \b1, \b1, \b2, \t1} 
+ 2 \HSB_{\b2, \b1, \b1, \b2, \b1} 
- 8 \HSB_{\b2, \b1, \b2, \b1, \t1} 
+ 4 \HSB_{\b2, \b1, \b2, \b1, \b1} \nonumber\\&
- 8 \HSB_{\b2, \b2, \b1, \b1, \t1} 
+ 4 \HSB_{\b2, \b2, \b1, \b1, \b1} 
- 4 \HSB_{\b3, \b1, \b1, \b1, \t1} 
+ 2 \HSB_{\b3, \b1, \b1, \b1, \b1}\Big]\nonumber\\&
+64\z2\Big[
16\Big(
-4 \HSB_{\b3, \t2} 
- 2 \HSB_{\b4, \t1} 
+ 4 \HSB_{\b1, \b2, \t2} 
+\HSB_{\b2, \t1, \b2} 
+ 5 \HSB_{\b2, \b2, \t1} 
+ 4 \HSB_{\b3, \t1, \b1} 
+5 \HSB_{\b3, \b1, \t1} \nonumber\\&
- 2 \HSB_{\b1, \b1, \b2, \t1} 
-2 \HSB_{\b1, \b2, \t1, \b1} 
- 2 \HSB_{\b1, \b2, \b1, \t1} 
-2 \HSB_{\b2, \t1, \b1, \b1} 
- 2 \HSB_{\b2, \b1, \t1, \b1} 
-2 \HSB_{\b2, \b1, \b1, \t1}
\Big)\nonumber\\&
-7 \HSH_{-5} 
+ 17 \HSH_{5} 
- 6 \HSH_{-4,-1} 
+ 12 \HSH_{-4, 1} 
- 10 \HSH_{-3,-2} 
+ 12 \HSH_{-3, 2} 
- 14 \HSH_{-2,-3}\nonumber\\& 
+ 2 \HSH_{-2, 3} 
+ 43 \HSH_{1,-4} 
- 14 \HSH_{1, 4} 
- 3 \HSH_{2,-3} 
- 11 \HSH_{2, 3} 
+ 8 \HSH_{3,-2} 
- 19 \HSH_{3, 2} 
+ 20 \HSH_{4,-1} \nonumber\\&
- 28 \HSH_{4, 1} 
+ 12 \HSH_{-3, 1,-1} 
- 12 \HSH_{-3, 1, 1} 
- 12 \HSH_{-2,-2,-1} 
+ 4 \HSH_{-2,-2, 1} 
+ 20 \HSH_{-2, 1,-2} \nonumber\\&
+ 2 \HSH_{-2, 1, 2} 
+ 12 \HSH_{-2, 2,-1} 
+ 2 \HSH_{-2, 2, 1} 
- 26 \HSH_{1,-2, 2} 
- 16 \HSH_{1, 1,-3} 
+ 8 \HSH_{1, 1, 3} \nonumber\\&
+ 16 \HSH_{1, 2,-2} 
+ 16 \HSH_{1, 2, 2} 
- 20 \HSH_{1, 3,-1} 
+ 32 \HSH_{1, 3, 1} 
+ 14 \HSH_{2,-2, 1} 
+ 2 \HSH_{2, 1,-2} \nonumber\\& 
+ 28 \HSH_{2, 1, 2} 
+ 2 \HSH_{2, 2,-1} 
+ 40 \HSH_{2, 2, 1} 
- 10 \HSH_{3, 1,-1} 
+ 56 \HSH_{3, 1, 1} 
- 24 \HSH_{-2, 1, 1,-1} \nonumber\\& 
+ 24 \HSH_{1,-2, 1, 1} 
- 24 \HSH_{1, 1,-2, 1} 
- 24 \HSH_{1, 1, 2, 1} 
- 48 \HSH_{1, 2, 1, 1} 
- 72 \HSH_{2, 1, 1, 1}
\Big],\label{z2}
\end{align}
where the first part in square brackets should be streakly speaking added to the rational part Eq.~(\ref{F4Rat}). Note, that the obtained result is not full, because there are combinations of the harmonic sums of weight~$5$ with double arguments $\HSB_{\vec{a}}$, which don't give contribution to the $\z2$ part, but only in the higher transcendental parts, such as $\z4$ part or $\z2\z3$ part.

\subsubsection{$\z3$ part}

For the reconstruction of the $\z3$ part we should take into account also the rational part, which appeared during the reconstruction of the $\z2$ part, i.e. the expression in the first square brackets of Eq.~(\ref{z2}). We know the same information as for $\z2$ part, while ansatz contains $20$ harmonic sums with the double argument and 54 usual harmonic sums. We found the following expression for the $\z3$ part of   $F^{(4)}$:
\begin{align}
F^{(4)}_{\z3}=&
32 \Big(
2816 \HSB_{\b2, \t2} 
- 304 \HSB_{\b1, \b1, \b2} 
- 192 \HSB_{\b1, \b2, \t1} 
- 304 \HSB_{\b1, \b2, \b1} 
- 192 \HSB_{\b2, \t1, \b1} 
- 192 \HSB_{\b2, \b1, \t1} 
- 304 \HSB_{\b2, \b1, \b1} \nonumber\\& 
+ 278 \HSH_{-4} 
- 44 \HSH_{4} 
+ 14 \HSH_{-3,-1} 
- 8 \HSH_{-3, 1} 
+ 20 \HSH_{-2,-2} 
- 30 \HSH_{-2, 2} 
- 312 \HSH_{1,-3} 
+ 28 \HSH_{1, 3} \nonumber\\& 
+ 189 \HSH_{2,-2} 
+ 12 \HSH_{2, 2} 
- 309 \HSH_{3,-1} 
+ 4 \HSH_{3, 1} 
- 28 \HSH_{-2, 1,-1} 
+ 20 \HSH_{-2, 1, 1} 
- 4 \HSH_{1,-2, 1} \nonumber\\& 
- 64 \HSH_{1, 1,-2} 
- 20 \HSH_{1, 1, 2} 
+ 4 \HSH_{1, 2, 1} 
+ 28 \HSH_{2, 1, 1}
\Big),\label{z3}
\end{align}
up to some combinations of the harmonic sums of weight $4$ with the double arguments $\HSB_{\vec{a}}$, which don't give contribution to the $\z3$ part, but only in the higher transcendental parts, such as $\z2\z3$ part.

\subsubsection{The rest parts}\label{Section:z4part}

For the less transcedental part we have found the following expressions for the poles, using the results for fixed values of $\Delta$, 
\begin{align}
&\left[F_4\right]_{\mathrm{weight}>3}=
-\frac{1568\pi^4}{9w^3}
+\frac{1}{w^2}
\Big(
-\frac{4192}{45}{\HSH}_1\pi^4
-\frac{1664}{3}\pi^2{\z3}
+22848{\z5}
\Big)\nonumber\\&\
+\frac{1}{w}
\Big(
\pi^4\Big(
\frac{416}{15}{\HSH}_2
+\frac{928}{45}{\HSH}_{-2}
-\frac{832}{45}{\HSH}_{1,1}
\Big)
-\frac{2048}{3}{\HSH}_1\pi^2{\z3}
+15936{\HSH}_1{\z5}
-\frac{44672\pi^6}{2835}
+640{\z3}^2
\Big)\nonumber\\&\
+\pi^4\Big(
\frac{448}{15}{\HSB}_{\b2,\t1}
+\frac{1424}{45}{\HSH}_3
-\frac{200}{9}{\HSH}_{-3}
-\frac{368}{45}{\HSH}_{-2,1}
+\frac{512}{45}{\HSH}_{1,-2}
-\frac{1264}{45}{\HSH}_{1,2}
-\frac{2272}{45}{\HSH}_{2,1}
\Big)\nonumber\\&\
+\pi^2{\z3}\Big(
+544{\HSH}_2
-512{\HSH}_{1,1}
-\frac{448}{3}{\HSH}_{-2}
\Big)
+{\z5}
\Big(7040{\HSH}_{1,1}
-3104{\HSH}_{-2}
-10240{\HSH}_2
\Big)\nonumber\\&\
-\frac{28936}{2835}\pi^6{\HSH}_1
+2240{\HSH}_1{\z3}^2
-\frac{212}{9}\pi^4{\z3}
+30198{\z7}
-560{\z5}\pi^2\nonumber\\&
%\end{align}
%
%for the part, which is proportional to $\pi^4$
%\begin{eqnarray}
%F^{(4)}_{\pi^4,\ w^1}&=&
+w\Bigg[\frac{4}{45}\pi^4 \Big(
-336 \HSB_{\b2,\t2}
-2016 \HSB_{\b3,\t1}
+672 \HSB_{\b1,\b2,\t1}
+672 \HSB_{\b2,\t1,\b1}
+672 \HSB_{\b2,\b1,\t1}\nonumber\\&
+169 \HSH_{-4}
-1129 \HSH_4
+72 \HSH_{-3,1}
+20 \HSH_{-2,-2}
+68 \HSH_{-2,2}
-128 \HSH_{1,-3}
+892 \HSH_{1,3}
-22 \HSH_{2,-2}
\nonumber\\&
+1166 \HSH_{2,2}
+1648 \HSH_{3,1}
+40 \HSH_{-2,1,1}
-336 \HSH_{1,1,2}
-840 \HSH_{1,2,1}
-1344 \HSH_{2,1,1}
\Big)\nonumber\\&
%\label{z4w}
%\end{eqnarray}
%and for all other transcedental contributions
%\begin{align}
%F^{(4)}_{\mathrm{weight}>4,\ w^1}=&\
+\pi^2{\z3}
\Big(
-\frac{512}{3}{\HSB}_{\b2,\t1}
+\frac{32}{3}
\Big(
12\big(-\HSH_{1,-2}
+\HSH_{1,2}
+2\HSH_{2,1}
\big)
+20\HSH_{-3}
-31\HSH_3
\Big)
\Big)\nonumber\\&
+{\z5}\Big(
1024
{\HSB}_{\b2,\t1}
+16\big(
64\HSH_{1,2}
-82\HSH_{-2,1}
-224\HSH_{1,-2}
+16\HSH_{2,1}
+347\HSH_{-3}
+148\HSH_3
\big)
\Big)\nonumber\\&
+64{\z3}^2\Big(
32\HSH_{1,1}
+11\HSH_{-2}
-49\HSH_2
\Big)
+\frac{8}{405}\pi^6\Big(
-195\HSH_{1,1}
+61\HSH_{-2}
+231\HSH_2
\Big)\nonumber\\&
-\frac{2212}{45}\pi^4{\z3}\HSH_1
-\frac{3176}{3}\pi^2{\z5}\HSH_1
+23354{\z7}\HSH_1
-\frac{2048}{19}{\hh53}
+\frac{30720}{19}{\hh71}\nonumber\\&\ 
+\frac{416}{3}\pi^2{\z3}^2
-\frac{26824}{19}{\z3}{\z5}
-\frac{9794059}{6463800}\pi^8\Bigg].\label{z5w}
\end{align}

The most general ansatz for the full analytical expression of the BFKL-pomeron at fourth order (NNNLLA) will consist of all harmonic sums with the single and double argument with weight $7$ and all transcendental numbers up to weight $7$ multiplied by all possible harmonic sums with the single and double argument with weight, which gives the total weight equals to $7$ in the sum with weight of transcendental number. All combinations of the transcendental numbers up to weight $7$ are the following:
\begin{align}
\mathrm{weight}=1:& \big\{\ln\!2\big\},\\
\mathrm{weight}=2:& \big\{\z2,\ln\!2^2\big\},\\
\mathrm{weight}=3:& \big\{\z3,\z2\ln\!2,\ln\!2^3\big\},\\
\mathrm{weight}=4:& \big\{\z4,\hh31,\z3\ln\!2,\z2\ln\!2^2,\ln\!2^4\big\},\\
\mathrm{weight}=5:& \big\{\z5,\hhh311,\z2 \z3,\z4\ln\!2,\hh31\ln\!2,\z3\ln\!2^2,\z2\ln\!2^3,\ln\!2^5\big\},\\
\mathrm{weight}=6:& \big\{\z6,\hhhh3111,\hh51,\z3^2,\z2\hh31,\z5\ln\!2,\hhh311\ln\!2,\z2\z3\ln\!2,\z4\ln\!2^2,\nonumber\\&\quad
\hh31\ln\!2^2,
\z3\ln\!2^3,\z2\ln\!2^4,\ln\!2^6\big\},\\
\mathrm{weight}=7:& \big\{\z7,\hhhhh31111,\hhh331,\hhh511,\z3\z4,\z3\hh31,\z2\z5,\z2\hhh311,\z6\ln\!2,\hhhh3111\ln\!2,\nonumber\\&\quad
\hh51\ln\!2,\z3^2\ln\!2,
\z2\hh31\ln\!2,\z5\ln\!2^2,\hhh311\ln\!2^2,\z2\z3\ln\!2^2,\z4\ln\!2^3,\nonumber\\&\quad
\hh31\ln\!2^3,\z3\ln\!2^4,\z2\ln\!2^5,\ln\!2^7\big\}.
%\mathrm{weight}=1:& \big\{\ln\!2\big\},\\
%\mathrm{weight}=2:& \big\{\pi^2,\ln\!2^2\big\},\\
%\mathrm{weight}=3:& \big\{\z3,\pi^2\ln\!2,\ln\!2^3\big\},\\
%\mathrm{weight}=4:& \big\{\pi^4,\hh31,\z3\ln\!2,\pi^2\ln\!2^2,\ln\!2^4\big\},\\
%\mathrm{weight}=5:& \big\{\z5,\hhh311,\z3\pi^2,\pi^4\ln\!2,\hh31\ln\!2,\z3\ln\!2^2,\pi^2\ln\!2^3,\ln\!2^5\big\},\\
%\mathrm{weight}=6:& \big\{\pi^6,\hhhh3111,\hh51,\z3^2,\hh31\pi^2,\z5\ln\!2,\hhh311\ln\!2,\z3\pi^2\ln\!2,\pi^4\ln\!2^2,\nonumber\\&\quad
%\hh31\ln\!2^2,
%\z3\ln\!2^3,\pi^2\ln\!2^4,\ln\!2^6\big\},\\
%\mathrm{weight}=7:& \big\{\z7,\hhhhh31111,\hhh331,\hhh511,\z3\pi^4,\z3\hh31,\z5\pi^2,\hhh311\pi^2,\pi^6\ln\!2,\hhhh3111\ln\!2,\nonumber\\&\quad
%\hh51\ln\!2,\z3^2\ln\!2,
%\hh31\pi^2\ln\!2,\z5\ln\!2^2,\hhh311\ln\!2^2,\z3\pi^2\ln\!2^2,\pi^4\ln\!2^3,\nonumber\\&\quad
%\hh31\ln\!2^3,\z3\ln\!2^4,\pi^2\ln\!2^5,\ln\!2^7\big\}.
\end{align}  
So, the most general ansatz will consist of 5209 terms.

Beyond the results from Eqs.~(\ref{F4Rat}), (\ref{z2}), (\ref{z3}) and (\ref{z5w}), the constraints are included the absence of the logarithmic terms $(\ln\!\,2)^i$ and terms with ${\mathrm{h}}_{\vec{\boldsymbol{a}}}$. Almost all terms from the ansatz are fixing {\it{except}} of terms, which are proportional to $\z4$, $\z2 z3$, $\z6$ and all similar with prefactor $\z2$. In the next subsection we show, haw to resolve this problem.

\subsection{Harmonic sums with one imagine index}

The resolution of this problem comes with the extension of the usual harmonic sums to the harmonic sums with one imagine index, for example
\begin{equation}
\HSH_{a_1,a_2,\ldots,%a_{n-1},
a_{n}i}(N)=
\sum_{i_1=1}^{N}\frac{\sign(a_1)^{i_1}}{i_1^{|a_1|}}
\sum_{i_2=1}^{i_1}\frac{\sign(a_2)^{i_2}}{i_2^{|a_2|}}\cdots
%\sum_{i_{n-1}=1}^{n-2}\frac{\sign(a_{n-1})^{i_{n-1}}}{i_{n-1}^{|a_{n-1}|}}
%\sum_{i_n=1}^{i_{n-1}}\frac{\Re e\left[(i)^{i_1}\right]}{i_1^{|a_n|}}
\sum_{i_n=1}^{i_{n-1}}\frac{\big(\sign({a_n})i\big)^{i_1}}{i_1^{|a_n|}}.\label{HSwithI}
\end{equation}
Studying the analytical continuation of the harmonic sums following Ref.~\cite{Kotikov:2005gr}, we have found, that the poles parts almost are the same (up to common factor) for the analytically continued harmonic sums, which differ by sign of the last index, or even multiplied by imagine unity, i.e.
\begin{equation}
AC\left[S_{\vec{a},k}\right]_{\mathrm{poles}}
\sim
AC\left[S_{\vec{a},-k}\right]_{\mathrm{poles}}
\sim
AC\left[S_{\vec{a},k\, i}\right]_{\mathrm{poles}}.
\sim
AC\left[S_{\vec{a},-k\, i}\right]_{\mathrm{poles}}
\label{pole_parts_rel}
\end{equation}
Moreover, in Eq.~(\ref{S_with_factor_I}) we added $i$ as a common factor for the harmonic sums with the double argument, but this modification is artificial and may cause some problem, for example, for large $k$ limit. The incorporation $i$ inside the harmonic sum is rather natural and automatically give the desired difference in sign between analytical continuation near negative and positive values.

Note also, that the harmonic sums with imagine indices appeared already during the computations of the anomalous dimension of composite operators in ABJM model with the help of QSC-approach~\cite{Lee:2017mhh,Lee:2019oml}

We have found, that it will more natural if we will perform the analytical continuation not for the harmonic sums with the imagine indices themselves, but for the following sums
\begin{equation}
\HSBI_{a_1,%a_2,
\ldots,%a_{n-1},
a_{n}i}\,(N)=
\sum_{i_1=0}^{\infty}\frac{\sign(a_1)^{N+i_1+1}}{(N+i_1+1)^{|a_1|}}
%\sum_{i_2=0}^{\infty}\frac{\sign(a_2)^{i_2+i_1+1+1}}{(i_2+i_1+1+1)^{|a_2|}}
\ \cdots
%\sum_{i_{n-1}=1}^{n-2}\frac{\sign(a_{n-1})^{i_{n-1}}}{i_{n-1}^{|a_{n-1}|}}
\sum_{i_n=0}^{\infty}\frac{
%\Re e\left[i^{{}^{{\sum_{k=1}^{n}(i_k+1)}}}\right]
\Big(\sign(a_n)\,i\Big)^{{}^{{\sum_{k=1}^{n}(N+i_k+1)}}}
%+
%(-i)^{{}^{{\sum_{k=1}^{n}(N+i_k+1)}}}
}
{\Bigg(N+{\displaystyle{\sum_{k=1}^{n}(i_k+1)}}\Bigg)^{|a_n|}}
%\eta_{a_1,a_2,\dots,a_k}(u)=\sum_{0\le n_1<\ldots< n_k<\infty}\frac{1}{(u+i n_1)^{a_1}%\cdots (u+i n_k)^{a_k}},\label{Eta_function_with_I},
\end{equation}
which is the extended version of $\eta$ function from Eq.~(\ref{Eta_function}) with the positive, negative and imagine indices.

We take as the starting point the part of Eq.~(\ref{F4Rat}), which has $\HSB_{\vec{\mathbf{a}},\t1,\mathbf{k}}$ and multiply the last index by $i$.
To perform the analytical continuation for such sums following Ref.~\cite{Kotikov:2005gr}, it was necessary to have the results for the relations between alternating multiple-zeta values (MZV) and the generalised MZV with the first imagine indices. To find these relations we evaluated numerically the multiple polylogarithm $\Li_{\vec{a}}$ with the help of \texttt{GiNaC} implementation of such computations from Ref.~\cite{Vollinga:2004sn} and using the standard relations between MZV and $\Li_{\vec{a}}$
\begin{equation}
\MZV_{a_1,a_2,\ldots,a_n}=(-1)^{n+1}\Li\big(\big\{|a_1|,|a_2|,\ldots,|a_n|\big\},
\big\{\sign(a_1),\sign(a_2),\ldots,\sign(a_n)\big\}\big),
\end{equation}
where $\sign(\pm k i)=\pm i$ and we wrote the $\Li_{\vec{a}}$ in the right-hand side as it is used in \texttt{GiNaC} code. The relations between the multiple polylogarithm $\Li_{\vec{a}}$, evaluated up to $10^{-6000}$, were found with the help of \texttt{MATHEMATICA} implementations of the \texttt{PSLQ} algorithm through the functions \texttt{FindIntegerNullVector}\footnote{For current purpose we dealt only with the $\Li_{\vec{a}}$ with first imagine indices up to weight $7$. The relations between all multiple polylogarithm of fourth-root of unity up to weight $7$ can be found in Ref.~\cite{Velizhanin:Li}. Our method for these computations is closed to Ref.~\cite{Henn:2015sem}.}.

We performed the analytical continuation of the harmonic sums with the last imagine index following the method, described in Ref.~\cite{Kotikov:2005gr}\footnote{The detailed description and the full results can be found in Ref.~\cite{Velizhanin:ACI}.} and using the obtained results for the relations between MZV with the first imagine index and the relations between the alternating MZV from Ref.~\cite{Blumlein:2009cf}. As the pole parts of the analytical continuation for the harmonic sums which differ by the sign of the last index (\ref{pole_parts_rel}), we suggested, that the general expression for the contribution, including the harmonic sums, or with the generalised $\eta$-function with the double argument~(\ref{Eta_function}) in our case, has the same form, as the part of Eq.~(\ref{F4Rat}), which contains $\HSB_{\boldsymbol{a_1},\ldots,\underline{a_k},\ldots,\boldsymbol{a_n}}$ (not $\HSB_{\vec{\boldsymbol{a}},\underline{a_n}}$) with replacement $\HSB_{\vec{\boldsymbol{a}},{\boldsymbol{a}_n}}\to\HSBI_{\vec{\boldsymbol{a}},{\boldsymbol{a}_n i}}$
\begin{align}
\mathbb{F}^{(4)}_{\mathrm{Rat}}=&
-4096 \Big(
 \HSBI_{\b2, \t1, \b4 i} 
+ 4 \HSBI_{\b3, \t1, \b3 i} 
+ 4 \HSBI_{\b4, \t1, \b2 i} 
- 2 \HSBI_{\b1, \b2, \t1, \b3 i} 
- 4 \HSBI_{\b1, \b3, \t1, \b2 i} 
\nonumber\\&
- 2 \HSBI_{\b2, \t1, \b1, \b3 i} 
- 2 \HSBI_{\b2, \b1, \t1, \b3 i} 
- 4 \HSBI_{\b3, \t1, \b1, \b2 i} 
- 4 \HSBI_{\b3, \b1, \t1, \b2 i} 
- 2 \HSBI_{\b2, \t1, \b2, \b2 i}  
\nonumber\\&
+ 2 \HSBI_{\b1, \b1, \b2, \t1, \b2 i} 
+ 2 \HSBI_{\b1, \b2, \t1, \b1, \b2 i} 
+ 2 \HSBI_{\b1, \b2, \b1, \t1, \b2 i} 
+ 2 \HSBI_{\b2, \b1, \t1, \b1, \b2 i} 
+ 2 \HSBI_{\b2, \b1, \b1, \t1, \b2 i} 
\Big),\label{RatResI}
\end{align}
where in general the result for the harmonic sums with double argument can be written as
\begin{equation}
\mathbb{F}^{(4)}=\mathbb{F}^{(4)}_{\mathrm{Rat}}
+\sum_{\vec{\boldsymbol{a}}}\zeta_a\mathbb{F}^{(4)}_{a}
\end{equation}
Analise this expression we proposed, that only the following term can be added to the general expression for $\mathbb{F}^{(4)}_{\z2}$
\begin{equation}
\mathbb{F}^{(4)}_{\z2}=
\mathbb{C}_{\z2}
%C_{\mathbb{F}^{(4)}_{\z2}} 
\HSBI_{\b2, \t1, \b2 i}.\label{z2ResI}
\end{equation}

Substitute our result for the analytical continuation of $\HSBI_{\vec{\boldsymbol{a}},{\boldsymbol{a}_n i}}$ we have found the following expansion for $\mathbb{F}^{(4)}$
\begin{eqnarray}
\mathbb{F}^{(4)}=\big[F^{(4)}\big]_{\mathrm{double\ argument}}=&&
\sum_{i=-4}^{1}\big[\mathbb{F}_4\big]_{\w^{i}}w^i
\end{eqnarray}
if we put $\mathbb{C}_{\z2}=-2048$:
\begin{eqnarray}
\big[\mathbb{F}_4\big]_{\w^{-4}}&=&-8192 \HSB_{\b2,\t1}+4096 \HSH_{3}-16384 \Ctl \pi +29696 \z3\label{F4HBSw4}\\
\big[\mathbb{F}_4\big]_{\w^{-3}}&=&
8192 \HSB_{\b2,\t2}
+49152 \HSB_{\b3,\t1}
-16384 \HSB_{\b1,\b2,\t1}
-16384 \HSB_{\b2,\t1,\b1}
-16384\HSB_{\b2,\b1,\t1}\nonumber\\&&
+4096 \HSH_{1,3}
+8192 \HSH_{2,2}
+4096 \HSH_{3,1}
-11264\HSH_{4}
-16384 \Ctl \pi  \HSH_{1}
+29696 \z3 \HSH_{1}\nonumber\\&&
+131072 \Li_{i,2,1}
-16384 \Ctl \ln2 \pi
+49152 \h31
+29696 \ln2 \z3
-256 \pi ^4\label{F4HBSw3}\\
\big[\mathbb{F}_4\big]_{\w^{-2}}&=&
-8192 \HSB_{\b2,\t3}
-49152 \HSB_{\b3,\t2}
-155648 \HSB_{\b4,\t1}
+16384 \HSB_{\b1,\b2,\t2}
+65536 \HSB_{\b1,\b3,\t1}
+32768\HSB_{\b2,\t1,\b2}\nonumber\\&&
+16384 \HSB_{\b2,\t2,\b1}
+16384 \HSB_{\b2,1,\t2}
+65536 \HSB_{\b2,\b2,\t1}
+65536 \HSB_{\b3,\t1,\b1}
+65536 \HSB_{\b3,\b1,\t1}\nonumber\\&&
-16384\HSB_{\b1,\b1,\b2,\t1}
-16384 \HSB_{\b1,\b2,\t1,\b1}
-16384 \HSB_{\b1,\b2,\b1,\t1}
-16384 \HSB_{\b2,\t1,\b1,\b1}
-16384 \HSB_{\b2,\b1,\t1,\b1}\nonumber\\&&
-16384\HSB_{\b2,\b1,\b1,\t1}
+\frac{1024}{3} \pi ^2\HSB_{\b2,\t1}
+27648 \HSH_{5}
-14336 \HSH_{1,4}
-20480\HSH_{2,3}
-20480 \HSH_{3,2}\nonumber\\&&
-14336 \HSH_{4,1}
+4096 \HSH_{1,1,3}
+8192 \HSH_{1,2,2}
+4096 \HSH_{1,3,1}
+8192\HSH_{2,1,2}
+8192 \HSH_{2,2,1}\nonumber\\&&
+4096 \HSH_{3,1,1}
-8192 \Ctl \pi  \HSH_{1,1}
+65536 \Li_{i,2,1} \HSH_{1}
+14848 \z3 \HSH_{1,1}
-16384 \Ctl^2 \HSH_{1}\nonumber\\&&
-8192 \Ctl \ln2 \pi \HSH_{1}
+8192 \Ctl \pi  \HSH_{2}
+24576 \h31 \HSH_{1}
+14848 \ln2 \z3 \HSH_{1}
-\frac{1088}{9} \pi ^4\HSH_{1}\nonumber\\&&
-\frac{256}{3} \pi ^2 \HSH_{3}
-14848 \z3 \HSH_{2}
+\frac{131072}{3} \ln2\Li_{i,2,1}
+540672 \Li_{i,4}
+65536 \Li_{i,2,2}\nonumber\\&&
+65536 \Li_{i,3,1}
-\frac{524288}{3} \Li_{i,2,1,1}
-16384 \Ctl^2 \ln2
-\frac{16384}{3} \Ctl \ln2^2 \pi 
-\frac{157696 \Ctl \pi ^3}{9}\nonumber\\&&
-\frac{10240 \hh31\ln2}{3}
-\frac{237568 \hhh311}{3}
+\frac{29696 \ln2^2 \z3}{3}
-\frac{6496 \ln2 \pi ^4}{45}\nonumber\\&&
-\frac{48512 \pi ^2\z3}{9}
+\frac{2113552 \z5}{3}\label{F4HBSw2}
\\
\big[\mathbb{F}_4\big]_{\w^{-1}}&=&
+8192 \HSB_{\b2,\t4}
+49152 \HSB_{\b3,\t3}
+155648 \HSB_{\b4,\t2}
+360448\HSB_{\b5,\t1}
-16384 \HSB_{\b1,\b2,\t3}\nonumber\\&&
-65536 \HSB_{\b1,\b3,\t2}
-147456 \HSB_{\b1,\b4,\t1}
-49152 \HSB_{\b2,\t1,\b3}
-32768 \HSB_{\b2,\t2,\b2}
-16384\HSB_{\b2,\t3,\b1}\nonumber\\&&
-16384 \HSB_{\b2,\b1,\t3}
-65536 \HSB_{\b2,\b2,\t2}
-147456 \HSB_{\b2,\b3,\t1}
-98304 \HSB_{\b3,\t1,\b2}
-65536 \HSB_{\b3,\t2,\b1}\nonumber\\&&
-65536\HSB_{\b3,\b1,\t2}
-147456 \HSB_{\b3,\b2,\t1}
-147456 \HSB_{\b4,\t1,\b1}
-147456 \HSB_{\b4,\b1,\t1}
+16384 \HSB_{\b1,\b1,\b2,\t2}\nonumber\\&&
+32768\HSB_{\b1,\b1,\b3,\t1}
+16384 \HSB_{\b1,\b2,\t1,\b2}
+16384 \HSB_{\b1,\b2,\t2,\b1}
+16384 \HSB_{\b1,\b2,\b1,\t2}
+32768 \HSB_{\b1,\b2,\b2,\t1}\nonumber\\&&
+32768\HSB_{\b1,\b3,\t1,\b1}
+32768 \HSB_{\b1,\b3,\b1,\t1}
+16384 \HSB_{\b2,\t1,\b1,\b2}
+16384 \HSB_{\b2,\t1,\b2,\b1}
+16384 \HSB_{\b2,\t2,\b1,\b1}\nonumber\\&&
+16384\HSB_{\b2,\b1,\t1,\b2}
+16384 \HSB_{\b2,\b1,\t2,\b1}
+16384 \HSB_{\b2,\b1,\b1,\t2}
+32768 \HSB_{\b2,\b1,\b2,\t1}
+32768 \HSB_{\b2,\b2,\t1,\b1}\nonumber\\&&
+32768\HSB_{\b2,\b2,\b1,\t1}
+32768 \HSB_{\b3,\t1,\b1,\b1}
+32768 \HSB_{\b3,\b1,\t1,\b1}
+32768 \HSB_{\b3,\b1,\b1,\t1}
+4096 \z3 \HSB_{\b2,\t1}\nonumber\\&&
-\frac{1024}{3} \HSB_{\b2,\t2} \pi ^2
-2048 \HSB_{\b3,\t1} \pi ^2
+\frac{2048}{3} \HSB_{\b1,\b2,\t1} \pi ^2
+\frac{2048}{3} \HSB_{\b2,\t1,\b1} \pi^2
+\frac{2048}{3} \HSB_{\b2,\b1,\t1} \pi ^2\nonumber\\&&
-43520 \HSH_{6}
+22528 \HSH_{1,5}
+28672 \HSH_{2,4}
+30720 \HSH_{3,3}
+28672 \HSH_{4,2}
+22528 \HSH_{5,1}\nonumber\\&&
-6144 \HSH_{1,1,4}
-10240\HSH_{1,2,3}
-10240 \HSH_{1,3,2}
-6144 \HSH_{1,4,1}
-10240 \HSH_{2,1,3}
-12288 \HSH_{2,2,2}\nonumber\\&&
-10240 \HSH_{2,3,1}
-10240\HSH_{3,1,2}
-10240 \HSH_{3,2,1}
-6144 \HSH_{4,1,1}
+\frac{1664}{3}\HSH_{4} \pi ^2
-\frac{1024}{3} \HSH_{1,3} \pi ^2\nonumber\\&&
-512 \HSH_{2,2} \pi ^2
-\frac{1024}{3} \HSH_{3,1} \pi ^2
+\frac{512}{3}\HSH_{1,1,2} \pi ^2
+\frac{512}{3} \HSH_{1,2,1} \pi ^2
+\frac{512}{3} \HSH_{2,1,1} \pi ^2
-512 \z3\HSH_{3}\nonumber\\&&
-512 \z3 \HSH_{1,2}
-512 \z3\HSH_{2,1}
+16384 \Ctl^2 \HSH_{2}
-16384 \Ctl^2 \HSH_{1,1}
-\frac{16}{5} \HSH_{2} \pi^4
+\frac{64}{9} \HSH_{1,1} \pi ^4\nonumber\\&&
+\frac{512}{45} \ln2 \HSH_{1} \pi ^4
-2240 \z3 \HSH_{1} \pi ^2
-30720 \hhh311 \HSH_{1}
-16384 \Ctl^2\ln2 \HSH_{1}
-10240 \hh31 \ln2 \HSH_{1}\nonumber\\&&
+172192 \z5 \HSH_{1}
+131072 \HSH_{1}\Li_{i,4}
-\frac{14336}{3} \Ctl \HSH_{1} \pi ^3
+\frac{39424}{9} \Ctl \z3 \pi 
-\frac{32768}{3} \Ctl^2 \ln2^2\nonumber\\&&
-10240 \hh31\ln2^2
+61440 \hhhh3111
-10240 \hhh311 \ln2
+\frac{15651360}{17} \ln2 \z5\nonumber\\&&
+\frac{360448}{9} \Li_{i,2,1} \pi^2
+\frac{917504}{3} \ln2 \Li_{i,4}
-\frac{21308833792}{459} \Li_{i,5}
-\frac{21280423936}{459} \Li_{2 i,4}\nonumber\\&&
-\frac{42511892480 }{1377}\Li_{3 i,3}
+\frac{31080448}{153}\Li_{4 i,2}
+\frac{131072}{3} \ln2 \Li_{i,2,2}\nonumber\\&&
+65536 \HSH_{1} \Li_{i,2,2}
-\frac{851968}{3} \Li_{i,2,3}
+\frac{65536}{3}\ln2 \Li_{i,3,1}
+65536 \HSH_{1} \Li_{i,3,1}\nonumber\\&&
-\frac{720896}{9} \Li_{i,3,2}
-\frac{524288}{3} \Li_{i,2,1,2}
-\frac{655360}{3}\Li_{i,2,2,1}
-262144 \Li_{i,3,1,1}\nonumber\\&&
-\frac{626988148}{20655} \pi ^6
-\frac{1984}{45} \ln2^2 \pi ^4
-\frac{93184}{9} \Ctl \ln2 \pi ^3
+\frac{29696}{27}\Ctl^2 \pi ^2\nonumber\\&&
+\frac{112640}{9} \hh31 \pi ^2
+\frac{333440}{27} \ln2 \z3 \pi ^2\label{F4HBSw1}
\\
\big[\mathbb{F}_4\big]_{\w^{0}}&=&
-8192 \HSB_{\b2,\t5}
-49152 \HSB_{\b3,\t4}
-155648 \HSB_{\b4,\t3}
-360448 \HSB_{\b5,\t2}
-696320 \HSB_{\b6,\t1}
+16384\HSB_{\b1,\b2,\t4}\nonumber\\&&
+65536 \HSB_{\b1,\b3,\t3}
+147456 \HSB_{\b1,\b4,\t2}
+262144 \HSB_{\b1,\b5,\t1}
+65536\HSB_{\b2,\t1,\b4}
+49152 \HSB_{\b2,\t2,\b3}\nonumber\\&&
+32768 \HSB_{\b2,\t3,\b2}
+16384 \HSB_{\b2,\t4,\b1}
+16384\HSB_{\b2,\b1,\t4}
+65536 \HSB_{\b2,\b2,\t3}
+147456 \HSB_{\b2,\b3,\t2}\nonumber\\&&
+262144 \HSB_{\b2,\b4,\t1}
+131072 \HSB_{\b3,\t1,\b3}
+98304 \HSB_{\b3,\t2,\b2}
+65536\HSB_{\b3,\t3,\b1}
+65536 \HSB_{\b3,\b1,\t3}\nonumber\\&&
+147456 \HSB_{\b3,\b2,\t2}
+262144 \HSB_{\b3,\b3,\t1}
+196608 \HSB_{\b4,\t1,\b2}
+147456 \HSB_{\b4,\t2,\b1}
+147456\HSB_{\b4,\b1,\t2}\nonumber\\&&
+262144 \HSB_{\b4,\b2,\t1}
+262144 \HSB_{\b5,\t1,\b1}
+262144 \HSB_{\b5,\b1,\t1}
-16384 \HSB_{\b1,\b1,\b2,\t3}
-32768\HSB_{\b1,\b1,\b3,\t2}\nonumber\\&&
-49152 \HSB_{\b1,\b1,\b4,\t1}
-16384 \HSB_{\b1,\b2,\t1,\b3}
-16384 \HSB_{\b1,\b2,\t2,\b2}
-16384 \HSB_{\b1,\b2,\t3,\b1}
-16384\HSB_{\b1,\b2,\b1,\t3}\nonumber\\&&
-32768 \HSB_{\b1,\b2,\b2,\t2}
-49152 \HSB_{\b1,\b2,\b3,\t1}
-32768 \HSB_{\b1,\b3,\t1,\b2}
-32768 \HSB_{\b1,\b3,\t2,\b1}
-32768\HSB_{\b1,\b3,\b1,\t2}\nonumber\\&&
-49152 \HSB_{\b1,\b3,\b2,\t1}
-49152 \HSB_{\b1,\b4,\t1,\b1}
-49152 \HSB_{\b1,\b4,\b1,\t1}
-16384 \HSB_{\b2,\t1,\b1,\b3}
-16384\HSB_{\b2,\t1,\b2,\b2}\nonumber\\&&
-16384 \HSB_{\b2,\t1,\b3,\b1}
-16384 \HSB_{\b2,\t2,\b1,\b2}
-16384 \HSB_{\b2,\t2,\b2,\b1}
-16384 \HSB_{\b2,\t3,\b1,\b1}
-16384\HSB_{\b2,\b1,\t1,\b3}\nonumber\\&&
-16384 \HSB_{\b2,\b1,\t2,\b2}
-16384 \HSB_{\b2,\b1,\t3,\b1}
-16384 \HSB_{\b2,\b1,\b1,\t3}
-32768 \HSB_{\b2,\b1,\b2,\t2}
-49152\HSB_{\b2,\b1,\b3,\t1}\nonumber\\&&
-32768 \HSB_{\b2,\b2,\t1,\b2}
-32768 \HSB_{\b2,\b2,\t2,\b1}
-32768 \HSB_{\b2,\b2,\b1,\t2}
-49152 \HSB_{\b2,\b2,\b2,\t1}
-49152\HSB_{\b2,\b3,\t1,\b1}\nonumber\\&&
-49152 \HSB_{\b2,\b3,\b1,\t1}
-32768 \HSB_{\b3,\t1,\b1,\b2}
-32768 \HSB_{\b3,\t1,\b2,\b1}
-32768 \HSB_{\b3,\t2,\b1,\b1}
-32768\HSB_{\b3,\b1,\t1,\b2}\nonumber\\&&
-32768 \HSB_{\b3,\b1,\t2,\b1}
-32768 \HSB_{\b3,\b1,\b1,\t2}
-49152 \HSB_{\b3,\b1,\b2,\t1}
-49152 \HSB_{\b3,\b2,\t1,\b1}
-49152\HSB_{\b3,\b2,\b1,\t1}\nonumber\\&&
-49152 \HSB_{\b4,\t1,\b1,\b1}
-49152 \HSB_{\b4,\b1,\t1,\b1}
-49152 \HSB_{\b4,\b1,\b1,\t1}
+\frac{1024}{3} \HSB_{\b2,\t3} \pi ^2
+2048 \HSB_{\b3,\t2} \pi ^2\nonumber\\&&
+\frac{19456}{3}\HSB_{\b4,\t1} \pi ^2
-\frac{2048}{3} \HSB_{\b1,\b2,\t2} \pi ^2
-\frac{8192}{3} \HSB_{\b1,\b3,\t1} \pi ^2
-\frac{4096}{3} \HSB_{\b2,\t1,\b2} \pi^2
-\frac{2048}{3} \HSB_{\b2,\t2,\b1} \pi ^2\nonumber\\&&
-\frac{2048}{3} \HSB_{\b2,\b1,\t2} \pi ^2
-\frac{8192}{3} \HSB_{\b2,\b2,\t1} \pi ^2
-\frac{8192}{3}\HSB_{\b3,\t1,\b1} \pi ^2
-\frac{8192}{3} \HSB_{\b3,\b1,\t1} \pi ^2
+\frac{2048}{3} \HSB_{\b1,\b1,\b2,\t1} \pi ^2\nonumber\\&&
+\frac{2048}{3} \HSB_{\b1,\b2,\t1,\b1} \pi^2
+\frac{2048}{3} \HSB_{\b1,\b2,\b1,\t1} \pi ^2
+\frac{2048}{3} \HSB_{\b2,\t1,\b1,\b1} \pi ^2
+\frac{2048}{3} \HSB_{\b2,\b1,\t1,\b1} \pi ^2\nonumber\\&&
+\frac{2048}{3}\HSB_{\b2,\b1,\b1,\t1} \pi ^2
-4096 \z3 \HSB_{\b2,\t2}
-16384 \z3\HSB_{\b3,\t1}
+4096 \z3 \HSB_{\b1,\b2,\t1}
+4096 \z3 \HSB_{\b2,\t1,\b1}\nonumber\\&&
+4096 \z3 \HSB_{\b2,\b1,\t1}
+\frac{448}{15} \HSB_{\b2,\t1} \pi ^4
-1152 \HSH_{-7}
+53120 \HSH_{7}
+1536 \HSH_{1,-6}
-25600 \HSH_{1,6}\nonumber\\&&
+1664 \HSH_{2,-5}
-30720\HSH_{2,5}
+1664 \HSH_{3,-4}
-33280 \HSH_{3,4}
+1536 \HSH_{4,-3}
-33280 \HSH_{4,3}\nonumber\\&&
+2048\HSH_{5,-2}
-30720 \HSH_{5,2}
-25600 \HSH_{6,1}
-1024 \HSH_{1,1,-5}
+6144\HSH_{1,1,5}
-1536 \HSH_{1,2,-4}\nonumber\\&&
+9216 \HSH_{1,2,4}
-1536 \HSH_{1,3,-3}
+10240\HSH_{1,3,3}
-2048 \HSH_{1,4,-2}
+9216 \HSH_{1,4,2}
+6144 \HSH_{1,5,1}\nonumber\\&&
-1536 \HSH_{2,1,-4}
+9216 \HSH_{2,1,4}
-2048 \HSH_{2,2,-3}
+11264 \HSH_{2,2,3}
-2560 \HSH_{2,3,-2}\nonumber\\&&
+11264 \HSH_{2,3,2}
+9216\HSH_{2,4,1}
-1536 \HSH_{3,1,-3}
+10240 \HSH_{3,1,3}
-2560 \HSH_{3,2,-2}\nonumber\\&&
+11264 \HSH_{3,2,2}
+10240 \HSH_{3,3,1}
-2048\HSH_{4,1,-2}
+9216 \HSH_{4,1,2}
+9216 \HSH_{4,2,1}\nonumber\\&&
+6144 \HSH_{5,1,1}
+1024 \HSH_{1,1,2,-3}
+1024\HSH_{1,1,3,-2}
+1024 \HSH_{1,2,1,-3}
+2048 \HSH_{1,2,2,-2}\nonumber\\&&
+1024 \HSH_{1,3,1,-2}
+1024 \HSH_{2,1,1,-3}
+2048\HSH_{2,1,2,-2}
+2048 \HSH_{2,2,1,-2}
+1024 \HSH_{3,1,1,-2}\nonumber\\&&
-1152 \HSH_{5} \pi ^2
-\frac{64}{3}\HSH_{-5} \pi ^2
+\frac{2048}{3}\HSH_{1,4} \pi ^2
+\frac{64}{3} \HSH_{2,-3} \pi ^2
+896 \HSH_{2,3} \pi ^2
+\frac{64}{3} \HSH_{3,-2} \pi ^2\nonumber\\&&
+896 \HSH_{3,2}\pi ^2
+\frac{2048}{3} \HSH_{4,1} \pi ^2
-256 \HSH_{1,1,3} \pi ^2
-\frac{1024}{3} \HSH_{1,2,2} \pi ^2
-256 \HSH_{1,3,1} \pi^2\nonumber\\&&
-\frac{1024}{3} \HSH_{2,1,2} \pi ^2
-\frac{1024}{3} \HSH_{2,2,1} \pi ^2
-256 \HSH_{3,1,1} \pi ^2
-6144 \Ctl \HSH_{4} \pi 
+6144 \Ctl \HSH_{1,3} \pi \nonumber\\&&
+6144 \Ctl \HSH_{2,2} \pi 
+6144 \Ctl\HSH_{3,1} \pi 
-4096 \Ctl \HSH_{1,1,2} \pi 
-4096 \Ctl \HSH_{1,2,1} \pi 
-4096 \Ctl \HSH_{2,1,1} \pi\nonumber\\&&
+15232 \z3\HSH_{4}
-13440 \z3 \HSH_{1,3}
-14720 \z3 \HSH_{2,2}
-13440 \z3 \HSH_{3,1}
+8960 \z3 \HSH_{1,1,2}\nonumber\\&&
+8960 \z3 \HSH_{1,2,1}
+8960 \z3\HSH_{2,1,1}
+\frac{4648}{45} \HSH_{3} \pi ^4
-\frac{1184}{15} \HSH_{1,2} \pi ^4
-\frac{1184}{15}\HSH_{2,1} \pi ^4\nonumber\\&&
+6144 \Ctl \ln2 \HSH_{3} \pi 
-4096 \Ctl \ln2 \HSH_{1,2} \pi
-4096 \Ctl \ln2 \HSH_{2,1} \pi 
+8192 \Ctl^2 \HSH_{1,2}\nonumber\\&&
+8192\Ctl^2 \HSH_{2,1}
-12288 \Ctl^2 \HSH_{3}
+12288 \hh31\HSH_{1,2}
+12288 \hh31 \HSH_{2,1}
-18432 \hh31 \HSH_{3}\nonumber\\&&
+7424 \ln2 \z3 \HSH_{1,2}
+7424 \ln2 \z3 \HSH_{2,1}
-11136 \ln2 \z3 \HSH_{3}
-49152 \HSH_{3}\Li_{i,2,1}\nonumber\\&&
+32768 \HSH_{1,2} \Li_{i,2,1}
+32768 \HSH_{2,1} \Li_{i,2,1}
+\frac{540680}{3} \z5\HSH_{2}
+17344 \z5 \HSH_{1,1}\nonumber\\&&
-\frac{3808}{9} \z3 \HSH_{2} \pi ^2
-\frac{2048}{3} \z3 \HSH_{1,1} \pi ^2
-\frac{35840}{9} \Ctl \HSH_{2} \pi ^3
-1024 \Ctl \HSH_{1,1} \pi ^3\nonumber\\&&
-\frac{8192}{3} \Ctl \ln2^2\HSH_{2} \pi 
-\frac{26624}{3} \hhh311 \HSH_{2}
+8192 \Ctl^2 \ln2\HSH_{2}
+\frac{25600}{3} \hh31 \ln2 \HSH_{2}\nonumber\\&&
+\frac{14848}{3} \ln2^2 \z3 \HSH_{2}
-32768 \HSH_{2} \Li_{i,2,2}
-32768\HSH_{2} \Li_{i,3,1}
-\frac{262144}{3} \HSH_{2} \Li_{i,2,1,1}\nonumber\\&&
+139264 \HSH_{2}\Li_{i,4}
+\frac{65536}{3} \ln2 \HSH_{2} \Li_{i,2,1}
-\frac{752}{9} \ln2 \HSH_{2} \pi ^4
+622592\HSH_{1} \Li_{2 i,4}\nonumber\\&&
+622592 \HSH_{1} \Li_{i,5}
-9120 \ln2 \z5 \HSH_{1}
-2048 \Ctl \z3 \HSH_{1} \pi 
-1024 \Ctl \ln2 \HSH_{1} \pi^3\nonumber\\&&
+\frac{1078736 \HSH_{1} \pi ^6}{2835}
+\frac{1245184}{3} \HSH_{1} \Li_{3 i,3}
-\frac{2048}{3} \Ctl^2 \HSH_{1} \pi ^2
+3840 \hh31 \HSH_{1} \pi ^2\nonumber\\&&
+1856 \ln2 \z3 \HSH_{1}\pi ^2
+8192 \HSH_{1} \Li_{i,2,1} \pi ^2
-\frac{149811200}{9} \ln2 \Li_{i,5}\nonumber\\&&
-\frac{151273472}{9} \ln2 \Li_{2 i,4}
-\frac{2665838 \ln2 \pi ^6}{243}
-\frac{68972 \Ctl \pi ^5}{45}
-\frac{44599 \z3 \pi^4}{180}\nonumber\\&&
-\frac{2048}{3} \Ctl \ln2^2 \pi ^3
-\frac{24320 \hhh311 \pi ^2}{3}
-\frac{2048}{3} \Ctl^2\ln2 \pi ^2
+\frac{1280}{3} \hh31 \ln2 \pi ^2\nonumber\\&&
+\frac{3712}{3} \ln2^2 \z3 \pi ^2
-\frac{4678163 \z5 \pi ^2}{36}
-\frac{190720}{9} \Li_{i,4} \pi^2
+\frac{16384}{3} \ln2 \Li_{i,2,1} \pi ^2\nonumber\\&&
+\frac{8192}{3} \Li_{i,2,2} \pi ^2
+\frac{8192}{3}\Li_{i,3,1} \pi ^2
-\frac{65536}{3} \Li_{i,2,1,1} \pi ^2
-2048 \Ctl \ln2 \z3 \pi \nonumber\\&&
-\frac{60544 \ln2 \z3^2}{3}
+4096 \Ctl^2 \z3
-7968 \hh31 \z3
+\frac{737800}{3} \ln2^2 \z5
+\frac{15131111}{8}\z7\nonumber\\&&
-\frac{304525312}{27} \ln2 \Li_{3 i,3}
+\frac{315392}{3}\ln2 \Li_{4 i,2}
-\frac{243712}{3} \Li_{i,-5,-1}
+731136 \Li_{i,-5,1}\nonumber\\&&
-\frac{487424}{3} \Li_{i,-4,-2}
+487424 \Li_{i,-4,2}
+286720\Li_{i,1,-5}
+286720 \Li_{i,1,5}\nonumber\\&&
+16384 \z3 \Li_{i,2,1}
-487424 \Li_{i,-6}\label{F4HBSw0}
%)
%
%\big[\mathbb{F}_4\big]_{\w^{-4}}&=&
%\\
\end{eqnarray}
The terms $\HSB_{\vec{\boldsymbol{a}}}$ with the double arguments in above expressions are in a full agreement with the corresponding terms from the analytical continuation of Eqs.~(\ref{F4Rat}), (\ref{z2}), (\ref{z3}) and our results for the less transcedental part~(\ref{z5w}), for which we did not find the general expression. Moreover, the expansion up to $\w^1$ for $\z4$ and less transcedental parts of $\mathbb{F}^{(4)}$
\begin{eqnarray}
\big[\mathbb{F}_4\big]_{\w^{1}}&=&
\left(-\frac{448}{15} \HSB_{\b2,\t2}
-\frac{896}{5} \HSB_{\b3,\t1}
+\frac{896}{15} \HSB_{\b1,\b2,\t1}
+\frac{896}{15} \HSB_{\b2,\t1,\b1}
+\frac{896}{15}  \HSB_{\b2,\b1,\t1}\right)\pi^4\nonumber\\&&
+1024 \HSB_{\b2,\t1}\z5
-\frac{512}{3} \HSB_{\b2,\t1}\pi^2\z3+\mathrm{other\ terms}
\end{eqnarray}
which coincide with the corresponding expressions from Eq.~(\ref{z5w}).

After fixing the part with the harmonic sums with double argument we can find the rest part of the general results for the usual harmonic sums, subtracting Eqs.~(\ref{F4HBSw4})-(\ref{F4HBSw0}) from Eqs.~

We found the following expressions for the rational part
\begin{align}
\frac{\mathbb{F}^{(4)}_{\mathrm{Rat}}}{32}=&
%32
%\Big(
5\HSH_7
+8\HSH_{-6,1}
+4\HSH_{-5,-2}
+20\HSH_{-5,2}
+4\HSH_{-4,-3}
-2\HSH_{-4,3}
+6\HSH_{-3,4}
-12\HSH_{-2,5}\nonumber\\&
+8\HSH_{2,-5}
+16\HSH_{3,-4}
-2\HSH_{3,4}
+28\HSH_{4,-3}
-4\HSH_{4,3}
+44\HSH_{5,-2}
-4\HSH_{5,2}
-16\HSH_{6,1}\nonumber\\&
-80\HSH_{-5,1,1}
+8\HSH_{-4,-2,1}
-8\HSH_{-4,1,-2}
-12\HSH_{-4,1,2}
-24\HSH_{-4,2,1}
+16\HSH_{-3,-3,1}\nonumber\\&
+4\HSH_{-3,-2,2}
-16\HSH_{-3,1,3}
-4\HSH_{-3,2,-2}
-16\HSH_{-3,3,1}
+24\HSH_{-2,-4,1}
+4\HSH_{-2,-3,2}\nonumber\\&
+8\HSH_{-2,-2,3}
-4\HSH_{-2,2,-3}
-8\HSH_{-2,3,-2}
-8\HSH_{-2,4,1}
-64\HSH_{1,-5,1}
-16\HSH_{1,-4,2}\nonumber\\&
-16\HSH_{1,-3,3}
-12\HSH_{1,-2,4}
-12\HSH_{1,2,-4}
-16\HSH_{1,3,-3}
-56\HSH_{1,4,-2}
+24\HSH_{2,-4,1}\nonumber\\&
-4\HSH_{2,-3,-2}
+16\HSH_{2,-3,2}
-4\HSH_{2,-2,-3}
-24\HSH_{2,1,-4}
-32\HSH_{2,2,-3}
-52\HSH_{2,3,-2}\nonumber\\&
-12\HSH_{3,-2,2}
-64\HSH_{3,1,-3}
-84\HSH_{3,2,-2}
-24\HSH_{4,-2,1}
-104\HSH_{4,1,-2}
+16\HSH_{5,1,1}\nonumber\\&
+48\HSH_{-4,1,1,1}
-16\HSH_{-3,-2,1,1}
-16\HSH_{-3,1,-2,1}
-16\HSH_{-2,-3,1,1}
+16\HSH_{-2,-2,-2,1}\nonumber\\&
-32\HSH_{-2,1,-3,1}
-8\HSH_{-2,1,-2,2}
+8\HSH_{-2,1,2,-2}
-16\HSH_{-2,2,-2,1}
+8\HSH_{-2,2,1,-2}\nonumber\\&
+64\HSH_{1,-4,1,1}
-16\HSH_{1,-3,1,2}
-32\HSH_{1,-3,2,1}
+32\HSH_{1,-2,1,3}
+32\HSH_{1,-2,3,1}\nonumber\\&
+16\HSH_{1,1,-4,1}
-16\HSH_{1,1,-3,2}
+32\HSH_{1,1,3,-2}
-32\HSH_{1,2,-3,1}
+48\HSH_{1,2,1,-3}
+64\HSH_{1,2,2,-2}\nonumber\\&
+16\HSH_{1,3,-2,1}
+128\HSH_{1,3,1,-2}
-64\HSH_{2,-3,1,1}
+8\HSH_{2,-2,1,-2}
-16\HSH_{2,-2,1,2}\nonumber\\&
-32\HSH_{2,-2,2,1}
-32\HSH_{2,1,-3,1}
+96\HSH_{2,1,1,-3}
+112\HSH_{2,1,2,-2}
+160\HSH_{2,2,1,-2}\nonumber\\&
+48\HSH_{3,-2,1,1}
+16\HSH_{3,1,-2,1}
+224\HSH_{3,1,1,-2}
+32\HSH_{-2,1,-2,1,1}
+32\HSH_{-2,1,1,-2,1}\nonumber\\&
+64\HSH_{1,-3,1,1,1}
+64\HSH_{1,1,-3,1,1}
+32\HSH_{1,1,-2,1,2}
+64\HSH_{1,1,-2,2,1}
-96\HSH_{1,1,2,1,-2}\nonumber\\&
-192\HSH_{1,2,1,1,-2}
+64\HSH_{2,-2,1,1,1}
-288\HSH_{2,1,1,1,-2}
-128\HSH_{1,1,-2,1,1,1}\,,\label{RatFin}
%\Big)
\end{align}
for the part, proportional to $\pi^2$
\begin{align}
\frac{3}{32}\mathbb{F}^{(4)}_{\pi^2}=&
%\frac{32}{3}
%\Big(
17\HSH_5
-7\HSH_{-5}
-6\HSH_{-4,-1}
+12\HSH_{-4,1}
-10\HSH_{-3,-2}
+12\HSH_{-3,2}
-14\HSH_{-2,-3}\nonumber\\&
+2\HSH_{-2,3}
+43\HSH_{1,-4}
-14\HSH_{1,4}
-3\HSH_{2,-3}
-11\HSH_{2,3}
+12\HSH_{3,-2}
-19\HSH_{3,2}\nonumber\\&
+18\HSH_{4,-1}
-28\HSH_{4,1}
+12\HSH_{-3,1,-1}
-12\HSH_{-3,1,1}
-12\HSH_{-2,-2,-1}
+4\HSH_{-2,-2,1}\nonumber\\&
+20\HSH_{-2,1,-2}
+2\HSH_{-2,1,2}
+12\HSH_{-2,2,-1}
+2\HSH_{-2,2,1}
-26\HSH_{1,-2,2}
-16\HSH_{1,1,-3}\nonumber\\&
+8\HSH_{1,1,3}
+8\HSH_{1,2,-2}
+16\HSH_{1,2,2}
-12\HSH_{1,3,-1}
+32\HSH_{1,3,1}
+14\HSH_{2,-2,1}
+2\HSH_{2,1,-2}\nonumber\\&
+28\HSH_{2,1,2}
+40\HSH_{2,2,1}
-12\HSH_{3,1,-1}
+56\HSH_{3,1,1}
-24\HSH_{-2,1,1,-1}
+24\HSH_{1,-2,1,1}\nonumber\\&
-24\HSH_{1,1,-2,1}
-24\HSH_{1,1,2,1}
-48\HSH_{1,2,1,1}
-72\HSH_{2,1,1,1}\,,\label{z2Fin}
%\Big)
\end{align}
and for the part, proportional to $\z3$
\begin{align}
\frac{\mathbb{F}^{(4)}_{\z3}}{32}=&
%32\Big(
103\HSH_{-4}
-32\HSH_4
14\HSH_{-3,-1}
-8\HSH_{-3,1}
+20\HSH_{-2,-2}
-30\HSH_{-2,2}
-100\HSH_{1,-3}\nonumber\\&
+16\HSH_{1,3}
+88\HSH_{2,-2}
-4\HSH_{2,2}
-70\HSH_{3,-1}
-8\HSH_{3,1}
-28\HSH_{-2,1,-1}
+20\HSH_{-2,1,1}\nonumber\\&
-4\HSH_{1,-2,1}
-64\HSH_{1,1,-2}
-12\HSH_{1,1,2}
+12\HSH_{1,2,1}
+36\HSH_{2,1,1}\label{z3Fin}
%\Big)
\end{align}

However, the reconstruction ended on the term, proportional to $\Ctl \pi$. The results for this contribution has the following form:
\begin{eqnarray}
\big[\mathbb{F}_4\big]_{\Ctl\pi}&=&
2048\bigg(\frac{8}{\w^4}
-8 \frac{\HSH_1}{\w^3}
+\frac{4}{\w^2}\Big(
 \HSH_2
-\HSH_{1,1}
\Big)
+\boldsymbol{0}\times\frac{1}{\w}\nonumber\\[2mm]&&
-3 \HSH_4
+3 \HSH_{1,3}
+3 \HSH_{2,2}
+3 \HSH_{3,1}
-2 \HSH_{1,1,2}
-2 \HSH_{1,2,1}
-2 \HSH_{2,1,1}\nonumber\\[2mm]&&
+\w\Big(
9/2\, \HSH_5
-4 \HSH_{1,4}
-4 \HSH_{2,3}
-4 \HSH_{3,2}
-4 \HSH_{4,1}
+2 \HSH_{1,1,3}
+2 \HSH_{1,2,2}\nonumber\\&&\qquad\
+2 \HSH_{1,3,1}
+2 \HSH_{2,1,2}
+2 \HSH_{2,2,1}
+2 \HSH_{3,1,1}
\Big)\bigg).\label{CtlpiBad}
\end{eqnarray}
We did not find any combination of the usual harmonic sums with weight $4$, which gives such result being analytically continued. Note, that if we will following the procedure of the analytical continuation from Ref.~\cite{Kotikov:2005gr} such contribution will absent due to cancellations between subexpressions, for example, for $S_{\b4,\t1,\b2\boldsymbol{i}}$ we have
\begin{equation}
\HSU_{4,1,2\boldsymbol{i}}=
-\mathfrak{S}_{4,1,2\boldsymbol{i}}
+\mathfrak{S}_{4,1}\zeta_{2i}
-\mathfrak{S}_{4}\zeta_{2i}\zeta_1
+\mathfrak{S}_{4}\zeta_{2i,1}
+\zeta_{2i}\zeta_1\zeta_4
-\zeta_{2i,1}\zeta_4
-\zeta_{2i}\zeta_{1,4}
+\zeta_{2i,1,4}
\end{equation}
and first and fourth terms contain expressions, proportional to $\Ctl\pi$, but with opposite signs.
The above relation provides us with the general method for the cancellation of such unwanted terms. In current case, for $\Ctl\pi$-contribution, we should take from Eq.~(\ref{RatResI}) the terms, which ended with indices $\t1$ and $\b2\boldsymbol{i}$, i.e. $\mathfrak{S}_{\vec{\boldsymbol{a}},\t1,\b2\boldsymbol{i}}$, and replaced $\HSBI_{\vec{\boldsymbol{a}}}$ by $\HSB_{\vec{\boldsymbol{a}}}$
\begin{align}
\mathbb{F}^{(4)}_{\Ctl\pi}=&
-4096 \Big(
% \HSB_{\b2, \t1, \b4 i} 
%+ 4 \HSB_{\b3, \t1, \b3 i} 
4 \HSB_{\b4} 
- 2 \HSB_{\b1, \b2} 
- 4 \HSB_{\b1, \b3} 
%\nonumber\\&
%- 2 \HSB_{\b2, \t1, \b1, \b3 i} 
- 2 \HSB_{\b2, \b1} 
%- 4 \HSB_{\b3, \t1, \b1, \b2 i} 
- 4 \HSB_{\b3, \b1} 
%- 2 \HSB_{\b2, \t1, \b2, \b2 i}  
%\nonumber\\&
+ 2 \HSB_{\b1, \b1, \b2} 
%+ 2 \HSB_{\b1, \b2, \t1, \b1, \b2 i} 
+ 2 \HSB_{\b1, \b2, \b1} 
%+ 2 \HSB_{\b2, \b1, \t1, \b1, \b2 i} 
+ 2 \HSB_{\b2, \b1, \b1} 
\Big),\label{CtlpiRes}
\end{align}
where $\HSB_{\vec{\boldsymbol{a}}}$ is defined as in Eq.~(\ref{S_with_factor_I}), that is with multiplication by $i$, what reflects the origin of these terms from the harmonic sums with the last imagine index (\ref{HSwithI}).
%where we replaced $\HSBI_{\vec{\boldsymbol{a}}}$ with $\HSB_{\vec{\boldsymbol{a}}}$.

For the weight $4$ contributions (proportional to $\hh31$, $\z4$ and others) we should take into account the following contribution
\begin{align}
\mathbb{\hat{F}}^{(4)}_{\mathrm{weight}=4}=&
4096 \Big(
%- 2 \HSB_{\b2, \t1, \b1, \b3 i} 
4 \HSB_{\b3} 
%- 2 \HSB_{\b2, \t1, \b2, \b2 i}  
%\nonumber\\&
- 2 \HSB_{\b1, \b2} 
- 2 \HSB_{\b2, \b1} 
\Big)\label{CtlpiRes}
\end{align}
and corresponding contribution from the already know parts, that is the rational part from Eq.~(\ref{RatFin}), $\z2$-part from Eq.~(\ref{z2Fin}), $\z3$-part from Eq.~(\ref{z3Fin}) and others, which will appear in the next steps.

Take into account $\mathbb{\hat{F}}^{(4)}_{\mathrm{weight}=4}$ we have found, that the result for $\hh31$ contribution is:
\begin{align}
\mathbb{F}^{(4)}_{\hh31}=-\frac{15}{8}\mathbb{\hat{F}}^{(4)}_{\mathrm{weight}=4}
+ 6144 \HSH_{-3} 
- 512 \HSH_{-2, -1} 
+ 512 \HSH_{-2, 1} 
+ 1536 \HSH_{1, 2} 
+ 1536 \HSH_{2, 1}.\label{h31Res}
\end{align}

Now we return to the contribution, which is proportional to $\z4$ and left uncomputed in Section~\ref{Section:z4part}. In this case we should take into account the contributions, which will appear after analytical continuation of Eq.~(\ref{RatFin}) and Eq.~(\ref{z2Fin}). Putting all together we have found for $\z4$-contribution
\begin{align}
\frac{1}{90}\mathbb{F}^{(4)}_{\z4}=&\frac{11}{1280}\mathbb{\hat{F}}^{(4)}_{\mathrm{weight}=4}
-\frac{2812}{45} \HSH_{-3}
+\frac{608}{15} \HSH_3
-\frac{2096}{45} \HSH_{-2,-1}
+\frac{304}{45} \HSH_{-2,1}
+\frac{496}{9} \HSH_{1,-2}\nonumber\\&
-\frac{56}{15} \HSH_{1,2}
+\frac{16}{3} \HSH_{2,-1}
-\frac{48}{5} \HSH_{2,1}
-\frac{32}{3}\HSH_{1,1,-1}.\label{z4Res}
\end{align}
So, we resolve our problem from Section~\ref{Section:z4part}.

For the weight $5$ contribution we should take into account the following term
\begin{align}
\mathbb{\hat{F}}^{(4)}_{\mathrm{weight}=5}=&
4096\, \HSB_{\b2}. \label{CtlpiRes}
\end{align}
With this term we obtained, for example, the contribution, which is proportional to $\hhh311$ in the form
\begin{align}
\mathbb{F}^{(4)}_{\hhh311}=&-\frac{63448}{3}\mathbb{\hat{F}}^{(4)}_{\mathrm{weight}=5}
+ 8192 \HSH_{-2} 
+ 1024 \HSH_{2} 
- 15360 \HSH_{1, -1}.\label{h311Res}
\end{align}

In general, the procedure for obtaining the contribution, which is proportional to $\zeta_{\vec{\boldsymbol{a}}}$ is the following:
\begin{itemize}
\item if $\zeta_{\vec{\boldsymbol{a}}}$ comes only from the analytical continuation of Eq.(\ref{RatResI}), then such contribution is pure proportional to 
$\mathbb{\hat{F}}^{(4)}_{\mathrm{weight}(\vec{\boldsymbol{a}})}$;
\item if $\zeta_{\vec{\boldsymbol{a}}}$ comes not only from the analytical continuation of Eq.(\ref{RatResI}), then the basis for such contribution is consist of all usual harmonic sums with weight $\vec{\boldsymbol{a}}$ and $\mathbb{\hat{F}}^{(4)}_{\mathrm{weight}(\vec{\boldsymbol{a}})}$ and we should take into account all other similar contributions;
\item if weight $\vec{\boldsymbol{a}}$ is equal to $6$, then the basis will consist of usual harmonic sums $\HSH_1$ and $\HSH_{-1}$.
\end{itemize}

The final result for the contribution proportional to $\zeta_{\vec{\boldsymbol{a}}}$ with weight more then $2$, except $\z3$ contribution, is the following:
\begin{align}
\mathbb{F}^{(4)}_{\mathrm{weight}>2}=&
-1024\Ctl\pi\Big(
\HSB_{\b4}
-2\HSB_{\b1,\b3}
-2\HSB_{\b2,\b2}
-2\HSB_{\b3,\b1}
+2\HSB_{\b1,\b1,\b2}
+2\HSB_{\b1,\b2,\b1}
+2\HSB_{\b2,\b1,\b1}\Big)\nonumber\\&
+\pi^2\ln2\Big(-256\HSH_{-4}
+256\HSH_4-128\HSH_{-3,-1}
+128\HSH_{-3,1}
-256\HSH_{-2,-2}
+256\HSH_{-2,2}\nonumber\\&
+128\HSH_{1,-3}
-128\HSH_{1,3}
+128\HSH_{3,-1}
-128\HSH_{3,1}
+256\HSH_{-2,1,-1}
-256\HSH_{-2,1,1}\Big)\nonumber\\&
+4096\Big(-\frac{15}{8}\hh31-\frac{29}{32}\ln2\z3
-4\Li_{i,2,1}+\frac{11}{1280}\pi^4+\frac{1}{2}\Ctl\ln2\pi\Big)
\Big(
\HSB_{\b3}
-\HSB_{\b1,\b2}
-\HSB_{\b2,\b1}
\Big)\nonumber\\&
+\pi^2\ln2^2\Big(-128\HSH_{-3}+128\HSH_3-128\HSH_{-2,-1}+128\HSH_{-2,1}\Big)
-2048\Ctl\pi\ln2 \HSH_{-3}\nonumber\\&
+\z3\ln2\Big(6400\HSH_{-3}
-2688\HSH_3
+896\HSH_{-2,-1}
-896\HSH_{-2,1}\Big)
+16384\Li_{i,2,1}\HSH_{-3}\nonumber\\&
+\hh31\Big(6144\HSH_{-3}-512\HSH_{-2,-1}+512\HSH_{-2,1}+1536\HSH_{1,2}+1536\HSH_{2,1}\Big)\nonumber\\&
+\Ctl^2\Big(4096\HSH_{1,-2}+4096\HSH_{2,-1}-8192\HSH_{1,1,-1}\Big)
+\pi^4
\Big(
-\frac{2812}{45}\HSH_{-3}
-\frac{2096}{45}\HSH_{-2,-1}
\nonumber\\&
+\frac{608}{15}\HSH_3
+\frac{304}{45}\HSH_{-2,1}
+\frac{496}{9}\HSH_{1,-2}
-\frac{56}{15}\HSH_{1,2}
+\frac{16}{3}\HSH_{2,-1}
-\frac{48}{5}\HSH_{2,1}
-\frac{32}{3}\HSH_{1,1,-1}
\Big)\nonumber\\&
+\Big(
\frac{14848}{3}\z3\ln2^2
-\frac{8192}{3}\Ctl\pi\ln2^2
+\frac{10240}{3}\hh31\ln2
-\frac{63488}{3}\hhh311
-\frac{1168}{15}\pi^4\ln2\nonumber\\&
+\frac{65536}{3}\Li_{i,2,1}\ln2
-\frac{10000}{9}\pi^2\z3
+\frac{794288}{3}\z5
+204800\Li_{i,4}
-\frac{262144}{3}\Li_{i,2,1,1}
\Big)
\HSB_\b2\nonumber\\&
+\pi^2\ln2^3\Big(\frac{128}{3}\HSH_2-\frac{128}{3}\HSH_{-2}\Big)
+\z3\ln2^2\Big(448\HSH_{-2}-448\HSH_2\Big)\nonumber\\&
+\hh31\ln2\Big(2048\HSH_{-2}+512\HSH_2-5120\HSH_{1,-1}\Big)
+\Ctl^2\ln2\Big(4096\HSH_2-8192\HSH_{1,1}\Big)\nonumber\\&
+\pi^4\ln2\Big(-\frac{2464}{45}\HSH_{-2}+\frac{2336}{45}\HSH_2+\frac{736}{45}\HSH_{1,-1}-\frac{32}{3}\HSH_{1,1}\Big)\nonumber\\&
+\hhh311\Big(8192\HSH_{-2}+1024\HSH_2-15360\HSH_{1,-1}\Big)
+\Li_{i,4}\Big(65536\HSH_{1,-1}-32768\HSH_{-2}\Big)
\nonumber\\&
+(\Li_{i,2,2}+\Li_{i,3,1})\Big(32768\HSH_{1,-1}-16384\HSH_{-2}\Big)
\nonumber\\&
+\z5\Big(-24880\HSH_{-2}+2428\HSH_2+52688\HSH_{1,-1}-6464\HSH_{1,1}\Big)
\nonumber\\&
+\pi^2\z3\Big(-\frac{1936}{3}\HSH_{-2}-24\HSH_2+\frac{2656}{3}\HSH_{1,-1}+\frac{640}{3}\HSH_{1,1}\Big)
\nonumber\\&
+\Ctl\pi^3\Big(\frac{8192}{3}\HSH_{-2}-\frac{17920}{9}\HSH_2+\frac{8192}{3}\HSH_{1,1}\Big)
+\Ctl^2\ln2^2\Big(-\frac{4096}{3}\HSH_{-1}-4096\HSH_1\Big)
\nonumber\\&
+\pi^4\ln2^2\Big(\frac{496}{45}\HSH_1-\frac{496}{15}\HSH_{-1}\Big)
+\hhh311\ln2\Big(10240\HSH_{-1}-15360\HSH_1\Big)
\nonumber\\&
+\pi^2\z3\ln2\Big(\frac{134464}{27}\HSH_{-1}-\frac{128}{3}\HSH_1\Big)
+\Ctl\pi^3\ln2\Big(512\HSH_1-\frac{45056}{9}\HSH_{-1}\Big)
\nonumber\\&
+\Li_{i,3,1}\ln2\Big(32768\HSH_1-\frac{65536}{3}\HSH_{-1}\Big)
+\Li_{i,2,2}\ln2\Big(32768\HSH_1-\frac{32768}{3}\HSH_{-1}\Big)
\nonumber\\&
+\z5\ln2\Big(\frac{609508\HSH_{-1}}{9}+52688\HSH_1\Big)
+\Li_{i,4}\ln2\Big(\frac{262144}{3}\HSH_{-1}+65536\HSH_1\Big)
\nonumber\\&
+30720\hhhh3111\HSH_{-1}
+\frac{19712}{9}\Ctl\pi\z3\HSH_{-1}
-\frac{20480}{3}\Li_{i,5}\HSH_{-1}
+\frac{217088}{9}\Li_{2i,4}\HSH_{-1}
\nonumber\\&
-\frac{425984}{3}\Li_{i,2,3}\HSH_{-1}
-\frac{360448}{9}\Li_{i,3,2}\HSH_{-1}
-\frac{262144}{3}\Li_{i,2,1,2}\HSH_{-1}
\nonumber\\&
-\frac{327680}{3}\Li_{i,2,2,1}\HSH_{-1}
-131072\Li_{i,3,1,1}\HSH_{-1}
+\hh31\pi^2\Big(\frac{668672}{9}\HSH_{-1}-16896\HSH_1\Big)
\nonumber\\&
+\z3^2\Big(\frac{549482\HSH_{-1}}{9}-7208\HSH_1\Big)
+\pi^2\Li_{i,2,1}\Big(\frac{167936}{9}\HSH_{-1}-4096\HSH_1\Big)
\nonumber\\&
+\pi^6\Big(\frac{228686}{2079}\HSH_{-1}+\frac{21702116}{93555}\HSH_1\Big)
+\Ctl^2\pi^2\Big(\frac{14848}{27}\HSH_{-1}+\frac{1024}{3}\HSH_1\Big)
\nonumber\\&
+\hh51\Big(14336\HSH_1-\frac{614656}{9}\HSH_{-1}\Big)
+\hh31\pi^2\Big(14336\HSH_1-\frac{614656}{9}\HSH_{-1}\Big)
\nonumber\\&
-5120\hh31\HSH_1\ln2^2
-\frac{16384}{3}\Ctl^2\ln2^3
-5120\hh31\ln2^3
-\frac{160}{3}\pi^4\ln2^3
\nonumber\\&
+5120\hhh311\ln2^2
+\frac{259424}{27}\pi^2\z3\ln2^2
+\frac{1717148}{9}\z5\ln2^2
+\frac{720896}{3}\Li_{i,4}\ln2^2
\nonumber\\&
+\frac{32768}{3}\Li_{i,2,2}\ln2^2
-\frac{32768}{3}\Li_{i,3,1}\ln2^2
-\frac{83968}{9}\Ctl\pi^3\ln2^2
+\frac{1039204}{9}\z3^2\ln2
\nonumber\\&
+61440\hhhh3111\ln2
-\frac{726656}{9}\hh51\ln2
+\frac{11776}{9}\Ctl\pi\z3\ln2
-\frac{1015808}{3}\Li_{i,5}\ln2
\nonumber\\&
-\frac{1028096}{9}\Li_{2i,4}\ln2
+\frac{286720}{9}\pi^2\Li_{i,2,1}\ln2
-\frac{851968}{3}\Li_{i,2,3}\ln2
\nonumber\\&
-\frac{720896}{9}\Li_{i,3,2}\ln2
-\frac{524288}{3}\Li_{i,2,1,2}\ln2
-\frac{655360}{3}\Li_{i,2,2,1}\ln2
\nonumber\\&
-262144\Li_{i,3,1,1}\ln2
+\frac{15687158}{93555}\pi^6\ln2
+\frac{48128}{27}\Ctl^2\pi^2\ln2
+\frac{104192}{9}\hh31\pi^2\ln2
\nonumber\\&
+\frac{65024}{17}\hhh331
+\frac{97280}{17}\hhh511
-1024\Ctl^2\z3
+8224\hh31\z3
+\frac{28887}{340}\pi^4\z3
\nonumber\\&
+\frac{79651219}{612}\pi^2\z5
-\frac{255141383}{136}\z7
+487424\Li_{i,-6}
+\frac{190720}{9}\pi^2\Li_{i,4}
\nonumber\\&
+\frac{243712}{3}\Li_{i,-5,-1}-731136\Li_{i,-5,1}
+\frac{487424}{3}\Li_{i,-4,-2}-487424\Li_{i,-4,2}
\nonumber\\&
-286720\Li_{i,1,-5}-286720\Li_{i,1,5}
+24576\z3\Li_{i,2,1}
-\frac{8192}{3}\pi^2\Li_{i,2,2}
\nonumber\\&
-\frac{8192}{3}\pi^2\Li_{i,3,1}
+\frac{65536}{3}\pi^2\Li_{i,2,1,1}
+\frac{21628}{9}\Ctl\pi^5
+\frac{24320}{3}\hhh311\pi^2.\label{ReFinalWeight3}
\end{align}
Note, that our procedure did not work for $\Ctl^2$ contribution, which consist of usual harmonic sums. This may indicate that this contribution will appear in the real computations.

The final result is given by the rational part with harmonic sums $\mathfrak{S}_{\vec{\boldsymbol{a}},k i}$ with last imagine index from Eq.~(\ref{RatResI}),
$\z2$-part with harmonic sums $\mathfrak{S}_{\vec{\boldsymbol{a}},k i}$ with last imagine index from Eq.~(\ref{z2ResI}) with $\mathbb{C}_{\z2}=-2048$, the rational part with the usual harmonic sums from Eq.~(\ref{RatFin}), the $\z2$-part with the usual harmonic sums from Eq.~(\ref{z2Fin}), the $\z3$-part with the usual harmonic sums from Eq.~(\ref{z3Fin}) and all other contributions, listed in Eq.~(\ref{ReFinalWeight3}).

\section{Non-zero conformal spin}\label{Section:ConfSpin}

Some times ago the quantum spectral curve approach was applied for the computations the BFKL intercept $j(n)=S(0,n)+1$ for arbitrary integer conformal spin $n$~\cite{Alfimov:2018cms}. The authors of~\cite{Alfimov:2018cms} compute a lot of results for the BFKL intercept for different values of conformal spin up to fourth order in the weak coupling expansion. Using the generalised maximal transcedentality principle~\cite{Kotikov:2002ab}\cite{Kotikov:2007cy} they reconstruct the general expression for the arbitrary conformal spin only up to third order. With the help of number theory and LLL-algorithm it is rather simple task to make this in fourth order using the results for fixed values from~\cite{Alfimov:2018cms}. There are 239 harmonic sums with transcedentality 7, which form the basis for fourth order results and odd values within range from 1 to 91 for BFKL intercept are available from Ref.~\cite{Alfimov:2018cms} (however, the final results for $n=4k+1$ and $n=4k-1$ are different, so, we have about 25 values). This is enough with large reserve for the applicability of the LLL-algorithm, and with standart procedure for such case we found the following general expression for the forth order rational part of the BFKL intercept for $n=4k-1$:\footnote{All other parts can be found in Ref.~\cite{Alfimov:2018cms}}
\bea
j_{{}_\mathrm{NNNLLA}}&=&-256 \bigg(2 \HSH_{-6,1}
+10 \HSH_{-5,2}
-\HSH_{-4,-3}
+\HSH_{-4,3}
-2 \HSH_{-3,-4}
+4 \HSH_{-3,4}
-4 \HSH_{-2,-5}\nonumber\\&&\quad
-2 \HSH_{-2,5}
-4 \HSH_{2,-5}
-6 \HSH_{3,-4}
-\HSH_{4,-3}
-\HSH_{4,3}
-2 \HSH_{5,2}
-4 \HSH_{6,1}
-20 \HSH_{-5,1,1}\nonumber\\&&\quad 
+2 \HSH_{-4,-2,1}
-6 \HSH_{-4,1,2}
-6 \HSH_{-4,2,1}
+4 \HSH_{-3,-3,1}
+2 \HSH_{-3,-2,2}
+2 \HSH_{-3,1,-3}\nonumber\\&&\quad
-4 \HSH_{-3,1,3}
-4 \HSH_{-3,3,1}
+6 \HSH_{-2,-4,1}
+2 \HSH_{-2,-3,2}
-2 \HSH_{-2,-2,-3}
+2 \HSH_{-2,-2,3}\nonumber\\&&\quad
+4 \HSH_{-2,1,-4}
+2 \HSH_{-2,2,-3}
-2 \HSH_{-2,4,1}
-16 \HSH_{1,-5,1}
-8 \HSH_{1,-4,2}
-2 \HSH_{1,-3,3}\nonumber\\&&\quad
-8 \HSH_{1,-2,4}
+8 \HSH_{1,2,-4}
+6 \HSH_{1,3,-3}
+6 \HSH_{2,-4,1}
+8 \HSH_{2,-3,2}
+2 \HSH_{2,-2,3}
+8 \HSH_{2,1,-4}\nonumber\\&&\quad
+8 \HSH_{2,2,-3}
-6 \HSH_{3,-2,2}
+6 \HSH_{3,1,-3}
-6 \HSH_{4,-2,1}
+4 \HSH_{5,1,1}
-8 \HSH_{2,1,1,-3}
+4 \HSH_{3,1,-2,1}
%+12 \HSH_{-4,1,1,1}
%-4 \HSH_{-3,-2,1,1}
\nonumber\\&&\quad
-4 \HSH_{-3,1,-2,1}
-4 \HSH_{-2,-3,1,1}
+4 \HSH_{-2,-2,-2,1}
-8 \HSH_{-2,1,-3,1}
-4 \HSH_{-2,1,-2,2}\nonumber\\&&\quad
-4 \HSH_{-2,1,1,-3}
-4 \HSH_{-2,2,-2,1}
+16 \HSH_{1,-4,1,1}
-8 \HSH_{1,-3,1,2}
-8 \HSH_{1,-3,2,1}
+8 \HSH_{1,-2,1,3}\nonumber\\&&\quad
+8 \HSH_{1,-2,3,1}
+4 \HSH_{1,1,-4,1}
-8 \HSH_{1,1,-3,2}
-4 \HSH_{1,1,-2,3}
-8 \HSH_{1,1,2,-3}
-8 \HSH_{1,2,-3,1}\nonumber\\&&\quad
-8 \HSH_{1,2,1,-3}
+4 \HSH_{1,3,-2,1}
-16 \HSH_{2,-3,1,1}
-8 \HSH_{2,-2,1,2}
-8 \HSH_{2,-2,2,1}
-8 \HSH_{2,1,-3,1}\nonumber\\&&\quad
%-8 \HSH_{2,1,1,-3}
%+4 \HSH_{3,1,-2,1}
+12 \HSH_{-4,1,1,1}
-4 \HSH_{-3,-2,1,1}
+12 \HSH_{3,-2,1,1}
+8 \HSH_{-2,1,-2,1,1}
+8 \HSH_{-2,1,1,-2,1}\nonumber\\&&\quad
+16 \HSH_{1,-3,1,1,1}
+16 \HSH_{1,1,-3,1,1}
+16 \HSH_{1,1,-2,1,2}
+16 \HSH_{1,1,-2,2,1}
+16 \HSH_{2,-2,1,1,1}\nonumber\\&&\quad
-32 \HSH_{1,1,-2,1,1,1}
+7 \HSH_{-7}+3 \HSH_7
\bigg)\label{CS}
\eea
We found, that the corresponding rational results for the conformal spin with $n=4k+1$ can be obtained easily with the substitution
\be
S_{i_1,i_2,\ldots,i_\ell}\to \left[\prod_{\;k=1}^\ell {\mathrm {sign}}(i_k)\right] S_{i_1,i_2,\ldots,i_\ell}\,.
\ee
In general the results for the conformal spin with $n=4k-1$ and with with $n=4k+1$ are related by means of the analytical continuation procedure from the odd argument of the harmonic sums to the even argument of the harmonic sums and vice versus. 

\section{Conclusion}

We found the general analytical expression for the BFKL-pomeron in the next-to-next-to-next-to-leading logarithm approximation (NNNLLA) in the planar $\mathcal{N}=4$ SYM theory. Using Quantum Spectral Curve approach we obtain result for the wide range of the fixed values of the quantum numbers with their expansion over small auxiliary parameter $\delta$. Using suggestions about possible set of the functions, which should enter into final expression, we reconstruct the general expression for the pole and regular parts from the corresponding pole and regular parts of the fixed values with the help of the number theory. Writing the more general ansatz, which consist of from the harmonic sums with single and double arguments and from the harmonic sums with last imagine index we obtained the final result in Eqs.~(\ref{RatResI}), (\ref{z2ResI}) with $\mathbb{C}_{\z2}=-2048$,  (\ref{RatFin}), (\ref{z2Fin}), (\ref{z3Fin}) and (\ref{ReFinalWeight3}).
We reconstructed also the rational part of the intercept function for arbitrary conformal spin at fourth order (\ref{CS}), which remained uncomputed in Ref.~\cite{Alfimov:2018cms}.

As we completely find the results in planar case, we want remind, that recently, the non-planar result for the anomalous dimension in $\mathcal{N}=4$ SYM theory was obtained for the arbitrary Lorenz spin of twist-two operator~\cite{Kniehl:2020rip,Kniehl:2021ysp}, which allow extract corresponding result in BFKL limit. This result is the following:
\begin{equation}
\gamma_{\mathrm{uni},\mathrm{np}}^{(3)}\overset{M=-1+\omega}{=}
-\frac{192 }{\omega^2}{\z2} {\z3}
+\frac{4 }{\omega}
\left(
31 {\z6}
-2 {\z3}^2
\right)
+2 \left(
20 {\z2} {\z5}
+280 {\z3} {\z4}
+69 {\z7}
\right)+\mathcal{O}(\omega)\,.\qquad\label{BFKLlimit}
\end{equation}
Obtained expressions implies that the BFKL equation receives a non-planar contribution in the next-to-next-to-leading logarithmic approximation (NNLLA), which is the third order of perturbation theory. However the information from Eq.~(\ref{BFKLlimit}) is not enough to find the general result in this case.

The new type of harmonic sums that appeared in our results necessitates taking such sums into account when calculating the multigluon amplitudes from which the BFKL equations obtained. Moreover, we found, that Catalan constant enter into the results for twist-2 operators.

\acknowledgments

I thanks Ivan Surnin for numerous fruitful discussions and for collaboration at the initial stages of this work, M.~Alfimov for the sharing the result for $c_{1,1}$ and V.S.~Fadin and A.~Shuvaev for useful discussions.
This research is supported by RFBR grants 19-02-00983-a.

\appendix
\section{Computed values}\label{Sec:CompVal}
Here we give our results for some values of $\Delta=N+\delta$.
Our results for $\Delta=7+\delta$ is
\begin{align}
\w_{7+\delta}^{\mathrm{NNNLLA}}=&\frac{20480}{\delta ^7}
+\frac{1}{\delta ^5}
\bigg(
2048 \pi ^2-\frac{195584}{9}
\bigg)
+\frac{1}{\delta ^4}
\bigg(
-16896 \z3
+\frac{2816 \pi^2}{3}
+\frac{2169088}{135}
\bigg)\nonumber\\&
+\frac{1}{\delta ^3}
\bigg(
\frac{10912 \pi ^4}{45}
-\frac{28160 \z3}{3}
-\frac{11072 \pi^2}{27}
-\frac{12914464}{2025}
\bigg)
+\frac{1}{\delta ^2}
\bigg(
\frac{256 \pi ^2 \z3}{3}
+\frac{37696 \z3}{9}\nonumber\\&
-27200 \z5
+\frac{23056 \pi ^4}{135}
+\frac{175168 \pi ^2}{135}
-\frac{195439808}{30375}
\bigg)
+\frac{1}{\delta }
\bigg(
1152 \z3^2
+\frac{11264 \pi ^2 \z3}{9}\nonumber\\&
-\frac{1967104 \z3}{135}
-29216 \z5
+\frac{72544 \pi ^6}{2835}
-\frac{1528 \pi ^4}{45}
-\frac{4156592 \pi ^2}{6075}
+\frac{2143119808}{151875}
\bigg)\nonumber\\&
-\frac{292 \pi ^4 \z3}{45}
+392 \pi ^2 \z3
+\frac{19501892 \z3}{2025}
+\frac{1760 \pi ^2 \z5}{3}
+\frac{1048 \z5}{9}
-31310 \z7
\nonumber\\&
+\frac{216964 \pi ^6}{8505}
+\frac{589589 \pi ^4}{6075}
+\frac{30500851 \pi^2}{91125}
-\frac{118737355147}{6834375}
-\frac{13024 \z3^2}{3}\nonumber\\&
+\delta  \bigg(
3512 \z3 \z5
-112 \pi ^2 \z3^2
-\frac{9152 \z3^2}{9}
+\frac{12166 \pi ^4 \z3}{135}
-\frac{71132 \pi ^2 \z3}{405}
+\frac{7619 \pi ^6}{25515}
\nonumber\\&
+\frac{17468 \pi ^2 \z5}{9}
-\frac{1780226 \z5}{135}
-\frac{128447 \z7}{3}
+\frac{608149 \pi ^8}{340200}
-\frac{1068481 \pi ^4}{20250}\nonumber\\&
-\frac{92191426 \z3}{30375}
-\frac{976481587 \pi ^2}{5467500}
+\frac{14246903001653}{820125000}\bigg)
\end{align}

%\begin{align}
%\w_{9+\delta}^{\mathrm{NNNLLA}}=&
%\frac{20480}{\delta ^7}
%+\frac{1}{\delta ^5}
%\bigg(
%\frac{193280}{9}
%-\frac{8192 \pi ^2}{3}
%\bigg)
%+\frac{1}{\delta ^4}
%\bigg(
%16896 \z3
%-\frac{3200 \pi ^2}{3}
%-\frac{250304}{21}
%\bigg)\nonumber\\&
%+\frac{1}{\delta ^3}
%\bigg(
%12800 \z3
%-\frac{1568 \pi ^4}{9}
%-\frac{10640 \pi^2}{27}
%+\frac{91195198}{19845}
%\bigg)
%+\frac{1}{\delta ^2}
%\bigg(
%-\frac{1664 \pi^2 \z3}{3}
%+\frac{14000 \z3}{9}\nonumber\\&
%+22848 \z5
%-\frac{5240 \pi^4}{27}
%-\frac{617572 \pi^2}{567}
%+\frac{15864960577}{2083725}
%\bigg)
%+\frac{1}{\delta }
%\bigg(
%640 \z3^2
%-\frac{12800 \pi ^2 \z3}{9}\nonumber\\&
%+\frac{749396 \z3}{63}
%+33200 \z5
%-\frac{44672 \pi ^6}{2835}
%-\frac{2452 \pi ^4}{81}
%+\frac{27139529 \pi ^2}{59535}
%-\frac{7544431812287}{583443000}
%\bigg)\nonumber\\&
%+\frac{14000 \z3^2}{3}
%-\frac{212 \pi ^4 \z3}{9}
%-\frac{15710 \pi ^2 \z3}{27}
%-\frac{498179681\z3}{79380}
%-560 \pi ^2 \z5
%+8190 \z5 \nonumber\\&
%+30198 \z7
%-\frac{36170 \pi^6}{1701}
%-\frac{740371 \pi ^4}{9720}
%-\frac{13409915711 \pi^2}{66679200}
%+\frac{4979134106860537}{367569090000}
%\end{align}

For $\Delta=9+\delta$ we have found
\begin{align}
\w_{9+\delta}^{\mathrm{NNNLLA}}=&
\frac{20480}{\delta^7}
+\frac{1}{\delta^5}\Big(\frac{193280}{9}-\frac{8192}{3}\pi^2\Big)
+\frac{1}{\delta^4}\Big(16896{\z3}-\frac{3200}{3}\pi^2-\frac{250304}{21}\Big)\nonumber\\&
+\frac{1}{\delta^3}\Big(12800{\z3}-\frac{1568}{9}\pi^4-\frac{10640}{27}\pi^2+\frac{91195198}{19845}\Big)
+\frac{1}{\delta^2}\Big(-\frac{1664}{3}\pi^2{\z3}\nonumber\\&
+\frac{14000}{9}{\z3}+22848{\z5}-\frac{5240}{27}\pi^4-\frac{617572}{567}\pi^2+\frac{15864960577}{2083725}\Big)\nonumber\\&
+\frac{1}{\delta}\Big(640{\z3}^2-\frac{12800}{9}\pi^2{\z3}+\frac{749396}{63}{\z3}+33200{\z5}-\frac{44672}{2835}\pi^6-\frac{2452}{81}\pi^4\nonumber\\&
+\frac{27139529}{59535}\pi^2-\frac{7544431812287}{583443000}\Big)
+\Big(\frac{14000{\z3}^2}{3}-\frac{212\pi^4{\z3}}{9}-\frac{15710\pi^2{\z3}}{27}\nonumber\\&
-\frac{498179681{\z3}}{79380}-560\pi^2{\z5}+8190{\z5}+30198{\z7}-\frac{36170\pi^6}{1701}-\frac{740371\pi^4}{9720}\nonumber\\&
-\frac{13409915711\pi^2}{66679200}+\frac{4979134106860537}{367569090000}\Big)
+{\delta}\Big(-\frac{2048{\hh53}}{19}+\frac{30720{\hh71}}{19}\nonumber\\&
+\frac{416\pi^2{\z3}^2}{3}+\frac{7880{\z3}^2}{9}-\frac{26824{\z3}{\z5}}{19}-\frac{2765\pi^4{\z3}}{27}+\frac{74311\pi^2{\z3}}{378}-\frac{5098462561{\z3}}{5556600}\nonumber\\&
-\frac{19850\pi^2{\z5}}{9}+\frac{7251613{\z5}}{756}+\frac{291925{\z7}}{6}-\frac{9794059\pi^8}{6463800}-\frac{8117\pi^6}{1458}+\frac{101274169\pi^4}{4082400}\nonumber\\&
+\frac{13963710989069\pi^2}{168031584000}-\frac{58331311924436775331}{4940128569600000}\Big)\nonumber\\&
+{\delta^2}\Big(-\frac{512000{\hh53}}{1539}+\frac{2560000{\hh71}}{513}-648{\z3}^3+100\pi^2{\z3}^2-\frac{2441683{\z3}^2}{756}\nonumber\\&
+\frac{796000{\z3}{\z5}}{171}-\frac{5713\pi^6{\z3}}{5670}-\frac{5785\pi^4{\z3}}{108}-\frac{955786049\pi^2{\z3}}{1905120}+\frac{365614901707651{\z3}}{56010528000}\nonumber\\&
-\frac{4919\pi^4{\z5}}{90}-\frac{9625\pi^2{\z5}}{12}-\frac{1752656477{\z5}}{635040}-\frac{1117\pi^2{\z7}}{2}+\frac{1569445{\z7}}{144}+\frac{163495{\z9}}{6}\nonumber\\&
-\frac{15679861\pi^8}{9307872}-\frac{8940551\pi^6}{2857680}-\frac{8972399761\pi^4}{762048000}+\frac{7061737981950977\pi^2}{35286632640000}
\nonumber\\&
+\frac{149340994999989268313}{19211611104000000}\Big)\nonumber\\&
+{\delta^3}\Big(-\frac{1590660288163465501505447}{1742877359354880000000}
+\mathrm{transcedental\ part}\Big)\nonumber\\&
+{\delta^4}\Big(-\frac{238013867699094016290064651}{27111425589964800000000}
+\mathrm{transcedental\ part}\Big)\label{N9}
\end{align}
while for $\Delta=13+\delta$
\begin{align}
%xxxxxxxxxx
\w_{13+\delta}^{\mathrm{NNNLLA}}=&
+\frac{20480}{\delta^7}
+\frac{1}{\delta^5}\Big(\frac{998144}{45}
-\frac{8192}{3}\pi^2\Big)
+\frac{1}{\delta^4}\Big(
16896{\z3}
-\frac{6272}{5}\pi^2
-\frac{1064053312}{111375}\Big)\nonumber\\&
+\frac{1}{\delta^3}\Big(
\frac{75264}{5}{\z3}
-\frac{1568}{9}\pi^4
-\frac{54544}{135}\pi^2
+\frac{694358600506}{385914375}\Big)
+\frac{1}{\delta^2}\Big(
-\frac{1664}{3}\pi^2{\z3}\nonumber\\&
+\frac{20720}{9}{\z3}
+22848{\z5}
-\frac{51352}{225}\pi^4
-\frac{487751716}{334125}\pi^2
+\frac{17453735274893639}{1337193309375}\Big)\nonumber\\&
+\frac{1}{\delta}\Big(640{\z3}^2-\frac{25088\pi^2{\z3}}{15}+\frac{326956924{\z3}}{22275}+\frac{195216{\z5}}{5}-\frac{44672\pi^6}{2835}-\frac{451948\pi^4}{10125}
\nonumber\\&
+\frac{281842719847}{1157743125}\pi^2
-\frac{592588073728926198271}{37066998535875000}\Big)
+\Big(5488{\z3}^2
-\frac{212}{9}\pi^4{\z3}
\nonumber\\&
-\frac{221858}{225}\pi^2{\z3}
-\frac{6202054022459}{1543657500}{\z3}
+\frac{1021762}{75}{\z5}
+30198{\z7}
-\frac{1325347153}{13365000}\pi^4
\nonumber\\&
-560\pi^2{\z5}
-\frac{50638}{2025}\pi^6
+\frac{3090695744712137}{25674111540000}\pi^2
+\frac{3187191687701634023147087}{256874299853613750000}\Big)\nonumber\\&
+{\delta}\Big(
-\frac{2048}{19}{\hh53}
+\frac{30720}{19}{\hh71}
+\frac{416}{3}\pi^2{\z3}^2
+\frac{60648}{25}{\z3}^2
-\frac{26824}{19}{\z3}{\z5}
-\frac{27097}{225}\pi^4{\z3}
\nonumber\\&
+\frac{43195139}{133650}\pi^2{\z3}
-\frac{15169289833832009}{2674386618750}{\z3}
-\frac{38906}{15}\pi^2{\z5}
+\frac{203477159}{17820}{\z5}
\nonumber\\&
+\frac{572173}{10}{\z7}
-\frac{9794059}{6463800}\pi^8
-\frac{20909}{2430}\pi^6
+\frac{238364927177}{26462700000}\pi^4
\nonumber\\&
-\frac{1262877773588119330097}{3558431859444000000}\pi^2
-\frac{671656972974035282101726707481}{113928889471074770400000000}\Big)
\nonumber\\&
+{\delta^2}\Big(-\frac{200704}{513}{\hh53}
+\frac{1003520}{171}{\hh71}
+\frac{312032}{57}{\z3}{\z5}
-\frac{4919}{90}\pi^4{\z5}
-\frac{673687}{8100}\pi^4{\z3}
\nonumber\\&
-648{\z3}^3
+\frac{588}{5}\pi^2{\z3}^2
-\frac{2024457697}{445500}{\z3}^2
-\frac{5713}{5670}\pi^6{\z3}
-\frac{30040081797739}{37047780000}\pi^2{\z3}
\nonumber\\&
+\frac{14639991567865673627257}{1186143953148000000}{\z3}
-\frac{3525053}{2700}\pi^2{\z5}
+\frac{13923976647761}{12349260000}{\z5}\nonumber\\&
-\frac{1117}{2}\pi^2{\z7}
+\frac{70846489}{3600}{\z7}
+\frac{163495}{6}{\z9}
-\frac{109759027}{55404000}\pi^8
-\frac{2297724853}{561330000}\pi^6\nonumber\\&
+\frac{11209223967823843}{1467092088000000}\pi^4
+\frac{5097453051063790504945507}{6164983196486730000000}\pi^2\nonumber\\&
-\frac{1347306361480214353376853626224229}{394763602017274079436000000000}\Big)
\nonumber\\&
+{\delta^3}\Big(
%-\frac{285496610438950532826018343{\z3}}{16439955190631280000000}-
%\frac{2117571267706470554744364736321\pi^2}{1367146673652897244800000000}
\frac{693117613075469914070767846953510394027}{43771388191675349927863680000000000}
+\mathrm{trans.\ part}\Big)\label{N13}
\end{align}

\section{Analytical continuation of ${F}^{(4)}$}\label{Sec:ACF4}

In this sections we list out all results for all poles, which we reconstructed from the computed fixed values and which were used to get Eqs.~(\ref{F4Rat}), (\ref{z2}) and (\ref{z3}).

For even values of $\Delta$ we found the following expression for the poles up to regular part
\begin{equation}
\mathrm{AC}\left[F^{(4)}\right]_{\mathrm{even}}=\sum_{i=-7}^{0}\hat{F}^{(4)}_{\w^{i}}
\end{equation}
with $\hat{F}^{(4)}_{\w^{i}}$ equal to
\begin{align}
\hat{F}^{(4)}_{\w^{-7}}=&\ 20480,\label{hFw7}\\
\hat{F}^{(4)}_{\w^{-6}}=&\ 0,\label{hFw6}\\
\hat{F}^{(4)}_{\w^{-5}}=&\ 
-\frac{8192\pi^2}{3}
-12288 \HSH_{-2}
+8192 \HSH_2
,\label{hFw5}\\
\hat{F}^{(4)}_{\w^{-4}}=&\ 
-8192\HSB_{2,0,1}
+9216 \HSH_{-3}
-15360 \HSH_3
+6144 \HSH_{1,-2}
+6144 \HSH_{1,2}\nonumber\\&
+12288 \HSH_{2,1}
-512 \HSH_1\pi^2
+16896\text{z3}
,\label{hFw4}\\
\hat{F}^{(4)}_{\w^{-3}}=&\ 
8192\HSB_{2,0,2}
+49152\HSB_{3,0,1}
-16384\HSB_{1,2,0,1}
-16384\HSB_{2,0,1,1}
-16384\HSB_{2,1,0,1}
-4096 \HSH_{-4}\nonumber\\&
+23040 \HSH_4
-4096 \HSH_{-3,1}
+2048 \HSH_{-2,-2}
-1024 \HSH_{-2,2}
-10240 \HSH_{1,-3}
-10240 \HSH_{1,3}\nonumber\\&
-4096 \HSH_{2,-2}
-17408 \HSH_{2,2}
-28672 \HSH_{3,1}
+8192 \HSH_{1,-2,1}
+12288 \HSH_{1,2,1}
+24576 \HSH_{2,1,1}\nonumber\\&
+\frac{1024}{3} \HSH_{-2}\pi^2
-\frac{256 \HSH_2\pi^2}{3}
+6144\z3 \HSH_1
-\frac{1568\pi^4}{9}
,\label{hFw3}\\
\hat{F}^{(4)}_{\w^{-2}}=&\ 
-8192\HSB_{2,0,3}
-49152\HSB_{3,0,2}
-155648\HSB_{4,0,1}
+16384\HSB_{1,2,0,2}
+65536\HSB_{1,3,0,1}\nonumber\\&
+32768\HSB_{2,0,1,2}
+16384\HSB_{2,0,2,1}
+16384\HSB_{2,1,0,2}
+65536\HSB_{2,2,0,1}
+65536\HSB_{3,0,1,1}\nonumber\\&
+65536\HSB_{3,1,0,1}
-16384\HSB_{1,1,2,0,1}
-16384\HSB_{1,2,0,1,1}
-16384\HSB_{1,2,1,0,1}
-16384\HSB_{2,0,1,1,1}\nonumber\\&
-16384\HSB_{2,1,0,1,1}
-16384\HSB_{2,1,1,0,1}
+\frac{1024}{3}\HSB_{2,0,1}\pi^2
-2560 \HSH_{-5}
-32256 \HSH_5
+8704 \HSH_{-4,1}\nonumber\\&
-2560 \HSH_{-3,-2}
+3584 \HSH_{-3,2}
-2560 \HSH_{-2,-3}
+2048 \HSH_{-2,3}
+12800 \HSH_{1,-4}
+17920 \HSH_{1,4}\nonumber\\&
+4608 \HSH_{2,-3}
+23040 \HSH_{2,3}
+4096 \HSH_{3,-2}
+34816 \HSH_{3,2}
+47104 \HSH_{4,1}
+1024 \HSH_{-2,1,-2}\nonumber\\&
-1024 \HSH_{-2,1,2}
-1024 \HSH_{-2,2,1}
-6144 \HSH_{1,-3,1}
-4096 \HSH_{1,-2,2}
+2048 \HSH_{1,1,-3}
-4096 \HSH_{1,1,3}\nonumber\\&
-14336 \HSH_{1,2,2}
-28672 \HSH_{1,3,1}
-3072 \HSH_{2,-2,1}
-26624 \HSH_{2,1,2}
-38912 \HSH_{2,2,1}
-53248 \HSH_{3,1,1}\nonumber\\&
-4096 \HSH_{1,1,-2,1}
+12288 \HSH_{1,1,2,1}
+24576 \HSH_{1,2,1,1}
+36864 \HSH_{2,1,1,1}
-\frac{2048}{3} \HSH_{-3}\pi^2\nonumber\\&
+\frac{1792 \HSH_3\pi^2}{3}
+256 \HSH_{-2,1}\pi^2
+\frac{1280}{3} \HSH_{1,-2}\pi^2
-\frac{1024}{3} \HSH_{1,2}\pi^2
-\frac{1792}{3} \HSH_{2,1}\pi^2
-2560\z3 \HSH_2
\nonumber\\&
-2816\z3 \HSH_{-2}
+1024\z3 \HSH_{1,1}
-\frac{4192}{45} \HSH_1\pi^4
-\frac{1664\z3\pi^2}{3}
+22848\z5
,\label{hFw2}\\
\hat{F}^{(4)}_{\w^{-1}}=&\ 
+8192\HSB_{2,0,4}
+49152\HSB_{3,0,3}
+155648\HSB_{4,0,2}
+360448\HSB_{5,0,1}
-16384\HSB_{1,2,0,3}\nonumber\\&
-65536\HSB_{1,3,0,2}
-147456\HSB_{1,4,0,1}
-49152\HSB_{2,0,1,3}
-32768\HSB_{2,0,2,2}
-16384\HSB_{2,0,3,1}\nonumber\\&
-16384\HSB_{2,1,0,3}
-65536\HSB_{2,2,0,2}
-147456\HSB_{2,3,0,1}
-98304\HSB_{3,0,1,2}
-65536\HSB_{3,0,2,1}\nonumber\\&
-65536\HSB_{3,1,0,2}
-147456\HSB_{3,2,0,1}
-147456\HSB_{4,0,1,1}
-147456\HSB_{4,1,0,1}
+16384\HSB_{1,1,2,0,2}\nonumber\\&
+32768\HSB_{1,1,3,0,1}
+16384\HSB_{1,2,0,1,2}
+16384\HSB_{1,2,0,2,1}
+16384\HSB_{1,2,1,0,2}
+32768\HSB_{1,2,2,0,1}\nonumber\\&
+32768\HSB_{1,3,0,1,1}
+32768\HSB_{1,3,1,0,1}
+16384\HSB_{2,0,1,1,2}
+16384\HSB_{2,0,1,2,1}
+16384\HSB_{2,0,2,1,1}\nonumber\\&
+16384\HSB_{2,1,0,1,2}
+16384\HSB_{2,1,0,2,1}
+16384\HSB_{2,1,1,0,2}
+32768\HSB_{2,1,2,0,1}
+32768\HSB_{2,2,0,1,1}\nonumber\\&
+32768\HSB_{2,2,1,0,1}
+32768\HSB_{3,0,1,1,1}
+32768\HSB_{3,1,0,1,1}
+32768\HSB_{3,1,1,0,1}
-\frac{1024}{3}\HSB_{2,0,2}\pi^2\nonumber\\&
-2048\HSB_{3,0,1}\pi^2
+\frac{2048}{3}\HSB_{1,2,0,1}\pi^2
+\frac{2048}{3}\HSB_{2,0,1,1}\pi^2
+\frac{2048}{3}\HSB_{2,1,0,1}\pi^2
+4096\z3\HSB_{2,0,1}\nonumber\\&
+9472 \HSH_{-6}
+38656 \HSH_6
-16384 \HSH_{-5,1}
+2816 \HSH_{-4,-2}
-7424 \HSH_{-4,2}
+4096 \HSH_{-3,-3}\nonumber\\&
-4608 \HSH_{-3,3}
+3840 \HSH_{-2,-4}
-2816 \HSH_{-2,4}
-17408 \HSH_{1,-5}
-23552 \HSH_{1,5}
-3328 \HSH_{2,-4}\nonumber\\&
-26624 \HSH_{2,4}
-3584 \HSH_{3,-3}
-35328 \HSH_{3,3}
-5376 \HSH_{4,-2}
-46592 \HSH_{4,2}
-56320 \HSH_{5,1}\nonumber\\&
+3072 \HSH_{-4,1,1}
-1024 \HSH_{-3,-2,1}
-2048 \HSH_{-3,1,-2}
+1024 \HSH_{-3,1,2}
+1024 \HSH_{-3,2,1}\nonumber\\&
-1024 \HSH_{-2,-3,1}
+1024 \HSH_{-2,-2,-2}
-3072 \HSH_{-2,1,-3}
+1024 \HSH_{-2,1,3}
-1536 \HSH_{-2,2,-2}\nonumber\\&
+1024 \HSH_{-2,2,2}
+1024 \HSH_{-2,3,1}
+7168 \HSH_{1,-4,1}
+1024 \HSH_{1,-3,2}
+4096 \HSH_{1,-2,3}
-2048 \HSH_{1,1,-4}\nonumber\\&
+6144 \HSH_{1,1,4}
-3072 \HSH_{1,2,-3}
+13312 \HSH_{1,2,3}
+1024 \HSH_{1,3,-2}
+22528 \HSH_{1,3,2}
+33792 \HSH_{1,4,1}\nonumber\\&
-2048 \HSH_{2,-3,1}
-512 \HSH_{2,-2,2}
-3072 \HSH_{2,1,-3}
+22528 \HSH_{2,1,3}
+30720 \HSH_{2,2,2}
+40960 \HSH_{2,3,1}\nonumber\\&
+4096 \HSH_{3,-2,1}
+1024 \HSH_{3,1,-2}
+40960 \HSH_{3,1,2}
+50176 \HSH_{3,2,1}
+61440 \HSH_{4,1,1}\nonumber\\&
+2048 \HSH_{-2,1,-2,1}
+2048 \HSH_{-2,1,1,-2}
+4096 \HSH_{1,-3,1,1}
+8192 \HSH_{1,1,-3,1}
+6144 \HSH_{1,1,-2,2}\nonumber\\&
-6144 \HSH_{1,1,2,2}
-12288 \HSH_{1,1,3,1}
+2048 \HSH_{1,2,-2,1}
-12288 \HSH_{1,2,1,2}
-18432 \HSH_{1,2,2,1}\nonumber\\&
-24576 \HSH_{1,3,1,1}
+4096 \HSH_{2,-2,1,1}
+2048 \HSH_{2,1,-2,1}
-18432 \HSH_{2,1,1,2}
-24576 \HSH_{2,1,2,1}\nonumber\\&
-30720 \HSH_{2,2,1,1}
-36864 \HSH_{3,1,1,1}
-8192 \HSH_{1,1,-2,1,1}
+\frac{2432}{3} \HSH_{-4}\pi^2
-1408 \HSH_4\pi^2\nonumber\\&
-\frac{1792}{3} \HSH_{-3,1}\pi^2
+\frac{512}{3} \HSH_{-2,-2}\pi^2
-\frac{512}{3} \HSH_{-2,2}\pi^2
-768 \HSH_{1,-3}\pi^2
+\frac{2816}{3} \HSH_{1,3}\pi^2\nonumber\\&
-256 \HSH_{2,-2}\pi^2
+1152 \HSH_{2,2}\pi^2
+\frac{5120}{3} \HSH_{3,1}\pi^2
+\frac{512}{3} \HSH_{-2,1,1}\pi^2
+512 \HSH_{1,-2,1}\pi^2\nonumber\\&
+\frac{512}{3} \HSH_{1,1,-2}\pi^2
-\frac{1024}{3} \HSH_{1,1,2}\pi^2
-\frac{2560}{3} \HSH_{1,2,1}\pi^2
-\frac{4096}{3} \HSH_{2,1,1}\pi^2
+4864\z3 \HSH_{-3}\nonumber\\&
-1536\z3 \HSH_3
-1792\z3 \HSH_{-2,1}
-2304\z3 \HSH_{1,-2}
+3072\z3 \HSH_{1,2}
+\frac{928}{45} \HSH_{-2}\pi^4
+\frac{416 \HSH_2\pi^4}{15}\nonumber\\&
-\frac{832}{45} \HSH_{1,1}\pi^4
+15936\z5 \HSH_1
-\frac{2048}{3}\HSH_1\z3 \pi^2
+640\z3^2
-\frac{44672\pi^6}{2835}
,\label{hFw1}\\
\hat{F}^{(4)}_{\w^{0}}=&\
-8192\HSB_{2,0,5}
-16384\z3\HSB_{3,0,1}
-49152\HSB_{3,0,4}
-155648\HSB_{4,0,3}
-360448\HSB_{5,0,2}\nonumber\\&
-696320\HSB_{6,0,1}
+16384\HSB_{1,2,0,4}
+65536\HSB_{1,3,0,3}
+147456\HSB_{1,4,0,2}
+262144\HSB_{1,5,0,1}\nonumber\\&
+65536\HSB_{2,0,1,4}
+49152\HSB_{2,0,2,3}
+32768\HSB_{2,0,3,2}
+16384\HSB_{2,0,4,1}
+16384\HSB_{2,1,0,4}\nonumber\\&
+65536\HSB_{2,2,0,3}
+147456\HSB_{2,3,0,2}
+262144\HSB_{2,4,0,1}
+131072\HSB_{3,0,1,3}
+98304\HSB_{3,0,2,2}\nonumber\\&
+65536\HSB_{3,0,3,1}
+65536\HSB_{3,1,0,3}
+147456\HSB_{3,2,0,2}
+262144\HSB_{3,3,0,1}
+196608\HSB_{4,0,1,2}\nonumber\\&
+147456\HSB_{4,0,2,1}
+147456\HSB_{4,1,0,2}
+262144\HSB_{4,2,0,1}
+262144\HSB_{5,0,1,1}
+262144\HSB_{5,1,0,1}\nonumber\\&
-16384\HSB_{1,1,2,0,3}
-32768\HSB_{1,1,3,0,2}
-49152\HSB_{1,1,4,0,1}
-16384\HSB_{1,2,0,1,3}
-16384\HSB_{1,2,0,2,2}\nonumber\\&
-16384\HSB_{1,2,0,3,1}
-16384\HSB_{1,2,1,0,3}
-32768\HSB_{1,2,2,0,2}
-49152\HSB_{1,2,3,0,1}
-32768\HSB_{1,3,0,1,2}\nonumber\\&
-32768\HSB_{1,3,0,2,1}
-32768\HSB_{1,3,1,0,2}
-49152\HSB_{1,3,2,0,1}
-49152\HSB_{1,4,0,1,1}
-49152\HSB_{1,4,1,0,1}\nonumber\\&
-16384\HSB_{2,0,1,1,3}
-16384\HSB_{2,0,1,2,2}
-16384\HSB_{2,0,1,3,1}
-16384\HSB_{2,0,2,1,2}
-16384\HSB_{2,0,2,2,1}\nonumber\\&
-16384\HSB_{2,0,3,1,1}
-16384\HSB_{2,1,0,1,3}
-16384\HSB_{2,1,0,2,2}
-16384\HSB_{2,1,0,3,1}
-16384\HSB_{2,1,1,0,3}\nonumber\\&
-32768\HSB_{2,1,2,0,2}
-49152\HSB_{2,1,3,0,1}
-32768\HSB_{2,2,0,1,2}
-32768\HSB_{2,2,0,2,1}
-32768\HSB_{2,2,1,0,2}\nonumber\\&
-49152\HSB_{2,2,2,0,1}
-49152\HSB_{2,3,0,1,1}
-49152\HSB_{2,3,1,0,1}
-32768\HSB_{3,0,1,1,2}
-32768\HSB_{3,0,1,2,1}\nonumber\\&
-32768\HSB_{3,0,2,1,1}
-32768\HSB_{3,1,0,1,2}
-32768\HSB_{3,1,0,2,1}
-32768\HSB_{3,1,1,0,2}
-49152\HSB_{3,1,2,0,1}\nonumber\\&
-49152\HSB_{3,2,0,1,1}
-49152\HSB_{3,2,1,0,1}
-49152\HSB_{4,0,1,1,1}
-49152\HSB_{4,1,0,1,1}
-49152\HSB_{4,1,1,0,1}\nonumber\\&
+\frac{448}{15}\HSB_{2,0,1}\pi^4
+\frac{1024}{3}\HSB_{2,0,3}\pi^2
+2048\HSB_{3,0,2}\pi^2
+\frac{19456}{3}\HSB_{4,0,1}\pi^2
-\frac{2048}{3}\HSB_{1,2,0,2}\pi^2\nonumber\\&
-\frac{8192}{3}\HSB_{1,3,0,1}\pi^2
-\frac{4096}{3}\HSB_{2,0,1,2}\pi^2
-\frac{2048}{3}\HSB_{2,0,2,1}\pi^2
-\frac{2048}{3}\HSB_{2,1,0,2}\pi^2
-\frac{8192}{3}\HSB_{2,2,0,1}\pi^2\nonumber\\&
-\frac{8192}{3}\HSB_{3,0,1,1}\pi^2
-\frac{8192}{3}\HSB_{3,1,0,1}\pi^2
+\frac{2048}{3}\HSB_{1,1,2,0,1}\pi^2
+\frac{2048}{3}\HSB_{1,2,0,1,1}\pi^2\nonumber\\&
+\frac{2048}{3}\HSB_{1,2,1,0,1}\pi^2
+\frac{2048}{3}\HSB_{2,0,1,1,1}\pi^2
+\frac{2048}{3}\HSB_{2,1,0,1,1}\pi^2
+\frac{2048}{3}\HSB_{2,1,1,0,1}\pi^2\nonumber\\&
-4096\z3\HSB_{2,0,2}
+4096\z3\HSB_{1,2,0,1}
+4096\z3\HSB_{2,0,1,1}
+4096\z3\HSB_{2,1,0,1}
-13120 \HSH_{-7}\nonumber\\&
-38976 \HSH_7
+24832 \HSH_{-6,1}
-2816 \HSH_{-5,-2}
+15104 \HSH_{-5,2}
-4992 \HSH_{-4,-3}
+7808 \HSH_{-4,3}\nonumber\\&
-6272 \HSH_{-3,-4}
+6272 \HSH_{-3,4}
-5888 \HSH_{-2,-5}
+2560 \HSH_{-2,5}
+22400 \HSH_{1,-6}
+24960 \HSH_{1,6}\nonumber\\&
+2816 \HSH_{2,-5}
+26624 \HSH_{2,5}
-256 \HSH_{3,-4}
+32256 \HSH_{3,4}
+4480 \HSH_{4,-3}
+40064 \HSH_{4,3}\nonumber\\&
+6272 \HSH_{5,-2}
+48256 \HSH_{5,2}
+54016 \HSH_{6,1}
-11264 \HSH_{-5,1,1}
+2048 \HSH_{-4,-2,1}
+2304 \HSH_{-4,1,-2}\nonumber\\&
-3840 \HSH_{-4,1,2}
-3840 \HSH_{-4,2,1}
+3072 \HSH_{-3,-3,1}
-1024 \HSH_{-3,-2,-2}
+1024 \HSH_{-3,-2,2}\nonumber\\&
+4608 \HSH_{-3,1,-3}
-2048 \HSH_{-3,1,3}
+2048 \HSH_{-3,2,-2}
-1024 \HSH_{-3,2,2}
-2048 \HSH_{-3,3,1}\nonumber\\&
+3072 \HSH_{-2,-4,1}
-1024 \HSH_{-2,-3,-2}
+1024 \HSH_{-2,-3,2}
-1536 \HSH_{-2,-2,-3}
+512 \HSH_{-2,-2,3}\nonumber\\&
+4864 \HSH_{-2,1,-4}
-768 \HSH_{-2,1,4}
+3072 \HSH_{-2,2,-3}
-768 \HSH_{-2,2,3}
+1280 \HSH_{-2,3,-2}\nonumber\\&
-768 \HSH_{-2,3,2}
-1280 \HSH_{-2,4,1}
-13312 \HSH_{1,-5,1}
-2560 \HSH_{1,-4,2}
-2048 \HSH_{1,-3,3}
-6144 \HSH_{1,-2,4}\nonumber\\&
+1024 \HSH_{1,1,-5}
-6144 \HSH_{1,1,5}
+6144 \HSH_{1,2,-4}
-10752 \HSH_{1,2,4}
+3072 \HSH_{1,3,-3}
-16384 \HSH_{1,3,3}\nonumber\\&
-1536 \HSH_{1,4,-2}
-23040 \HSH_{1,4,2}
-30720 \HSH_{1,5,1}
+4864 \HSH_{2,-4,1}
+6144 \HSH_{2,-3,2}
+768 \HSH_{2,-2,3}\nonumber\\&
+6144 \HSH_{2,1,-4}
-16896 \HSH_{2,1,4}
+4608 \HSH_{2,2,-3}
-22016 \HSH_{2,2,3}
-512 \HSH_{2,3,-2}
-28160 \HSH_{2,3,2}\nonumber\\&
-35328 \HSH_{2,4,1}
+512 \HSH_{3,-3,1}
-1536 \HSH_{3,-2,2}
+3072 \HSH_{3,1,-3}
-28672 \HSH_{3,1,3}
-512 \HSH_{3,2,-2}\nonumber\\&
-34304 \HSH_{3,2,2}
-40960 \HSH_{3,3,1}
-6400 \HSH_{4,-2,1}
-1536 \HSH_{4,1,-2}
-41472 \HSH_{4,1,2}
-47616 \HSH_{4,2,1}\nonumber\\&
-54272 \HSH_{5,1,1}
+3072 \HSH_{-4,1,1,1}
-1024 \HSH_{-3,-2,1,1}
-3072 \HSH_{-3,1,-2,1}
-2048 \HSH_{-3,1,1,-2}\nonumber\\&
-1024 \HSH_{-2,-3,1,1}
+1024 \HSH_{-2,-2,-2,1}
-4096 \HSH_{-2,1,-3,1}
-2048 \HSH_{-2,1,-2,2}\nonumber\\&
-3072 \HSH_{-2,1,1,-3}
-1024 \HSH_{-2,1,2,-2}
-2048 \HSH_{-2,2,-2,1}
-1024 \HSH_{-2,2,1,-2}
-2048 \HSH_{1,-4,1,1}\nonumber\\&
-4096 \HSH_{1,-3,1,2}
-4096 \HSH_{1,-3,2,1}
+2048 \HSH_{1,-2,1,3}
+2048 \HSH_{1,-2,3,1}
-8192 \HSH_{1,1,-4,1}\nonumber\\&
-10240 \HSH_{1,1,-3,2}
-6144 \HSH_{1,1,-2,3}
-2048 \HSH_{1,1,2,-3}
+3072 \HSH_{1,1,2,3}
+6144 \HSH_{1,1,3,2}\nonumber\\&
+9216 \HSH_{1,1,4,1}
-6144 \HSH_{1,2,-3,1}
-3072 \HSH_{1,2,-2,2}
-2048 \HSH_{1,2,1,-3}
+6144 \HSH_{1,2,1,3}\nonumber\\&
+9216 \HSH_{1,2,2,2}
+12288 \HSH_{1,2,3,1}
+12288 \HSH_{1,3,1,2}
+15360 \HSH_{1,3,2,1}
+18432 \HSH_{1,4,1,1}\nonumber\\&
-10240 \HSH_{2,-3,1,1}
-4096 \HSH_{2,-2,1,2}
-4096 \HSH_{2,-2,2,1}
-6144 \HSH_{2,1,-3,1}
-3072 \HSH_{2,1,-2,2}\nonumber\\&
-2048 \HSH_{2,1,1,-3}
+9216 \HSH_{2,1,1,3}
+12288 \HSH_{2,1,2,2}
+15360 \HSH_{2,1,3,1}
-1024 \HSH_{2,2,-2,1}\nonumber\\&
+15360 \HSH_{2,2,1,2}
+18432 \HSH_{2,2,2,1}
+21504 \HSH_{2,3,1,1}
-1024 \HSH_{3,-2,1,1}
+18432 \HSH_{3,1,1,2}\nonumber\\&
+21504 \HSH_{3,1,2,1}
+24576 \HSH_{3,2,1,1}
+27648 \HSH_{4,1,1,1}
+2048 \HSH_{-2,1,-2,1,1}
+2048 \HSH_{-2,1,1,-2,1}\nonumber\\&
+4096 \HSH_{1,-3,1,1,1}
+12288 \HSH_{1,1,-3,1,1}
+8192 \HSH_{1,1,-2,1,2}
+8192 \HSH_{1,1,-2,2,1}\nonumber\\&
+4096 \HSH_{1,2,-2,1,1}
+4096 \HSH_{2,-2,1,1,1}
+4096 \HSH_{2,1,-2,1,1}
-8192 \HSH_{1,1,-2,1,1,1}\nonumber\\&
-704 \HSH_{-5}\pi^2
-\frac{448}{3}\z3 \HSH_{-2}\pi^2
+544\z3 \HSH_2\pi^2
+\frac{6688 \HSH_5\pi^2}{3}
+704 \HSH_{-4,1}\pi^2
-\frac{512}{3} \HSH_{-3,-2}\pi^2\nonumber\\&
+448 \HSH_{-3,2}\pi^2
-\frac{640}{3} \HSH_{-2,-3}\pi^2
+\frac{64}{3} \HSH_{-2,3}\pi^2
+\frac{2752}{3} \HSH_{1,-4}\pi^2
-512\z3 \HSH_{1,1}\pi^2\nonumber\\&
-\frac{4352}{3} \HSH_{1,4}\pi^2
+448 \HSH_{2,-3}\pi^2
-\frac{4736}{3} \HSH_{2,3}\pi^2
+\frac{448}{3} \HSH_{3,-2}\pi^2
-\frac{6208}{3} \HSH_{3,2}\pi^2\nonumber\\&
-\frac{8128}{3} \HSH_{4,1}\pi^2
-\frac{1280}{3} \HSH_{-3,1,1}\pi^2
+\frac{256}{3} \HSH_{-2,-2,1}\pi^2
-\frac{128}{3} \HSH_{-2,1,2}\pi^2
-\frac{128}{3} \HSH_{-2,2,1}\pi^2\nonumber\\&
-\frac{2560}{3} \HSH_{1,-3,1}\pi^2
-512 \HSH_{1,-2,2}\pi^2
-256 \HSH_{1,1,-3}\pi^2
+512 \HSH_{1,1,3}\pi^2
-\frac{256}{3} \HSH_{1,2,-2}\pi^2\nonumber\\&
+\frac{2816}{3} \HSH_{1,2,2}\pi^2
+1536 \HSH_{1,3,1}\pi^2
-\frac{896}{3} \HSH_{2,-2,1}\pi^2
-\frac{256}{3} \HSH_{2,1,-2}\pi^2
+\frac{4352}{3} \HSH_{2,1,2}\pi^2\nonumber\\&
+\frac{5888}{3} \HSH_{2,2,1}\pi^2
+2560 \HSH_{3,1,1}\pi^2
+512 \HSH_{1,-2,1,1}\pi^2
+\frac{512}{3} \HSH_{1,1,-2,1}\pi^2
-512 \HSH_{1,1,2,1}\pi^2\nonumber\\&
-1024 \HSH_{1,2,1,1}\pi^2
-1536 \HSH_{2,1,1,1}\pi^2
-6336\z3 \HSH_{-4}
+4224\z3 \HSH_4
+2432\z3 \HSH_{-3,1}\nonumber\\&
-768\z3 \HSH_{-2,-2}
+1664\z3 \HSH_{-2,2}
+2432\z3 \HSH_{1,-3}
-6272\z3 \HSH_{1,3}
+1408\z3 \HSH_{2,-2}\nonumber\\&
-3328\z3 \HSH_{2,2}
-128\z3 \HSH_{3,1}
-1024\z3 \HSH_{-2,1,1}
-256\z3 \HSH_{1,-2,1}
+3584\z3 \HSH_{1,1,2}\nonumber\\&
+512\z3 \HSH_{1,2,1}
-2560\z3 \HSH_{2,1,1}
-\frac{200}{9} \HSH_{-3}\pi^4
+\frac{1424}{45} \HSH_3\pi^4
-\frac{368}{45} \HSH_{-2,1}\pi^4
+\frac{512}{45} \HSH_{1,-2}\pi^4\nonumber\\&
-\frac{1264}{45} \HSH_{1,2}\pi^4
-\frac{2272}{45} \HSH_{2,1}\pi^4
-3104\z5 \HSH_{-2}
-10240\z5 \HSH_2
+7040\z5 \HSH_{1,1}
+2240\z3^2 \HSH_1\nonumber\\&
-\frac{28936}{2835} \HSH_1\pi^6
-560\z5\pi^2
-\frac{212\z3\pi^4}{9}
+30198\z7\label{hFw0}
\end{align}

For odd values of $\Delta$ we found the following expression for the poles up to the lowest pole $\w^{-1}$
\begin{equation}
\mathrm{AC}\left[F^{(4)}\right]_{\mathrm{odd}}=\sum_{i=-7}^{-1}\check{F}^{(4)}_{\w^{i}}
\end{equation}
with 
\begin{align}
\check{F}^{(4)}_{\w^{-7}}=&\ 20480,\label{cFw7}\\
\check{F}^{(4)}_{\w^{-6}}=&\ 0,\label{cFw6}\\
\check{F}^{(4)}_{\w^{-5}}=&\ 
+12288S_{-2}
-8192S_2
2048\pi^2
,\label{cFw5}\\
\check{F}^{(4)}_{\w^{-4}}=&\ 
8192\HSB_{2,0,1}
-9216S_{-3}
+17408S_3
-6144S_{1,-2}
-6144S_{1,2}
-12288S_{2,1}\nonumber\\&
+512S_1\pi^2
-16896\z3
,\label{cFw4}\\
\check{F}^{(4)}_{\w^{-3}}=&\ 
-8192\HSB_{2,0,2}
-49152\HSB_{3,0,1}
+16384\HSB_{1,2,0,1}
+16384\HSB_{2,0,1,1}
+16384\HSB_{2,1,0,1}\nonumber\\&
+6144\HSH_{-4}
+4096\HSH_{-3,1}
+2048\HSH_{-2,-2}
-1024\HSH_{-2,2}
+10240\HSH_{1,-3}
+10240\HSH_{1,3}\nonumber\\&
+2048\HSH_{2,-2}
+17408\HSH_{2,2}
+28672\HSH_{3,1}
-8192\HSH_{1,-2,1}
-12288\HSH_{1,2,1}
-24576\HSH_{2,1,1},\nonumber\\&
+\frac{1024}{3}\HSH_{-2}\pi^2
-\frac{256\HSH_2\pi^2}{3}
-5120\z3\HSH_1-28160\HSH_4
\frac{10912\pi^4}{45}
,\label{cFw3}\\
\check{F}^{(4)}_{\w^{-2}}=&\ 
+8192\HSB_{2,0,3}
+49152\HSB_{3,0,2}
+155648\HSB_{4,0,1}
-16384\HSB_{1,2,0,2}
-65536\HSB_{1,3,0,1}\nonumber\\&
-32768\HSB_{2,0,1,2}
-16384\HSB_{2,0,2,1}
-16384\HSB_{2,1,0,2}
-65536\HSB_{2,2,0,1}
-65536\HSB_{3,0,1,1}\nonumber\\&
-65536\HSB_{3,1,0,1}
+16384\HSB_{1,1,2,0,1}
+16384\HSB_{1,2,0,1,1}
+16384\HSB_{1,2,1,0,1}
+16384\HSB_{2,0,1,1,1}\nonumber\\&
+16384\HSB_{2,1,0,1,1}
+16384\HSB_{2,1,1,0,1}
-\frac{1024}{3}\pi^2\HSB_{2,0,1}
-2560\HSH_{-5}
+40448\HSH_5
-6656\HSH_{-4,1}\nonumber\\&
-2560\HSH_{-3,-2}
-512\HSH_{-3,2}
-2560\HSH_{-2,-3}
+2048\HSH_{-2,3}
-12800\HSH_{1,-4}
-17920\HSH_{1,4}\nonumber\\&
-1536\HSH_{2,-3}
-23040\HSH_{2,3}
-2048\HSH_{3,-2}
-34816\HSH_{3,2}
-47104\HSH_{4,1}
+1024\HSH_{-2,1,-2}\nonumber\\&
-1024\HSH_{-2,1,2}
-1024\HSH_{-2,2,1}
+6144\HSH_{1,-3,1}
+4096\HSH_{1,-2,2}
-2048\HSH_{1,1,-3}
+4096\HSH_{1,1,3}\nonumber\\&
+14336\HSH_{1,2,2}
+28672\HSH_{1,3,1}
+1024\HSH_{2,-2,1}
+26624\HSH_{2,1,2}
+38912\HSH_{2,2,1}
+53248\HSH_{3,1,1}\nonumber\\&
+4096\HSH_{1,1,-2,1}
-12288\HSH_{1,1,2,1}
-24576\HSH_{1,2,1,1}
-36864\HSH_{2,1,1,1},
-256\HSH_{-3}\pi^2\nonumber\\&
-\frac{1792}{3}\HSH_3\pi^2
+256\HSH_{-2,1}\pi^2
-\frac{1280}{3}\HSH_{1,-2}\pi^2
+\frac{1024}{3}\HSH_{1,2}\pi^2
+\frac{1792}{3}\HSH_{2,1}\pi^2\nonumber\\&
-2816\z3\HSH_{-2}
+3072\z3\HSH_2
-1024\z3\HSH_{1,1}
\frac{4192}{45}\HSH_1\pi^4
-27200\z5
+\frac{256\z3\pi^2}{3}
,\label{cFw2}\\
\check{F}^{(4)}_{\w^{-1}}=&\ 
-8192\HSB_{2,0,4}
-49152\HSB_{3,0,3}
-155648\HSB_{4,0,2}
-360448\HSB_{5,0,1}
+16384\HSB_{1,2,0,3}\nonumber\\&
+65536\HSB_{1,3,0,2}
+147456\HSB_{1,4,0,1}
+49152\HSB_{2,0,1,3}
+32768\HSB_{2,0,2,2}
+16384\HSB_{2,0,3,1}\nonumber\\&
+16384\HSB_{2,1,0,3}
+65536\HSB_{2,2,0,2}
+147456\HSB_{2,3,0,1}
+98304\HSB_{3,0,1,2}
+65536\HSB_{3,0,2,1}\nonumber\\&
+65536\HSB_{3,1,0,2}
+147456\HSB_{3,2,0,1}
+147456\HSB_{4,0,1,1}
+147456\HSB_{4,1,0,1}
-16384\HSB_{1,1,2,0,2}\nonumber\\&
-32768\HSB_{1,1,3,0,1}
-16384\HSB_{1,2,0,1,2}
-16384\HSB_{1,2,0,2,1}
-16384\HSB_{1,2,1,0,2}
-32768\HSB_{1,2,2,0,1}\nonumber\\&
-32768\HSB_{1,3,0,1,1}
-32768\HSB_{1,3,1,0,1}
-16384\HSB_{2,0,1,1,2}
-16384\HSB_{2,0,1,2,1}
-16384\HSB_{2,0,2,1,1}\nonumber\\&
-16384\HSB_{2,1,0,1,2}
-16384\HSB_{2,1,0,2,1}
-16384\HSB_{2,1,1,0,2}
-32768\HSB_{2,1,2,0,1}
-32768\HSB_{2,2,0,1,1}\nonumber\\&
-32768\HSB_{2,2,1,0,1}
-32768\HSB_{3,0,1,1,1}
-32768\HSB_{3,1,0,1,1}
-32768\HSB_{3,1,1,0,1}
\frac{1024}{3}\pi^2\HSB_{2,0,2}\nonumber\\&
+2048\pi^2\HSB_{3,0,1}
-\frac{2048}{3}\pi^2\HSB_{1,2,0,1}
-\frac{2048}{3}\pi^2\HSB_{2,0,1,1}
-\frac{2048}{3}\pi^2\HSB_{2,1,0,1}
-4096\z3\HSB_{2,0,1}\nonumber\\&
-1792\HSH_{-6}
-50944\HSH_6
+12288\HSH_{-5,1}
+2816\HSH_{-4,-2}
+3328\HSH_{-4,2}
+4096\HSH_{-3,-3}\nonumber\\&
-512\HSH_{-3,3}
+3840\HSH_{-2,-4}
-1792\HSH_{-2,4}
+17408\HSH_{1,-5}
+23552\HSH_{1,5}
+256\HSH_{2,-4}\nonumber\\&
+26624\HSH_{2,4}
+512\HSH_{3,-3}
+35328\HSH_{3,3}
+3840\HSH_{4,-2}
+46592\HSH_{4,2}
+58368\HSH_{5,1}\nonumber\\&
-3072\HSH_{-4,1,1}
-1024\HSH_{-3,-2,1}
-2048\HSH_{-3,1,-2}
+1024\HSH_{-3,1,2}
+1024\HSH_{-3,2,1}\nonumber\\&
-1024\HSH_{-2,-3,1}
-1024\HSH_{-2,-2,-2}
-3072\HSH_{-2,1,-3}
+1024\HSH_{-2,1,3}
-1536\HSH_{-2,2,-2}\nonumber\\&
+1024\HSH_{-2,2,2}
+1024\HSH_{-2,3,1}
-7168\HSH_{1,-4,1}
-1024\HSH_{1,-3,2}
-4096\HSH_{1,-2,3}\nonumber\\&
+2048\HSH_{1,1,-4}
-6144\HSH_{1,1,4}
+3072\HSH_{1,2,-3}
-13312\HSH_{1,2,3}
-1024\HSH_{1,3,-2}
-22528\HSH_{1,3,2}\nonumber\\&
-33792\HSH_{1,4,1}
+4096\HSH_{2,-3,1}
+1536\HSH_{2,-2,2}
+3072\HSH_{2,1,-3}
-22528\HSH_{2,1,3}
-30720\HSH_{2,2,2}\nonumber\\&
-40960\HSH_{2,3,1}
-2048\HSH_{3,-2,1}
-1024\HSH_{3,1,-2}
-40960\HSH_{3,1,2}
-50176\HSH_{3,2,1}
-61440\HSH_{4,1,1}\nonumber\\&
+2048\HSH_{-2,1,-2,1}
+2048\HSH_{-2,1,1,-2}
-4096\HSH_{1,-3,1,1}
-8192\HSH_{1,1,-3,1}
-6144\HSH_{1,1,-2,2}\nonumber\\&
+6144\HSH_{1,1,2,2}
+12288\HSH_{1,1,3,1}
-2048\HSH_{1,2,-2,1}
+12288\HSH_{1,2,1,2}
+18432\HSH_{1,2,2,1}\nonumber\\&
+24576\HSH_{1,3,1,1}
-4096\HSH_{2,-2,1,1}
-2048\HSH_{2,1,-2,1}
+18432\HSH_{2,1,1,2}
+24576\HSH_{2,1,2,1}\nonumber\\&
+30720\HSH_{2,2,1,1}
+36864\HSH_{3,1,1,1}
+8192\HSH_{1,1,-2,1,1}
+\frac{256}{3}\HSH_{-4}\pi^2
+\frac{4352\HSH_4\pi^2}{3}\nonumber\\&
-\frac{256}{3}\HSH_{-3,1}\pi^2
-\frac{512}{3}\HSH_{-2,2}\pi^2
+768\HSH_{1,-3}\pi^2
-\frac{2816}{3}\HSH_{1,3}\pi^2
+\frac{1024}{3}\HSH_{2,-2}\pi^2\nonumber\\&
-1152\HSH_{2,2}\pi^2
-\frac{5120}{3}\HSH_{3,1}\pi^2
+\frac{512}{3}\HSH_{-2,1,1}\pi^2
-512\HSH_{1,-2,1}\pi^2
-\frac{512}{3}\HSH_{1,1,-2}\pi^2\nonumber\\&
+\frac{1024}{3}\HSH_{1,1,2}\pi^2
+\frac{2560}{3}\HSH_{1,2,1}\pi^2
+\frac{4096}{3}\HSH_{2,1,1}\pi^2
+2560\z3\HSH_{-3}
+1024\z3\HSH_3\nonumber\\&
-1792\z3\HSH_{-2,1}
+2304\z3\HSH_{1,-2}
-3072\z3\HSH_{1,2}
+\frac{2336}{45}\HSH_{-2}\pi^4
-\frac{1088\HSH_2\pi^4}{45}\nonumber\\&
+\frac{832}{45}\HSH_{1,1}\pi^4
-15936\z5\HSH_1
+\frac{2048}{3}\z3\HSH_1\pi^2
+1152\z3^2
+\frac{72544}{2835}\pi^6
,\label{cFw1}
\end{align}

%\paragraph{Note added.} This is also a good position for notes added after the paper has been written.


%\bibliography{NNNLABFKLv1.bib}
%
%\end{document}

\end{document}